\RequirePackage{snapshot}
\pdfoutput=1
\documentclass{sig-alternate-2013}

\usepackage{subfigure}
\usepackage{graphicx}
\usepackage{epsfig}
\usepackage{psfrag}
\usepackage{amssymb}
\usepackage{color}
\usepackage{algorithm}
\usepackage[noend]{algorithmic}
\usepackage{url}
\usepackage{booktabs}
\usepackage{multirow}
\usepackage{epstopdf}
\usepackage{cite}
\usepackage{amsmath}
\usepackage{fancyhdr}
\usepackage{hhline}
\usepackage{balance}

\usepackage[table]{xcolor}
\usepackage[normalem]{ulem}
\usepackage{soul}
\definecolor{gray}{rgb}{0.745,0.745,0.745}
\newcommand{\grayCol}{gray!45}
\newcommand{\whiteCol}{white}

\newcommand{\comment}[1]{{\textbf{\em$[$#1$]$}}}

\newtheorem{theorem}{Theorem}

\newtheorem{observation}{Observation}

\newtheorem{definition}{Definition}

\newenvironment{myitemize}{\begin{list}{$\bullet$}{\addtolength{\itemsep}{-0.09in}\addtolength{\topsep}{-0.05in}\renewcommand{\leftmargin}{0.23in}}}{\end{list}}

\def\TextUlineRLB{%
  \def\\{\relax}%
  \protected@edef\tmp{\OrigTitle}%
  \expandafter\uline\expandafter{\tmp}%
}

\usepackage{etoolbox}
\makeatletter
\def\@copyrightspace{\relax}
\makeatother

\newboolean{JSACsingle}
\setboolean{JSACsingle}{false}

\newboolean{JSACsubmission}
\setboolean{JSACsubmission}{false}

\newboolean{JSACdouble}
\setboolean{JSACdouble}{false}

\newboolean{IncludeAppendix}
\setboolean{IncludeAppendix}{true}

\def\sharedaffiliation{%
\end{tabular}
\begin{tabular}{c}}
\ifIncludeAppendix
\long\def\symbolfootnote[#1]#2{\begingroup%
\def\thefootnote{\fnsymbol{footnote}}\footnote[#1]{#2}\endgroup}
\definecolor{blue}{rgb}{0,0,0}

\fi

\newcommand*{\refname}{Bibliography}

\begin{document}

\ifJSACsubmission
\ifJSACsingle
\title{Movers and Shakers: Kinetic Energy Harvesting for the Internet of Things}
\else
\title{Movers and Shakers: \\ Kinetic Energy Harvesting for the Internet of Things}
\fi
\else
\title{Movers and Shakers: \\ Kinetic Energy Harvesting for the Internet of Things\vspace{-.3in}}
\fi

\ifJSACsubmission
\ifJSACsingle
\author{Maria Gorlatova, John Sarik, Guy Grebla, Mina Cong, Ioannis Kymissis,
  \\Gil Zussman \vspace{-50pt}
\thanks{M. Gorlatova, J. Sarik, G. Grebla, M. Cong, I. Kymissis, and
   G. Zussman are with
   the Department of Electrical Engineering, Columbia
   University, New York, NY 10027. E-mail: \{mag2206, jcs2160, gg2519, mc3415\}@columbia.edu, \{johnkym@ee,
     gil@ee\}.columbia.edu}
 \thanks{Preliminary version of this paper will appear in Proc. ACM
   SIGMETRICS'14~\cite{MoversShakers2014}.} 
}
\fi
\ifJSACdouble
\author{Maria Gorlatova, John Sarik, Guy Grebla, Mina Cong, Ioannis Kymissis, Gil Zussman \thanks{M. Gorlatova, J. Sarik, G. Grebla, M. Cong, I. Kymissis, and
   G. Zussman are with
   the Department of Electrical Engineering, Columbia
   University, New York, NY 10027. E-mail: \{mag2206, jcs2160, gg2519, mc3415\}@columbia.edu, \{johnkym@ee,
     gil@ee\}.columbia.edu}
 \thanks{Preliminary version of this paper will appear in Proc. ACM SIGMETRICS'14~\cite{MoversShakers2014}.} 
}
\fi
\else

\numberofauthors{6}
    \author{
      \sharedaffiliation
    Maria Gorlatova, John Sarik, Guy Grebla, Mina Cong, Ioannis Kymissis, Gil
    Zussman \vspace{.05in}\\
     \affaddr{Department of Electrical Engineering}  \\
      \affaddr{Columbia University }   \\
      \affaddr{New York, NY, 10027, USA } \vspace{.05in}\\
\{mag2206@, jcs2160@,  guy@ee., mc3415@, johnkym@ee., gil@ee.\}columbia.edu
      \vspace*{-.15in}
          }
%

\fi


\maketitle

\begin{abstract}
Numerous energy harvesting wireless devices that will serve as building blocks for the Internet of Things (IoT) are currently under development.
However, there is still only limited understanding of the properties of various energy sources and their impact on energy harvesting adaptive algorithms. Hence, we focus on \emph{characterizing the kinetic (motion) energy
that can be harvested by a wireless 
node with an IoT form factor} and on
\emph{developing energy allocation algorithms} for such nodes. In this paper,
we describe methods for estimating harvested energy from acceleration
traces. To characterize the energy availability associated with 
\ifJSACsubmission
\emph{specific human activities} 
\else
specific human activities
\fi
(e.g., relaxing, walking, cycling), we analyze a motion dataset with over 40 participants.
Based on acceleration measurements that we collected for over 200 hours, we
study energy generation processes associated with 
\ifJSACsubmission
\emph{day-long human routines}.
\else
day-long human routines.
\fi
 We also briefly summarize our experiments with 
\ifJSACsubmission
\emph{moving objects}.
\else
moving objects.
\fi
We develop energy allocation algorithms that \emph{take into account
practical IoT node design considerations}, and evaluate the algorithms using the collected measurements.
Our observations provide insights into the design of motion energy harvesters,
IoT nodes, and energy harvesting adaptive algorithms.
\end{abstract}

\ifJSACsubmission
\vspace{1mm} \noindent {\bf Keywords:} Energy harvesting; motion energy;
measurements; low-power networking; Algorithms; Internet of Things.
\else




\fi

\section{Introduction}


Advances in the areas of solar, kinetic, 
and thermal energy harvesting
as well as in low-power wireless communications 
will soon enable the realization of \emph{self-sustainable wireless devices} \cite{paradiso2005esm,Gorlatova_Enhants_wircom,Kansal06Journ,yerva2012grafting}. 
These devices can compose networks of rechargeable sensors \cite{Kansal06Journ,yerva2012grafting}, active tags \cite{Gorlatova_Enhants_wircom}, or computational RFIDs \cite{Gummeson10}.
\ifJSACsubmission
Such networks will serve as building blocks for 
emerging Internet-of-Things (IoT) applications, including supply chain management 
and wearable computing.
\else
Such networks will serve as building blocks for 
emerging Internet-of-Things (IoT) applications (e.g., supply chain management 
and wearable computing).
\fi


Two promising energy sources for IoT nodes are~\emph{light and motion}.\footnote{The power available from \emph{RF harvesting} is \emph{100~times less than the power available from indoor light~\cite{Vullers2009684}}. 
\emph{Thermal gradients} can provide substantial power in industrial applications, but 
 are currently impractical for non-industrial IoT applications.}
Accordingly, extensive effort has been dedicated to the design of solar cells and kinetic energy harvesters (e.g.,~\cite{kymissis98,ApplePatentShake,nPowerPEG,Mitcheson2008}). Moreover, the design of energy harvesting-adaptive communication and networking algorithms recently gained extensive attention~\cite{Kansal06Journ,Chen2012Asymptotically,Gorlatova_TMC2013,liu_infocom2010,Wang2013WhenSimplicity}.
To complement these efforts, \cite{Gorlatova_TMC2013,yerva2012grafting,yun2011design} collected traces and studied the impact of the 
\ifJSACsubmission
\emph{energy source properties}
\else
energy source properties
\fi
on higher layer algorithms.
However, there is still only limited understanding of motion energy availability
and its impact on the design 
of both hardware (energy harvesters, energy storage components) and algorithms.
Hence, we focus on \emph{characterizing the kinetic (motion) energy that can be harvested by an 
IoT node 
and on the impact of the energy characteristics on harvesting adaptive algorithms}. Self-sustainable IoT nodes powered by motion will be implemented in ultra-low-power architectures. 
Thus, we additionally focus on \emph{developing algorithms that take practical IoT node design considerations into account}.

Everyday activities such as walking can generate substantial power~\cite{starner1996human}. Therefore, many harvesters are under development, including shoe inserts that harvest energy from footfalls~\cite{kymissis98}
and mobile phone chargers integrated in backpacks~\cite{nPowerPEG} or phones~\cite{ApplePatentShake}. While there are several ways of harvesting motion energy, we focus on 
\ifJSACsubmission
\emph{inertial energy harvesters}, 
\else
inertial energy harvesters, 
\fi
since their form factor fits IoT applications. An inertial harvester suitable for a small wireless device (e.g., under 5cm x 5cm, and weighing less than 2~grams) can generate 100--200~$\mu$W from  walking~\cite{huang2011human,von2006optimization}, which is sufficient for many applications.\footnote{This is comparable to the power a solar cell of similar size can harvest from indoor light~\cite{Gorlatova_TMC2013,yerva2012grafting}.} However, the harvesting level changes dynamically as
illustrated in Fig.~\ref{fig:AccToEnergy} that shows the power harvesting level corresponding to a 
device carried by a walking person. 

In inertial harvesters, the output power is maximized when the harvester resonant frequency is ``matched" to the motion frequency~\cite{Mitcheson2008} (see Section~\ref{sect:perActivityEnergy} for details).
Human motion is a combination of low frequency vibrations ($<$10~Hz) that vary from activity to activity and from person to person. Therefore, characterizing the properties of the harvested power requires 
\ifJSACsubmission
\emph{an in-depth study of human motion (e.g., the frequencies associated with different motions) and human mobility patterns}. 
\else
an in-depth study of human motion (e.g., the frequencies associated with different motions) and human mobility patterns. 
\fi
Namely, characterizing kinetic energy harvesting is substantially more complex than characterizing light energy harvesting (e.g., \cite{Gorlatova_TMC2013}).



We first describe methods for collecting
\ifJSACsubmission
\emph{motion acceleration traces} 
\else
motion acceleration traces
\fi
and the methods of~\cite{von2006optimization,Mitcheson2008,yun2011design} for estimating harvested power from the traces. 
 Our study is based on traces that we collected using SparkFun ADXL345 boards\footnote{Although smartphones include accelerometers, we use dedicated sensing units, since the phones' accelerometers have a limited range, restricted sampling rate control, and high energy consumption (that hinders day-scale trace collection).}
and traces collected in~\cite{xue2010naturalistic} using similar devices.
While the traces in \cite{xue2010naturalistic} were collected to examine activity recognition,
we use them to estimate the amount of energy that could be harvested.

%

We 
examine the energy availability associated with 
\ifJSACsubmission
\emph{specific human motions}, 
\else
specific human motions,
\fi
such as walking, running, 
and cycling. Unlike previous studies that obtained 
estimates based
on small numbers of participants~\cite{buren2003kinetic,von2006optimization,huang2011human}, we use a motion dataset \emph{with over 40 participants}~\cite{xue2010naturalistic},
obtaining extensive and general kinetic energy characterization for common human motions. The study demonstrates the range of motion frequencies and harvested powers for different participants and activities, and \emph{uniquely demonstrates the importance of human physical parameters for energy harvesting}. For example, the taller half of the participants can harvest on average 20\% more power than the shorter half.

\ifJSACsubmission
The \emph{short duration} traces in \cite{xue2010naturalistic} are for \emph{specific motions}.
\else
The short duration traces in \cite{xue2010naturalistic} are for specific motions.
\fi
In order to study the energy generation processes associated with \emph{day-scale human routines} (as opposed to specific motions), we conducted a 
measurement campaign with 5~participants over a total of 25~days. We collected traces with over 200~hours of acceleration information for normal human routines.
The traces provide important input for IoT node design (e.g., for determining the battery capacity and harvester size necessary for self-sustainable operation) and for algorithm design (as will be discussed below).
Hence, \emph{we share the collected dataset in~\cite{KineticMeasurements} and via CRAWDAD~\cite{CrawdadMoversShakers}}.\footnote{To the best of our knowledge, this is the first publicly available long-term human motion acceleration dataset.} 
We analyze the traces and show that the power availability from normal routines and from indoor lights are comparable. We also demonstrate that the power generation process associated with human motion is highly variable. 
We compare this process with i.i.d.\ and Markov processes, demonstrating the importance of evaluating 
algorithms with real world traces. 

We note that the primary goal of collecting and analyzing traces is to set a reasonable upper bound on the available energy and to study the energy availability dynamics. Commercially available kinetic energy harvesters \cite{Mide, SmartMaterial, Piezo} are optimized for harvesting energy from machine vibrations above 40Hz. Therefore, these harvesters would generate essentially no energy when subjected to human motion. In general, measuring acceleration is preferable to measuring the energy harvested by a particular harvester, since the traces can be used to calculate how much energy any past, present, or future harvester would generate.
\ifJSACsubmission
\else
While our results are based on the assumption that a harvester is modeled as a mass-spring system, the acceleration traces can be applied to any such future kinetic energy harvester (e.g., conductive droplet sliding on electret film \cite{Yang2012power}, reverse electrowetting \cite{Krupenkin2011reverse}).
\fi


As the IoT will incorporate many objects, we additionally briefly present results regarding measurements with a variety of \emph{moving objects}. For example, we measured the power that can be harvested from everyday activities such as writing with a pencil and 
opening a door. We also collected measurements for objects in transit. We shipped a FedEx box with a measurement unit across the U.S., placed a unit in a checked-in luggage during a 3 hour flight, and carried units on cars and trains. 
We confirm that, as expected based on inertial harvesters' 
\ifJSACsubmission
\emph{filter properties}
\else
filter properties
\fi
 (see Section~\ref{sect:InertialHarvModel}),
the energy availability is low for many common non-periodic motions. We additionally demonstrate that the energy availability is low for many 
\ifJSACsubmission
\emph{high-amplitude periodic object motions}. 
\else
high-amplitude periodic object motions. 
\fi
For example, we show that inertial harvesters can harvest little energy from opening and closing a door, opening cabinet drawers, and spinning a swivel chair.

Next, we develop \emph{energy allocation algorithms} for wireless IoT nodes.
Due to the high variability of energy obtained from motion,
IoT nodes that harvest this energy will implement algorithms that control the node's 
\ifJSACsubmission
\emph{energy spending rates}
\else
energy spending rates
\fi
\cite{liu_infocom2010,Chen2012Asymptotically,Kansal06Journ,Gorlatova_TMC2013,Gunduz2012,noh-efficient}.
 The spending rates will provide inputs for determining node transmission power, duty cycle, sensing rate, or communication rate.
We formulate an optimization problem
of a node whose objective is to  \emph{maximize the utility of its energy allocations}, 
 and develop algorithms for solving it. The problem formulation and the algorithms take into account \emph{realistic properties of an ultra-low-power IoT node}.

In particular, IoT nodes that are powered by the motion energy 
will likely to be
implemented in \emph{ultra-low-power architectures}.
As such, they will support only a limited number of possible energy spending rates,
and their energy use patterns may call for considering various possible utility functions.
Moreover, these nodes will 
likely to use \emph{capacitors}~\cite{Gorlatova_TMC2013,ting-mobisys09,Gummeson10}, rather
than \emph{batteries}, as their energy storage components. This is due to the
fact that capacitors can be charged and discharged many more times than batteries, which is an important
feature for nodes powered by the widely varying motion energy. Additionally, capacitors are more environmentally friendly than batteries~\cite{Gummeson10},
and are therefore more suitable for human-facing IoT applications such as wearable computing.
 To the best of our knowledge, these aspects of IoT node modeling have not been jointly considered before.

For solving the energy allocation problem, we develop an optimal offline algorithm, an efficient approximation scheme, and an online algorithm which is optimal in certain cases.
We evaluate the algorithms using the collected measurement traces. The evaluation results demonstrate that the approximation and online algorithms perform well and highlight the importance of designing algorithms that take into account the energy storage properties of the IoT nodes.


To summarize, the main contributions of this paper are: (i) insights into
energy availability from human motion, based on a dataset with a large number
of participants, (ii) collection of a dataset of long-term human
motion and a study of the corresponding energy generation processes,
and (iii) energy allocation algorithms that take practical IoT node design considerations into account. 
The collected motion traces are already available online~\cite{KineticMeasurements}. The paper contributes to the understanding of motion energy
harvesting availability and properties, and  provides insights that are important for the design of motion energy harvesters, IoT nodes, and energy harvesting adaptive algorithms.

The paper is organized as follows. Section~\ref{sect:SourcesOverview} summarizes the related work and Section~\ref{sect:KineticEnergy} describes the 
harvester model, the 
measurements, the procedures for
determining the 
harvester
parameters, and the wireless node model. Section~\ref{sect:perActivityEnergy} focuses on common human motions and Section~\ref{sect:DailyEnergy} focuses on our measurement campaign and day-scale human motion measurements. Section~\ref{sect:Intuition} provides brief comments regarding motion of objects. Section~\ref{sect:EnergyAlgorithms} describes our algorithms and provides the results of algorithm evaluations with the collected measurements. Section~\ref{sect:Conclusions} concludes the paper.
\ifIncludeAppendix
\else
Due to space constraints, the proofs are omitted and appear in \cite{MoversShakersReport}.
\fi

\section{Related Work}
\label{sect:SourcesOverview}

To the best of our knowledge, our experiments with long-term activities (Section~\ref{sect:ActivityDuration}) and with object motion (Section~\ref{sect:Intuition}) are unique. Below we briefly summarize the related work for our
other contributions.

Previous 
studies that examined 
energy of \emph{particular human motions} had a small number of participants
(10 in~\cite{huang2011human} and 8 in~\cite{von2006optimization,yun2011design}). Additionally, with the exception of~\cite{yun2011design}, they 
examined short intervals 
of
walking and running on a treadmill at a constant pace. We examine a dataset~\cite{xue2010naturalistic} with over 40 participants performing a set of several unrestricted motions\footnote{The properties of restricted 
and unrestricted human motions are known to differ~\cite{orendurff2008humans}.}
and \emph{labeled with human physical parameters}. 
To the best of our knowledge, this is the first publicly available acceleration dataset collected for a large number of participants. It \emph{was not previously used for an energy study}. 


\emph{Day-scale} human motion acceleration traces were previously collected for 8 participants over 3 days and examined in~\cite{yun2011design}, which established energy budgets for wearable nodes using assumptions suitable for larger electronic devices. The data collected in~\cite{yun2011design} is not publicly available. 
We collect day-scale data that in some cases has more information per participant, 
examine the traces under assumptions suitable for small IoT nodes, and characterize energy harvesting process variability and properties that \emph{have not been considered before}.

 Many energy harvesting adaptive communication and networking algorithms 
 have been recently developed (e.g.,~\cite{Kansal06Journ,Chen2012Asymptotically,liu_infocom2010,Gunduz2012,Wang2013WhenSimplicity,noh-efficient,YS12,OKYSA11,NKM13,ODE13}).
 We consider a wireless node model and develop algorithms
 that capture several practical 
IoT node design aspects: 
(i) \emph{discrete}, rather than continuous~\cite{Chen2012Asymptotically,Gorlatova_TMC2013,liu_infocom2010,Gunduz2012,Wang2013WhenSimplicity}, energy spending rates; (ii) \emph{general}, rather than concave~\cite{Chen2012Asymptotically,Gorlatova_TMC2013,liu_infocom2010,Gunduz2012,Wang2013WhenSimplicity} or linear~\cite{Kansal06Journ}, utility functions; and (iii) use of a \emph{capacitor}~\cite{Gorlatova_TMC2013,ting-mobisys09}, rather than a battery~\cite{Chen2012Asymptotically,liu_infocom2010,Gunduz2012,Wang2013WhenSimplicity}, as an energy storage component. 
These aspects have not been jointly considered before.
Existing algorithms are typically evaluated with 
light~\cite{Kansal06Journ,ting-mobisys09,Chen2012Asymptotically} or wind~\cite{Chen2012Asymptotically} energy traces. We evaluate the algorithms with the collected day-scale human
motion energy measurements.

\section{Models \& Measurement Setup}
\label{sect:KineticEnergy}
Our motion energy study is based on recorded \emph{acceleration} traces which
are processed, following the methods developed
in~\cite{von2006optimization,Mitcheson2008,yun2011design}, to determine the
energy generated by an inertial harvester.
Our algorithms are developed based on a model that extends existing models~\cite{liu_infocom2010,Chen2012Asymptotically,Kansal06Journ,Gorlatova_TMC2013,Gunduz2012}
to capture important 
IoT node design considerations. In this section, we describe the kinetic energy
harvester model, the collection of acceleration measurements, the procedures
for
determining the 
harvester
parameters, and the wireless 
node model. 
The notation is summarized in Table~\ref{table:symbols_used}.

\ifJSACsubmission
\else
\setlength{\textfloatsep}{7pt}
\fi

\ifJSACsingle

\else

\begin{table}[t]
\centering
\small
\vspace{-0.25 cm}
\caption{Nomenclature \label{table:symbols_used}}
\begin{tabular}{|l|p{6.1cm}|} \hline 
$m$  & Harvester proof mass [ kg ]\tabularnewline
$Z_L$ & Harvester proof mass displacement limit [ m ] \tabularnewline
$k$  & Harvester spring constant [ kg$\cdot$s$^2$ ] \tabularnewline
$b$  & Harvester damping factor [ kg/s ] \tabularnewline
$f_{r}$ & Harvester resonant frequency [ Hz ] \tabularnewline
$f_{m}$ & Dominant motion frequency [ Hz ] \tabularnewline
$a(t)$  & Acceleration [ m/s$^2$ ]\tabularnewline
$D$ & Absolute deviation of acceleration [ m/s$^2$ ]\tabularnewline
\ifJSACsubmission
$P(t)$ & Power [ W ] \tabularnewline
$z(t)$  & Proof mass displacement [ m ]\tabularnewline
\else
$P(t)$, $z(t)$ & Power [ W ] and proof mass displacement [ m ]\tabularnewline
\fi
$i$, $K$ & Time slot index and a number of time slots \tabularnewline 
$s(i)$ & Energy spending rate [ J/slot ] \tabularnewline
$\mathcal{S}$ & Set of feasible $s(i)$
  values \tabularnewline
\ifJSACsubmission
$r(i)$ & Data rate [ Kb/s ] \tabularnewline
$U(s(i))$ & Utility function \tabularnewline
\else
$r(i)$, $U(s(i))$ & Data rate [ Kb/s ] and utility function \tabularnewline
\fi
$B(i)$ & Energy storage level [ J ] \tabularnewline
$e(i)$ & Environmental energy level [ J ] \tabularnewline
$Q(e(i), B(i))$ & Energy harvesting rate [ J/slot ] \tabularnewline
$L(i, B(i))$ & Energy loss (leakage) rate [ J/slot ] \tabularnewline
$\eta(i, B(i))$ & Energy conversion efficiency [ dimensionless ] \tabularnewline
$C$ & Energy storage capacity [ J ] \tabularnewline
\hline
\end{tabular}
\vspace*{-0.2 cm}
\end{table}
\fi

\subsection{Inertial Harvester Model} \label{sect:InertialHarvModel}

\ifJSACsingle
\else
\begin{figure}[t] 
\centering
\subfigure[\label{fig:InertialGeneratorSchematic}]
{\includegraphics[height = 2.4cm]{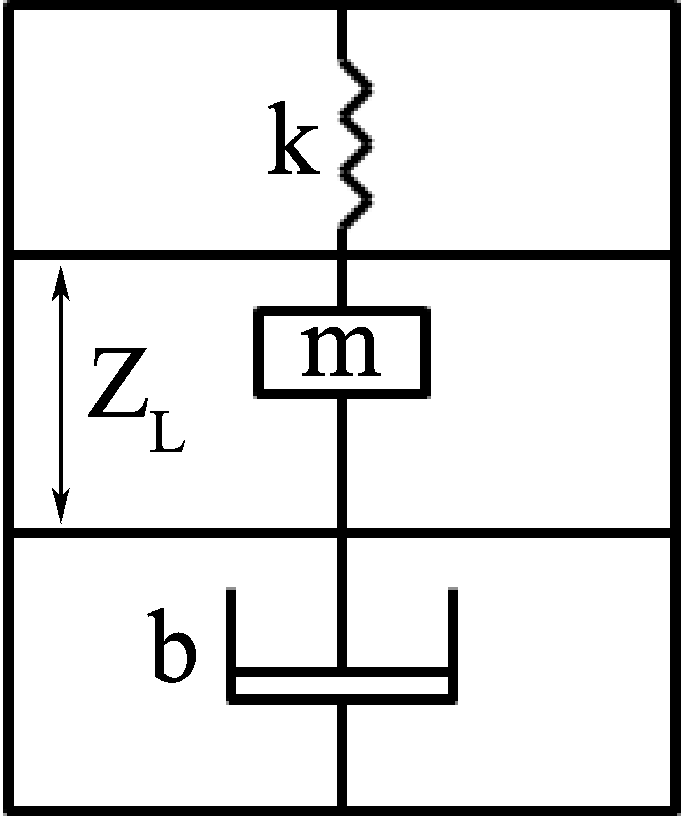}}
\subfigure[\label{fig:FilterViewHarvesters}]
{\includegraphics[height = 2.6cm]{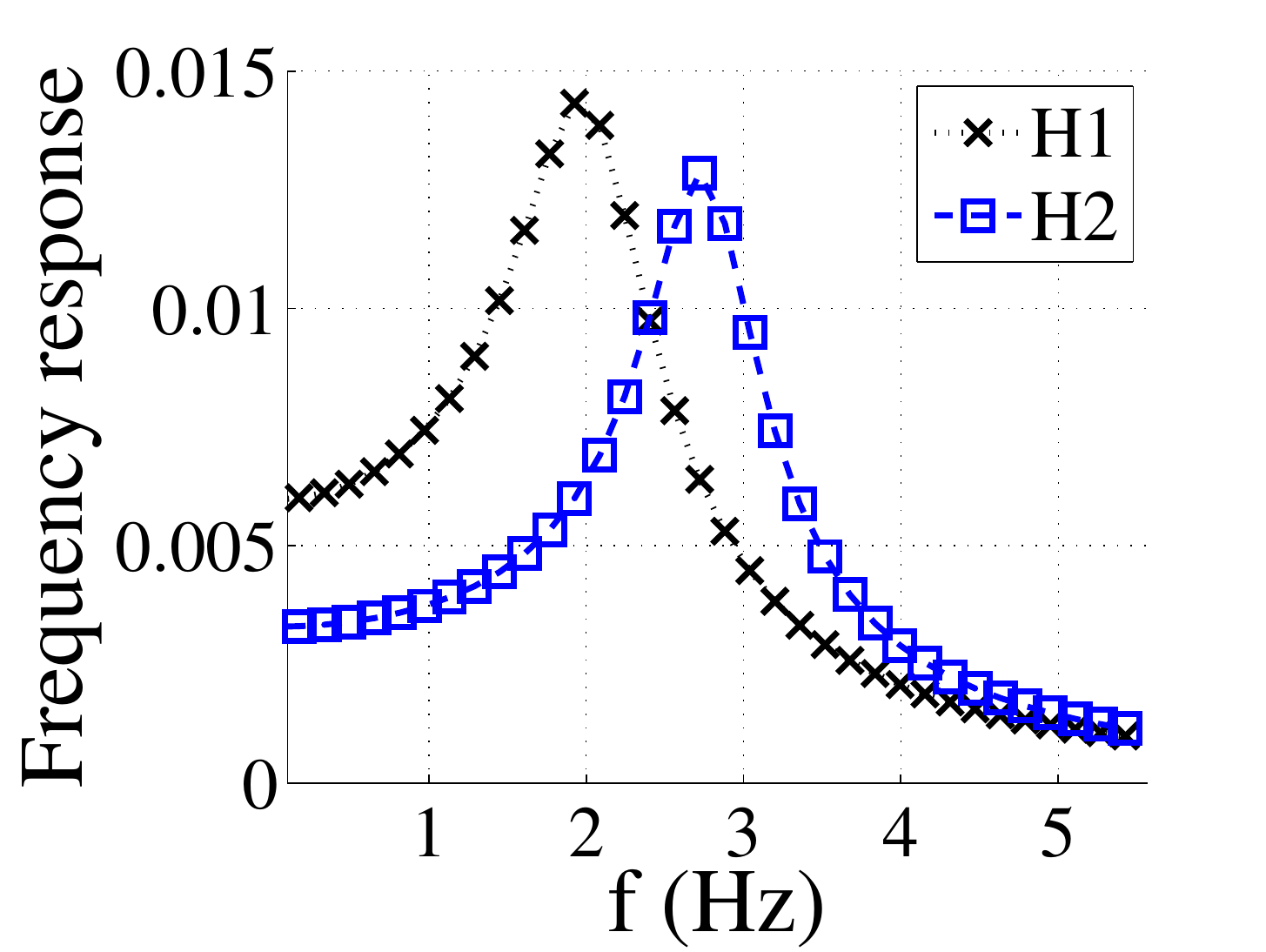}}
\ifJSACdouble
\else
\vspace{-0.35cm}
\fi
\caption{(a) A \emph{second-order mass-spring system model} of a harvester with proof mass $m$, proof mass displacement limit $Z_L$, spring constant $k$, and damping factor $b$, and (b) the frequency response magnitude for harvesters $H1$ and $H2$.}
\end{figure}
\fi

An inertial harvester can be modeled as a \mbox{second-order} \mbox{mass-spring} system
with a harvester proof mass $m$, proof mass displacement limit $Z_L$, spring constant $k$, and spring damping factor $b$~\cite{von2006optimization,Mitcheson2008}. Fig.~\ref{fig:InertialGeneratorSchematic} demonstrates such a harvester model.

Two important \emph{harvester design parameters} are $m$ and $Z_L$. The harvester output power, $P$,  increases linearly with $m$~\cite{beeby2006energy}, and is non-decreasing (but generally \mbox{non-linear}) in $Z_L$. Yet, $m$ and $Z_L$ are limited by the harvester weight and size considerations, which ultimately depend on the application.
We use the following values that are consistent with the IoT restrictions on the size and weight of a node,
and correspond to one of the 
configurations examined in~\cite{von2006optimization}: (i) 
$m$ = $1\cdot10^{-3}$~kg and (ii) 
$Z_L= 10$~mm.

The other two model parameters, $k$ and $b$, are tuned to optimize the energy harvested for given motion properties. The parameter $k$ determines the \emph{harvester resonant frequency},
$f_r= 2\pi\sqrt{k/m}.$
To maximize power output, \emph{the resonant frequency, $f_r$, should match, reasonably closely, the dominant frequency of motion, $f_m$}.

Jointly, $k$ and $b$ determine the \emph{harvester quality factor}, $Q = \sqrt{km}/b$,
which determines the spectral width of the harvester. A harvester with a small~$Q$ harvests a wide range of frequencies with a low peak value, while a harvester with a large~$Q$ is finely tuned to its resonant frequency $f_r$. The role of $f_r$ and $Q$ can be observed in
Fig.~\ref{fig:FilterViewHarvesters}, which shows the magnitude of the frequency response 
of two different harvesters, denoted by $H1$ and $H2$. For $H1$, $f_r=2.06$~Hz (which corresponds to a typical frequency of human walking) and $Q = 2.35$ ($k=0.17$, $b=0.0055$). For $H2$, $f_r=2.77$~Hz (which corresponds to a typical frequency of human running) and $Q=3.87$ ($k=0.30$, $b=0.0045$).

\subsection{Collecting Motion Information \label{sect:KineticCollection}}
\ifJSACsingle
\else
\begin{figure}[t]
\centering
\subfigure[ 
\label{fig:ADXLPhoto}]
{\includegraphics[height=2.4cm]{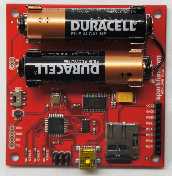}}\hspace{+0.07in}
\subfigure[\label{fig:BoardPlacementsTag}]
{\includegraphics[height=2.4cm]{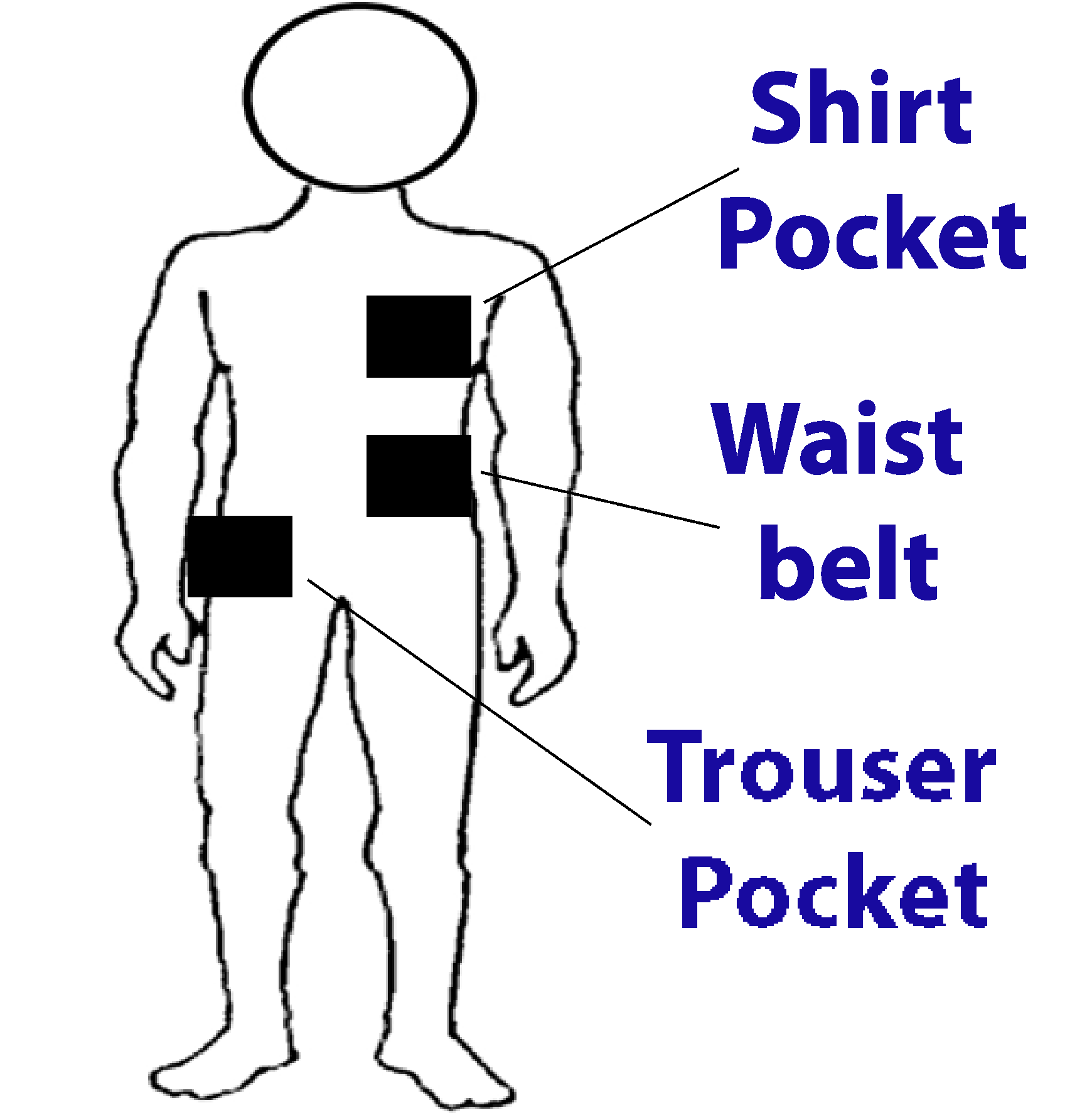}}
\ifJSACdouble
\else
\vspace{-0.35cm}
\fi
\caption{Acceleration measurement unit and placements: (a) our sensing unit 
based on a SparkFun ADXL345 
board, and (b) the sensing unit placements in a multi-participant human motion characterization study~\cite{xue2010naturalistic}.} 
\end{figure}
\fi

In Sections 4-6, we examine measurements that we collected and measurements provided in a triaxial acceleration
dataset of common human motions~\cite{xue2010naturalistic}.
Our measurements were obtained with sensing units based on 
SparkFun ADXL345 evaluation
boards (see Fig.~\ref{fig:ADXLPhoto}). Each unit includes an ADXL345 \mbox{tri-axis}
accelerometer, an Atmega328P microcontroller, and a microSD card for data logging.
The sensing units record acceleration along the $x$, $y$, and $z$ axes, $a_x(t)$, $a_y(t)$, $a_z(t)$, with a $100~$Hz sampling frequency. We conducted multiple experiments with multiple sensing
unit placements, as described in Sections~\ref{sect:DailyEnergy} and \ref{sect:Intuition}.

\ifJSACsingle
\noindent
\begin{minipage}[c]{\textwidth}
\begin{minipage}[l]{0.48\textwidth}
\begin{minipage}[l]{\textwidth}
\begin{figure}[H] 
\centering
\subfigure[\label{fig:InertialGeneratorSchematic}]
{\includegraphics[height=2.0cm]{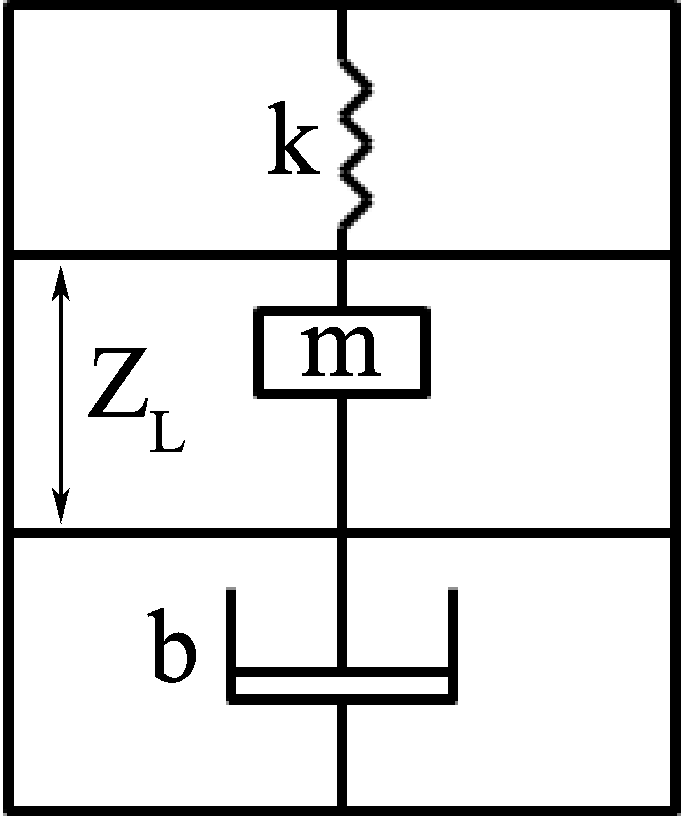}}
\subfigure[\label{fig:FilterViewHarvesters}]
{\includegraphics[height=2.2cm]{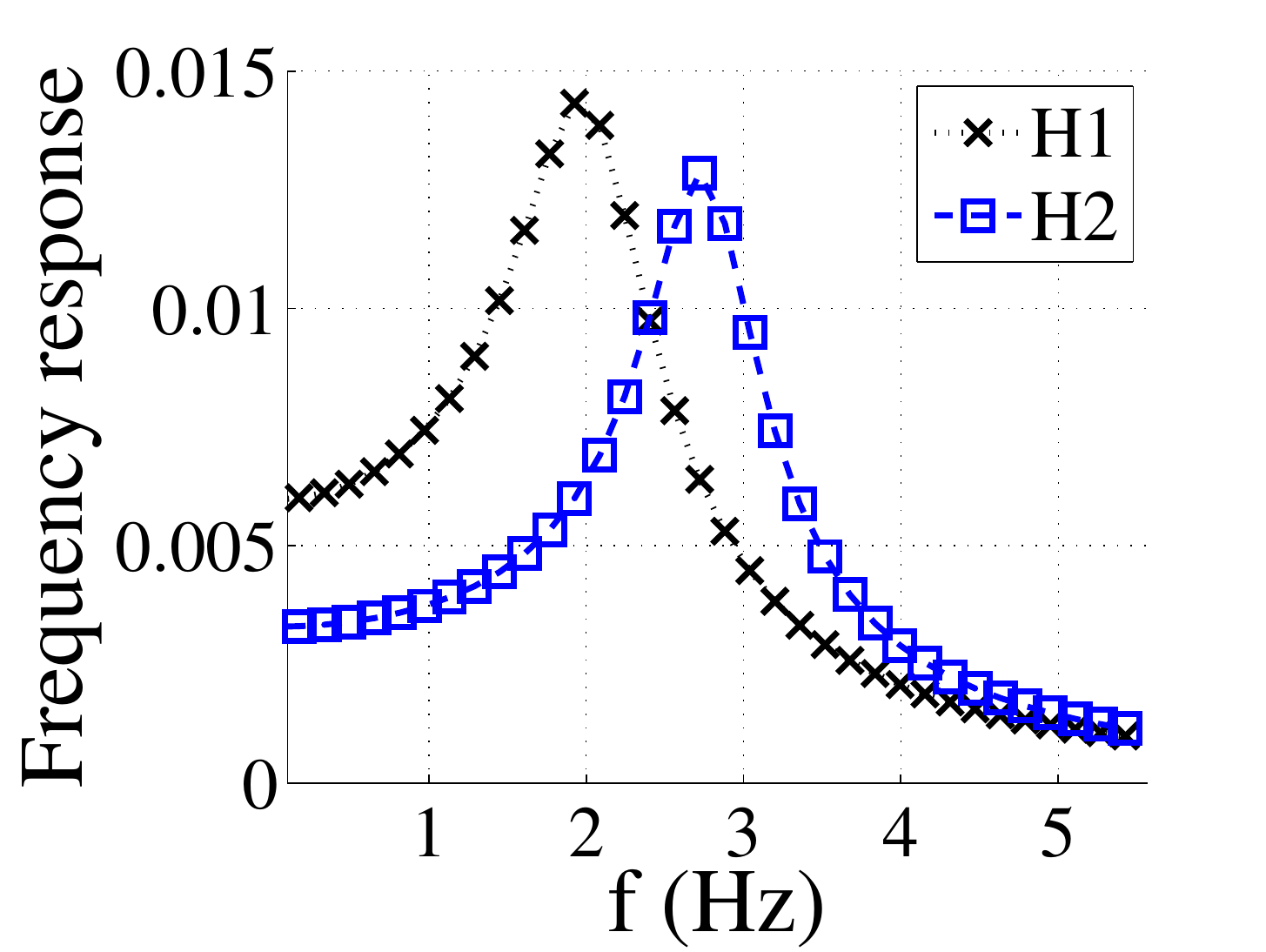}}
\vspace{-0.35cm}
\caption{(a) A \emph{second-order mass-spring system model} of a harvester with proof mass $m$, proof mass displacement limit $Z_L$, spring constant $k$, and damping factor $b$, and (b) the frequency response magnitude for harvesters $H1$ and $H2$.}
\end{figure}
\end{minipage}
\begin{minipage}[r]{\textwidth}
\begin{figure}[H]
\centering
\subfigure[ 
\label{fig:ADXLPhoto}]
{\includegraphics[height=2.2cm]{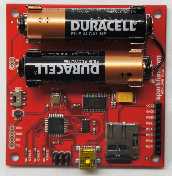}}\hspace{+0.07in}
\subfigure[\label{fig:BoardPlacementsTag}]
{\includegraphics[height=2.2cm]{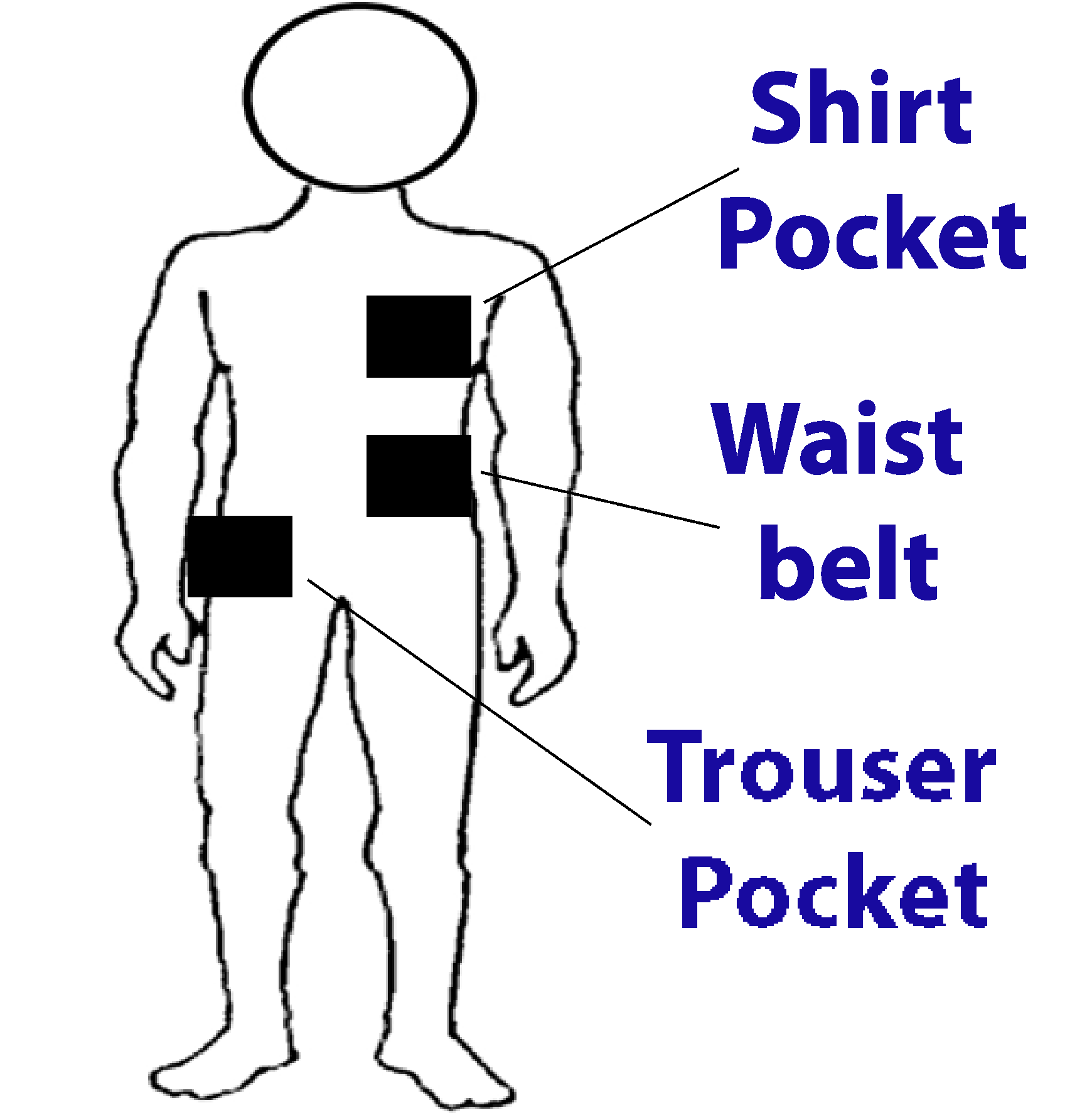}}
\vspace{-0.35cm}
\caption{Acceleration measurement unit and placements: (a) our sensing unit 
based on a SparkFun ADXL345 
board, and (b) the sensing unit placements in a multi-participant human motion characterization study~\cite{xue2010naturalistic}.} 
\end{figure}
\end{minipage}
\end{minipage}
\hfill
\begin{minipage}[r]{0.52\textwidth}
\begin{spacing}{1}
\begin{table}[H]
\centering
\scriptsize
\vspace{-0.25 cm}
\caption{Nomenclature \label{table:symbols_used}}
\begin{tabular}{|l|p{5.5cm}|} \hline 
$m$  & Harvester proof mass [ kg ]\tabularnewline
$Z_L$ & Harvester proof mass displacement limit [ m ] \tabularnewline
$k$  & Harvester spring constant [ kg$\cdot$s$^2$ ] \tabularnewline
$b$  & Harvester damping factor [ kg/s ] \tabularnewline
$f_{r}$ & Harvester resonant frequency [ Hz ] \tabularnewline
$f_{m}$ & Dominant motion frequency [ Hz ] \tabularnewline
$a(t)$  & Acceleration [ m/s$^2$ ]\tabularnewline
$D$ & Absolute deviation of acceleration [ m/s$^2$ ]\tabularnewline
$P(t)$ & Power [ W ] \tabularnewline
$z(t)$  & Proof mass displacement [ m ]\tabularnewline
$i$, $K$ & Time slot index and a number of time slots \tabularnewline 
$s(i)$ & Energy spending rate [ J/slot ] \tabularnewline
\textcolor{blue}{$\mathcal{S}$} & \textcolor{blue}{Set of feasible $s(i)$
  values} \tabularnewline
$r(i)$ & Data rate [ Kb/s ] \tabularnewline
$U(s(i))$ & Utility function \tabularnewline
$B(i)$ & Energy storage level [ J ] \tabularnewline
$e(i)$ & Environmental energy level [ J ] \tabularnewline
$Q(e(i), B(i))$ & Energy harvesting rate [ J/slot ] \tabularnewline
$L(i, B(i))$ & Energy loss (leakage) rate [ J/slot ] \tabularnewline
$\eta(i, B(i))$ & Energy conversion efficiency [ dimensionless ] \tabularnewline
$C$ & Energy storage capacity [ J ] \tabularnewline
\hline
\end{tabular}
\vspace*{-0.2 cm}
\end{table}
\end{spacing}
\end{minipage}
\end{minipage}

\fi

The dataset of~\cite{xue2010naturalistic} was obtained using an ADXL330 \mbox{tri-axis} accelerometers with a $100$~Hz sampling frequency. The measurements of~\cite{xue2010naturalistic} were conducted with sensing unit placements corresponding to a shirt pocket, waist belt, and trouser pocket, as shown in Fig.~\ref{fig:BoardPlacementsTag}.
In all the measurements, the orientation of the sensing unit is not controlled. We examine $a(t) =$ $\sqrt{a_x(t)^2 + a_y(t)^2 + a_z(t)^2}$, the overall magnitude of the acceleration.  
Due to the earth gravity of 9.8~m/s$^2$ (\mbox{``1$g$''}), the measured
acceleration includes a constant component that we filter out 
\ifJSACsingle
(as in~\cite{von2006optimization,yun2011design}, we use a \mbox{3$^{\textrm{rd}}$ order} Butterworth \mbox{high-pass} filter with a 0.1~Hz cutoff frequency).
\else
(similarly to~\cite{von2006optimization,yun2011design}, we use a \mbox{3$^{\textrm{rd}}$ order} Butterworth \mbox{high-pass} filter with a 0.1~Hz cutoff frequency).
\fi

We examine two motion properties of the measurements: the \emph{average absolute deviation of the acceleration}, $D$, and the dominant frequency of motion, $f_m$. $D$ quantifies the variability in the $a(t)$ value and is a measure of the ``amount of motion''. It is calculated as $D = \frac{1}{T} \sum_T (a(t)  - \overline{a}(t))$, where $\overline{a}(t)$ denotes the average of $a(t)$ over time interval $T$. We obtain $f_m$ by determining the maximum spectral component of the Fourier Transform of $a(t)$.


\label{sect:DataCollection}



\ifJSACsingle
\else
\begin{figure}[t]
\centering
\subfigure[] {\includegraphics[width=0.9\columnwidth]{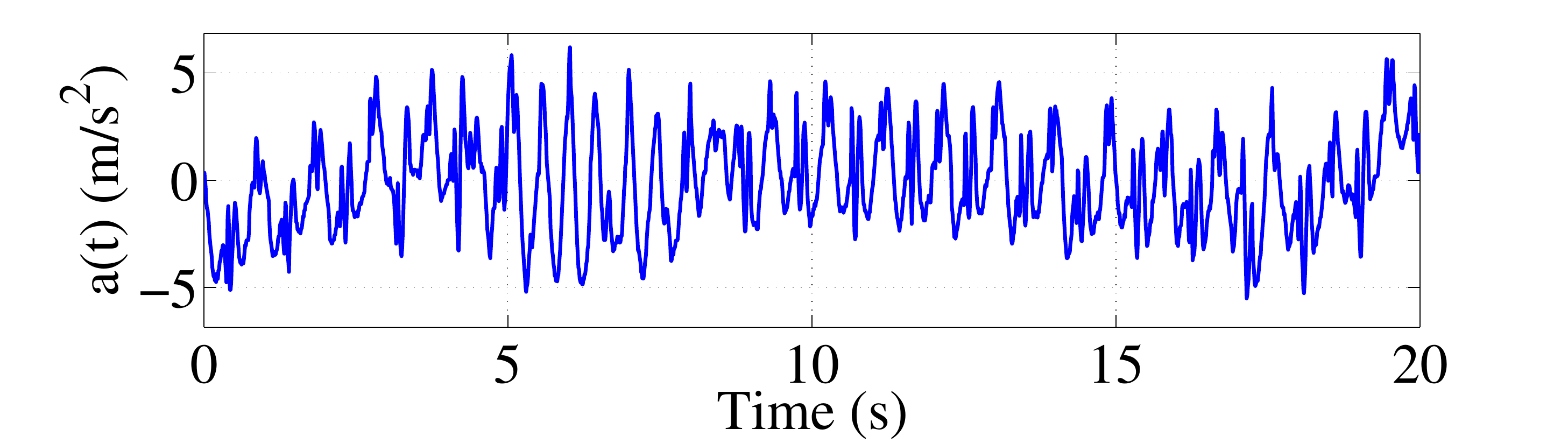}
\label{fig:AccToEnergyAcc}}
\subfigure[]
{\includegraphics[width=0.9\columnwidth]{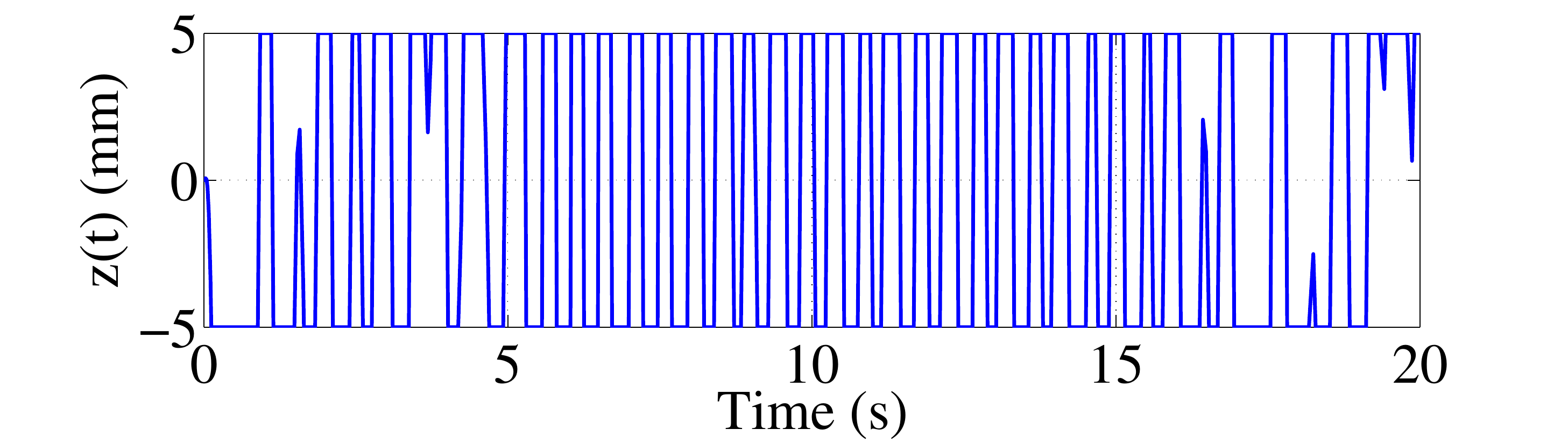}
\label{fig:AccToEnergyDispl}}
\subfigure[]
{\includegraphics[width=0.9\columnwidth]{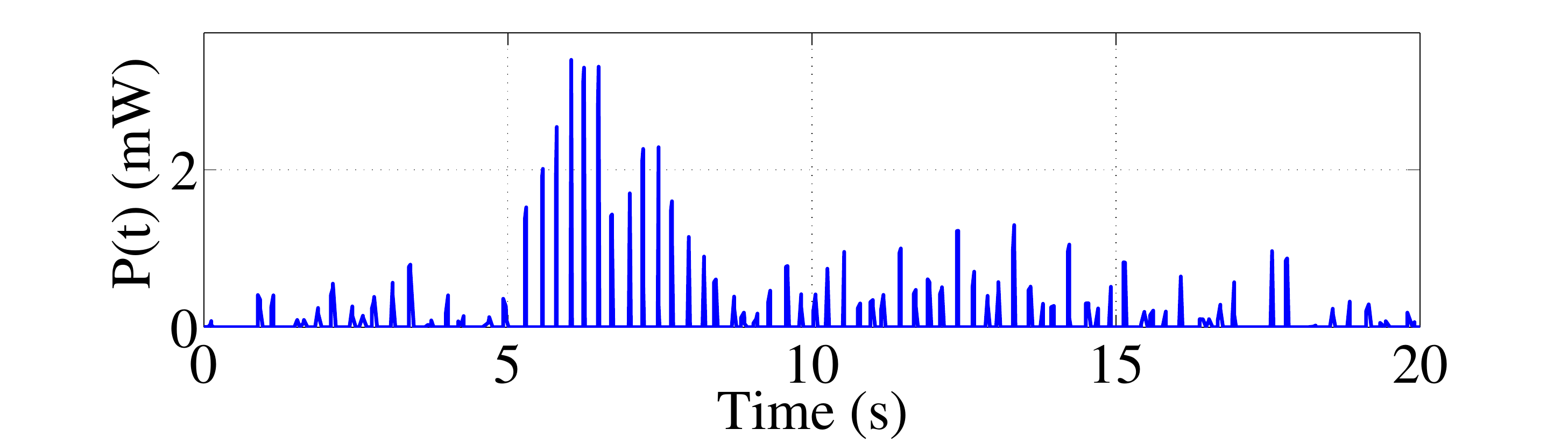}
\label{fig:AccToEnergyPower}}
\ifJSACdouble
\else
\vspace*{-0.35cm}
\fi
\caption{Demonstration of obtaining the power generated by a harvester, $P(t)$, from the recorded acceleration, $a(t)$: (a) $a(t)$ recorded by a person walking, (b) the corresponding harvester proof mass displacement, $z(t)$, and (c) the resulting $P(t)$ for harvester $H1$
($k=0.17$, $b = 0.0055$). \label{fig:AccToEnergy}}
\end{figure}
\fi

\subsection{Harvesting Rates and Data Rates} \label{sect:AccelerationToPower}

We calculate the power generated by a harvester, $P(t)$, subjected to acceleration $a(t)$, using the following procedure based on the methods developed in~\cite{yun2011design}. We first convert $a(t)$ to proof mass displacement, $z(t)$, using the Laplace-domain transfer function 
\ifJSACsingle
$z(t) = \mathcal{L}^{-1} \{z(s)\} = \frac{a(s)}{s^2 + (2\pi f_r/Q) s +
  (2\pi f_r)^2}.$
\else
\begin{equation}
\notag  z(t) = \mathcal{L}^{-1} \{z(s)\} = \frac{a(s)}{s^2 + (2\pi f_r/Q) s +
  (2\pi f_r)^2}.\end{equation}
\fi
Next, to account for~$Z_L$, we limit~$z(t)$ using a Simulink limiter block. 
The power $P(t)$ generated by the harvester is then determined as 
$P(t) = b(dz(t)/dt)^2.$
The average of $P(t)$ is denoted by $\overline{P}$.

\ifJSACsingle
\noindent
\begin{minipage}[l]{\textwidth}

\begin{minipage}[l]{0.62\textwidth}
\begin{figure}[H]
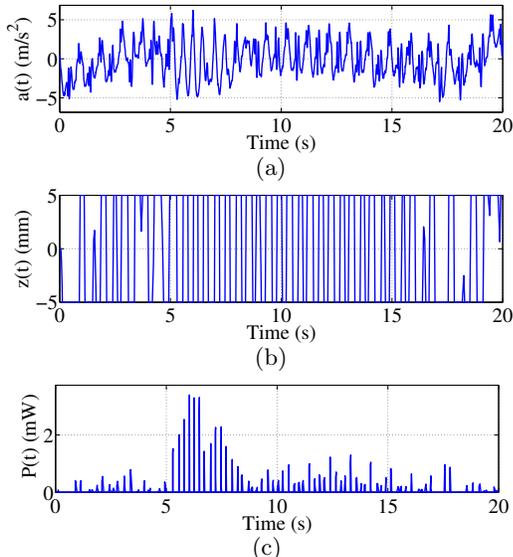

\centering
\subfigure[] {\includegraphics[width=0.47\textwidth]{fig/acceleration/AccelerationExample.eps}
\label{fig:AccToEnergyAcc}}
\subfigure[]
{\includegraphics[width=0.47\textwidth]{fig/acceleration/DisplacementExample.eps}
\label{fig:AccToEnergyDispl}}
\subfigure[]
{\includegraphics[width=0.47\textwidth]{fig/acceleration/PowerExample.eps}
\label{fig:AccToEnergyPower}}
\caption{Demonstration of obtaining the power generated by a harvester, $P(t)$, from the recorded acceleration, $a(t)$: (a) $a(t)$ recorded by a person walking, (b) the corresponding harvester proof mass displacement, $z(t)$, and (c) the resulting $P(t)$ for harvester $H1$
($k=0.17$, $b = 0.0055$). \label{fig:AccToEnergy}}
\end{figure}
\end{minipage}
\hfill
\begin{minipage}[r]{0.35\textwidth}
\begin{figure}[H] 
\centering
\includegraphics[width=0.6\textwidth]{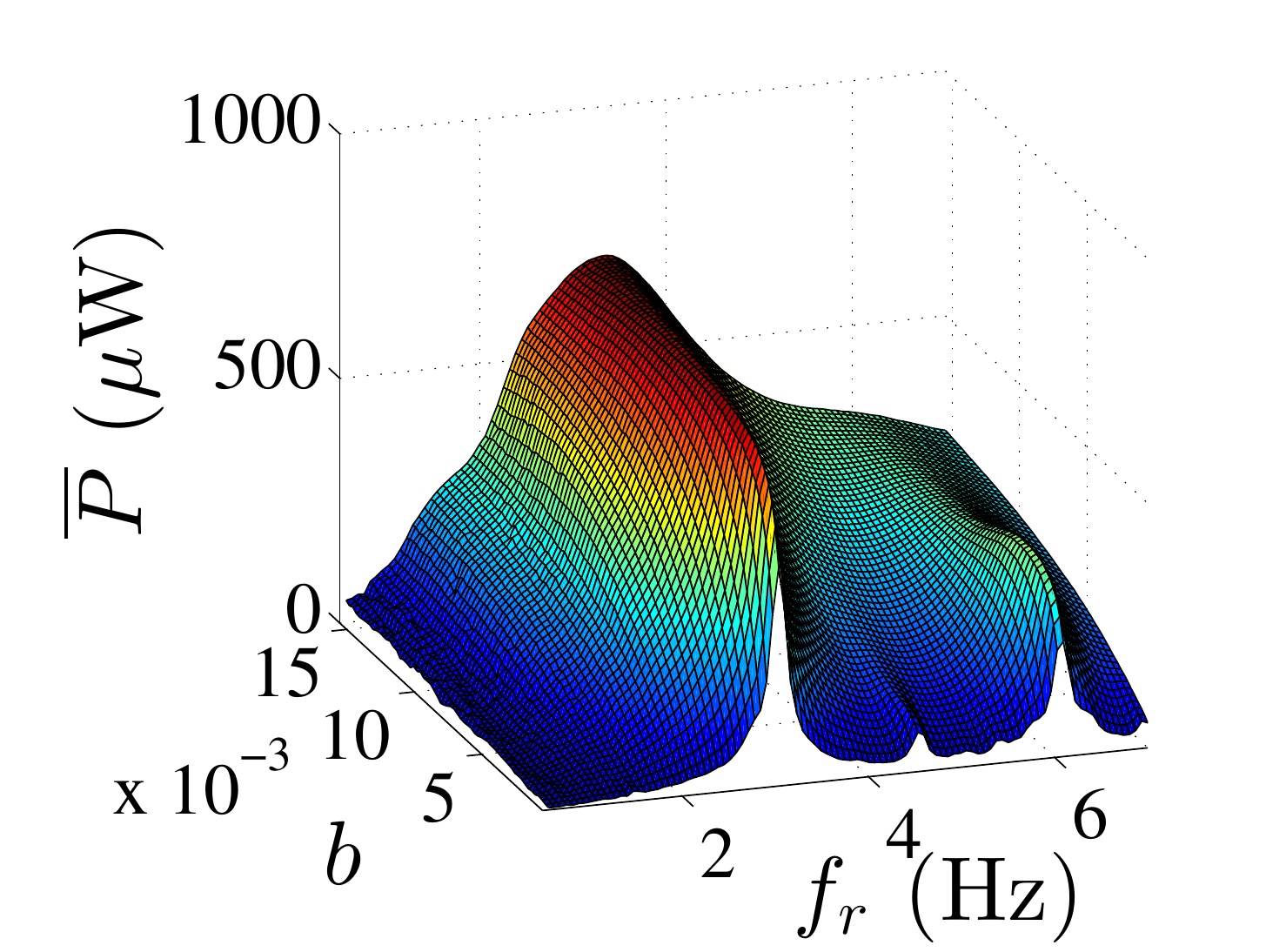}
\caption{The average power generated by a harvester, $\overline{P}$, from the same motion (human running) for different combinations of harvester resonant frequencies, $f_r$, 
and damping factors, $b$.\label{fig:ParameterSearchSpace}}
\end{figure}
\end{minipage}
\end{minipage}
\fi

We implemented this procedure in MATLAB and Sim\-u\-link. Fig.~\ref{fig:AccToEnergy} shows an example of obtaining $P(t)$ for a particular $a(t)$. The $a(t)$ values were recorded by a sensing unit carried by a walking person (Fig.~\ref{fig:AccToEnergy}(a)), and the $z(t)$ and $P(t)$ values were obtained using the procedure described above for the harvester $H1$.


To characterize the performance of wireless IoT nodes, we calculate the data
rates, $\overline{r}$, that a node would be able to maintain when harvesting
the generated $\overline{P}$. 
The harvester energy conversion efficiency, $\eta_h$, depends
  on various factors~\cite{Mide} (e.g., selected regulated output and
  temperature). While perfectly optimized energy harvesting systems obtain
  energy conversion efficiency values between 30\% and 90\%~\cite{LTC3588}, we
  use $\eta_h=20\%$ which is more realistic for practical systems where the
  harvester cannot be continuously aligned with the axis that generates the
  maximum output throughout the day.
Similar to~\cite{Gorlatova_TMC2013}, we assume that the communication cost is $c_{\textrm{tx}}=$1~nJ/bit for ultra-low-power transceivers appropriate for IoT nodes. Hence, $\overline{r}=\eta_h\overline{P}/c_{\textrm{tx}}=2\cdot10^{5}~ \overline{P}$~(Kb/s).

\subsection{Optimizing the Harvester Parameters}
\label{sect:ObtainingOptimalHarvester}

\ifJSACsingle
\else
\begin{figure}[t] 
\centering
\includegraphics[width=0.485\columnwidth]{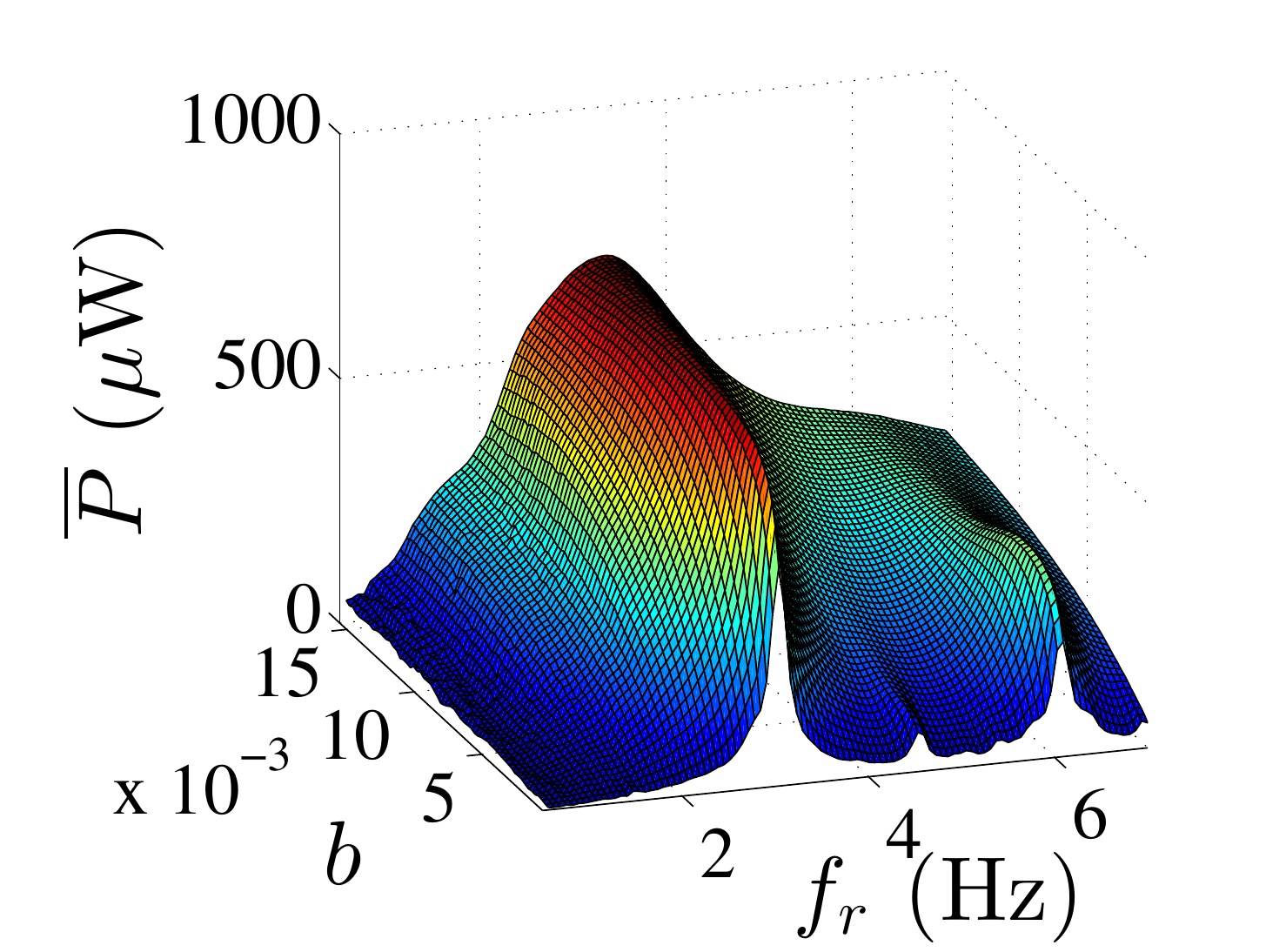}
\vspace*{-0.35cm}
\caption{The average power generated by a harvester, $\overline{P}$, from the same motion (human running) for different combinations of harvester resonant frequencies, $f_r$, 
and damping factors, $b$.\label{fig:ParameterSearchSpace}}
\end{figure}
\fi

Finding the optimal harvester parameters $k$ and $b$ is difficult because it requires optimizing over a multi-dimensional surface of unknown geometry~\cite{von2006optimization}. For example,
Fig.~\ref{fig:ParameterSearchSpace} shows the average power ($\overline{P}$) values calculated from a set of $a(t)$ measurements (corresponding to 
a person running) for different $f_r$ and $b$ combinations. To determine the optimal harvester parameters for short $a(t)$ samples, we implemented an \emph{exhaustive search} algorithm. The algorithm considers a large number of $k$ and $b$ combinations, obtains the corresponding $\overline{P}$ (using the procedure described in Section  \ref{sect:AccelerationToPower}), and chooses the $k$ and $b$ combination that maximizes $\overline{P}$.

The exhaustive search algorithm is time-consuming even for relatively short $a(t)$ samples. 
For longer $a(t)$ samples, 
we implemented a simplified procedure developed in~\cite{yun2011design}. The procedure first determines the $k$ value that matches the harvester's $f_r$ to the dominant frequency in the $a(t)$ sample, $f_m$. 
Specifically, the procedure selects $k$ such that
$k = m f_r^2/(2 \pi )^2$ $=(m f_m^2)/(2 \pi )^2. $
It then considers a relatively large number of $b$ values 
and selects the $b$ that maximizes $\overline{P}$. 

\subsection{Wireless Node Model}
\label{sect:ModelDesignConsiderations}

We model an ultra-low-power IoT node that harvests 
energy,
stores it in an energy storage device, 
and uses it 
to communicate wirelessly (e.g., a wearable node may be communicating with a human-carried mobile phone).
We assume that the time is slotted and
denote the slot index by $i$ and the number of slots by $K$. We will develop
algorithms that control the node energy spending rates, $s(i)$, which can provide inputs for determining node transmission power, duty cycle, sensing rate, or communication rate.
An IoT node is likely to support only a restricted 
number of modes of operation (i.e., sleep, idle), transmission power levels\footnote{For example, the ultra-low-power Chipcon CC1000, 
Chipcon CC2420, 
and Nordic NRF24L01 
RF transceivers support, correspondingly, only 32, 8, and 4 transmission power
levels.}, and transmission rates, thereby supporting only a finite set
$\mathcal{S}$ of
$s(i)$ values. We thus 
\emph{restrict $s(i)$ as $s(i) \in \mathcal{S} \cup \{0\}$} (note that $s(i)$
is typically modeled as a continuous variable~\cite{Gunduz2012,zafer2009calculus,Gorlatova_TMC2013}). 
This complicates the energy allocation problems, as we will demonstrate in Section~\ref{sect:ResourceAllocationProblem}.

We formulate an optimization problem for a single node which
  maximizes \emph{the sum of the utilities of its per-time-slot energy
    allocations}.  This problem is important, for example, in networks where
  nodes transmit mostly ID information~\cite{Gorlatova2013Prototyping} to a
  common gateway\footnote{Single node energy allocation problems were studied
    in 
\cite{Gorlatova_TMC2013,Wang2013WhenSimplicity,Gunduz2012}
under simpler models. In
    Section \ref{sect:ResourceAllocationProblem} we show that even for a
    single node, the considered optimization problem is NP-hard. The extension
    to the case of multiple nodes is a subject for future research.}.  We
consider a utility function $U(s(i))$ that corresponds to the \emph{data rate
  $r(i)$ obtained when the energy spending rate is $s(i)$}\footnote{The model
  allows $U(s(i))$ to account for other considerations as well (e.g., the
  number of activations of nodes' sensors when the energy spending rate is
  $s(i))$.}. 
 The node may achieve different
$r(i)$ in a slot $i$ by transmitting different number of packets, changing the transmission
power, or changing the packet size. Thus the utility function, $U$, may be 
\emph{concave} (when the node changes its transmission
power~\cite{zafer2009calculus,Wang2013WhenSimplicity,Gunduz2012}), 
\emph{linear} (when it transmits different number of packets), 
\emph{convex} (when it changes the packet size under
certain 
settings~\cite{Domingo2011}), or 
\emph{not concave and not convex} (when it changes a combination of the parameters). Correspondingly, \emph{we place no restrictions on $U(s(i))$
  except that it can be computed efficiently}.

An IoT node may use a \emph{battery} or a \emph{capacitor} as its energy
storage device. For a slot $i$, $B(i)$ is the node energy storage level, $e(i)$
is the environmental energy available to the node, and $L(i,B(i))$ is the energy
loss (leakage) from the storage. $Q(e(i), B(i))$ is the energy harvested by the node; its dependency on $B(i)$ is characteristic of capacitor-based nodes~\cite{Gorlatova_TMC2013,merrett2008empirical}.
$\eta(i,B(i))$ is the \emph{energy conversion efficiency} and $C$ is the storage capacity.
Between time slots, the energy storage evolves as
\ifJSACsingle
\begin{align}
  B(i) = \textrm{min}\{&B(i-1) + Q(e(i-1), B(i-1)) - L(i-1, B(i-1))-
  \frac{s(i-1)} {\eta(i-1, B(i-1))},C\}. \notag
\end{align}
\else
\begin{align}
  B(i) = \textrm{min}\{&B(i-1) + Q(e(i-1), B(i-1)) - L(i-1, \notag\\
& B(i-1))- s(i-1)/\eta(i-1, B(i-1)),C\}. \notag
\end{align}
\fi
$\eta(i,B(i))$ depends on the difference between the energy storage voltage, $V_{\textrm{out}}(i)$, and the node's operating voltage, $V_{\textrm{op}}$. Within a \emph{battery}'s operating region, $V_{\textrm{out}}(i)$ is nearly constant. For a \emph{capacitor}, $V_{\textrm{out}}(i)$ depends on $B(i)$~\cite{Gorlatova_TMC2013,merrett2008empirical}. 
We define two 
node models: for a \emph{battery model}, $\eta(i, B(i))=1$, while for a \emph{capacitor model}, $\eta(i, B(i))$ is a non-linear function. 
The 
$\eta(i,B(i))$ 
that we use in the performance evaluations is described in Section~\ref{sect:AlgsPerformance}.

\section{Human Motion}
\label{sect:perActivityEnergy}

We now examine a dataset with over 40 participants performing 7 common 
motions in unconstrained environments. We emphasize that this dataset, 
previously used to examine techniques for 
activity recognition~\cite{xue2010naturalistic}, \emph{has not been used for energy characterization}. 
We first introduce the study. Then, we characterize the energy availability for different motions, the variability in motion properties among sensing unit placements and  participants, and the dependence of energy availability on the participant's physical parameters.

\ifJSACsingle
\else
\begin{figure}[t] 
\centering
\subfigure[\label{fig:DisplacementChineseData}]
{\includegraphics[width=0.95\columnwidth]{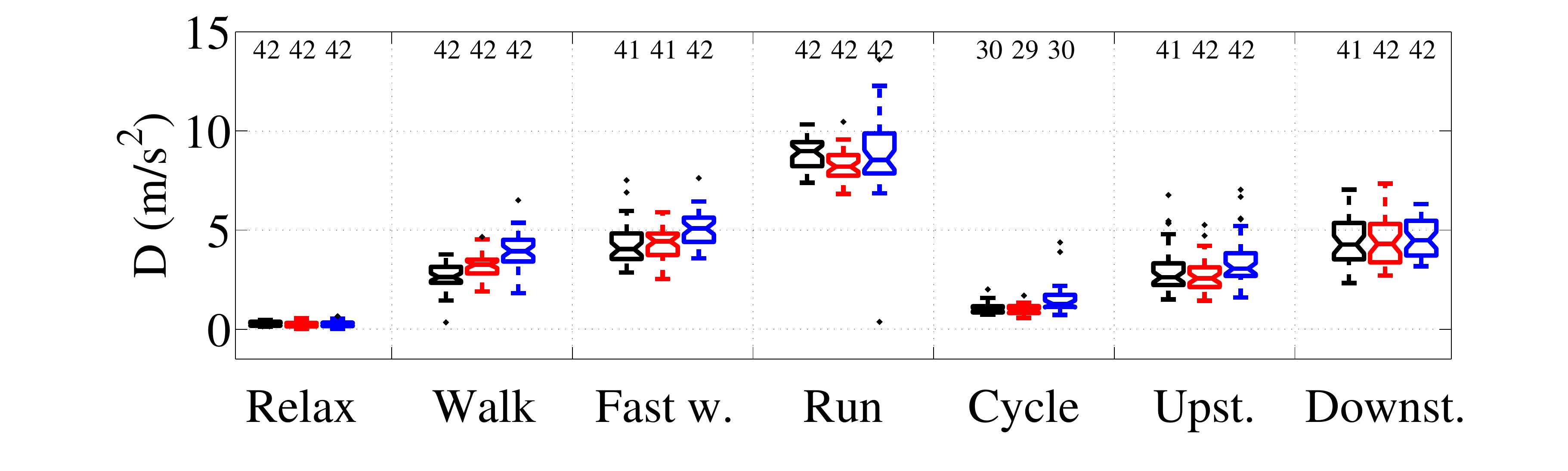}}
\subfigure[\label{fig:FreqChineseData}]
{\includegraphics[width=0.95\columnwidth]{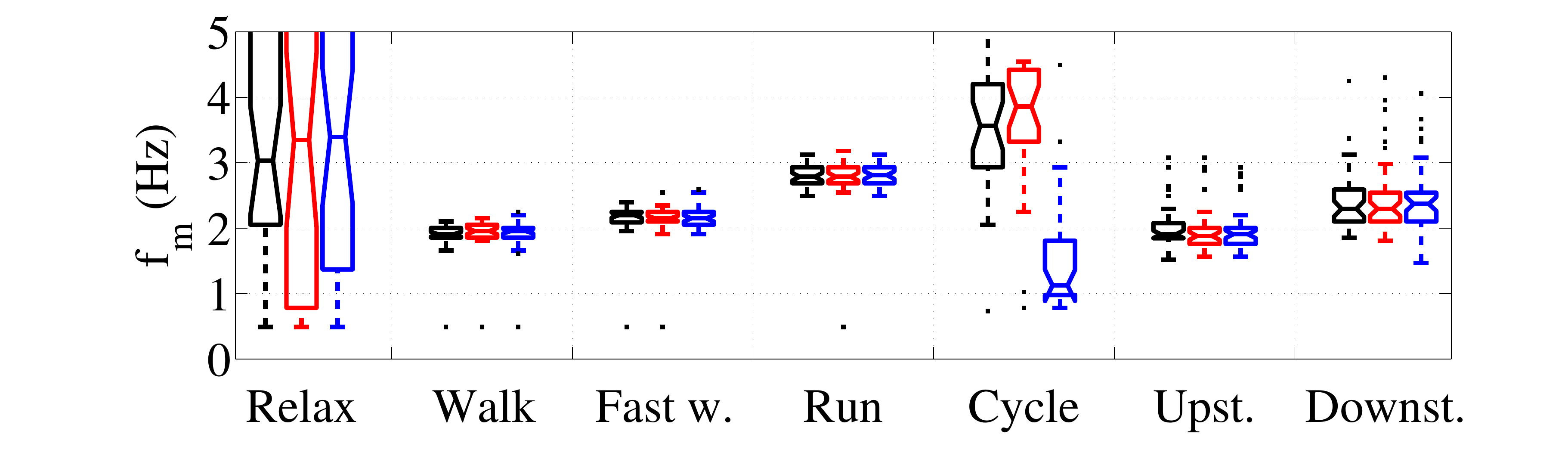}}
\subfigure[\label{fig:PowerChineseData}]
{\includegraphics[width=0.95\columnwidth]{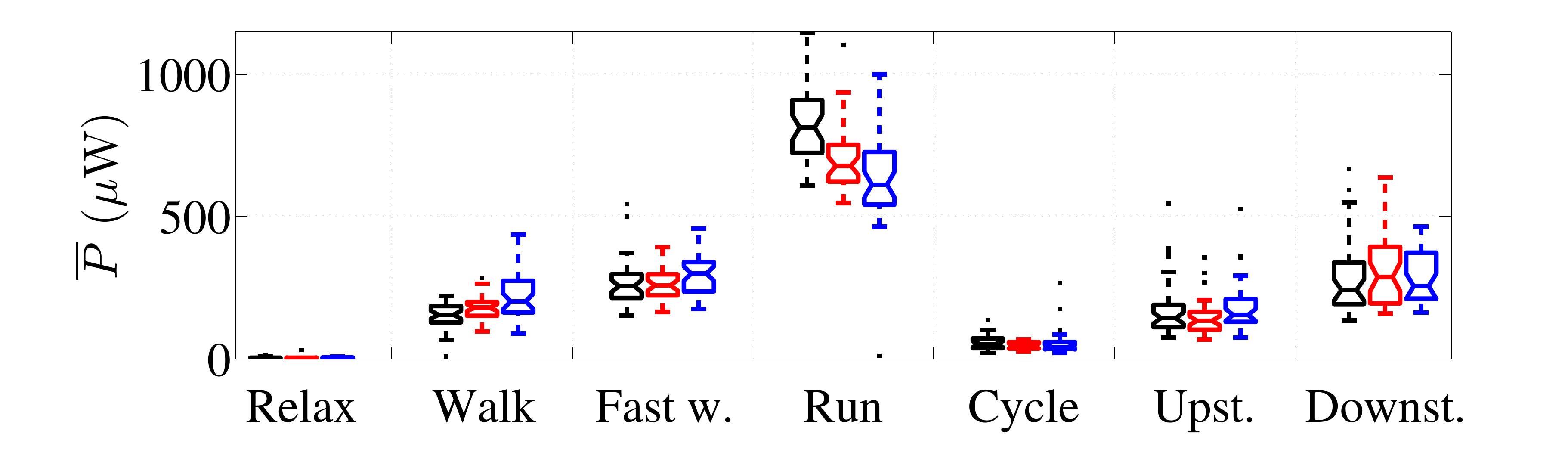}}
\vspace*{-0.35cm}
\caption{\label{fig:ChineseDataResults} Characterization of kinetic energy for
  common human activities, based on a 40-participant study: (a) average
  absolute deviation of acceleration, $D$, (b) dominant motion frequency,
  $f_m$, and (c) power harvested by an optimized 
\ifJSACsubmission
inertial harvester, $\overline{P}$.
\else
harvester, $\overline{P}$.
\vspace*{-0.2cm}
\fi
}
\end{figure}
\fi

\begin{table*}[th!]
\centering
\caption{Energy budgets and data rates based on measurements of common human activities.
\label{table:InstantaneosDataRates}}
\scriptsize
\ifJSACsingle
\begin{tabular}{|p{1.00cm}|p{2.3cm}|p{1.7cm}|p{1.7cm}|p{1.5cm}|p{1.5cm}|p{1.5cm}|p{1.4cm}|} \hline
\else
\begin{tabular}{|p{1.00cm}|p{2.3cm}|p{1.7cm}|p{1.7cm}|p{2.0cm}|p{2.0cm}|p{2.0cm}|p{1.4cm}|} \hline
\fi
    Activity &  Sensing unit  & \# subjects & Median $f_m$  & \multicolumn{3}{c|}{ $\overline{P}$ ($\mu$W)} & Median $r$ \\ \hhline{~~~~---~}
    &  placement &  & (Hz) & $25^{\textrm{th}}$ percentile & Median & $75^{\textrm{th}}$ percentile  & (Kb/s) \\
    \hhline{|-|-|-|-|-|-|-|-}
 \multirow{3}{*}{Relaxing}& \cellcolor{\grayCol} Trouser pocket & \cellcolor{\grayCol} 42 & \cellcolor{\grayCol} N/A & \cellcolor{\grayCol} 1.0 & \cellcolor{\grayCol} 3.1 & \cellcolor{\grayCol} 4.8 & \cellcolor{\grayCol} 0.6\\ \hhline{~-------}
                          & \cellcolor{\whiteCol} Waist belt & \cellcolor{\whiteCol} 42 & \cellcolor{\whiteCol} N/A & \cellcolor{\whiteCol} 0.3 & \cellcolor{\whiteCol} 2.4 & \cellcolor{\whiteCol} 4.8 & \cellcolor{\whiteCol} 0.5\\ \hhline{~|-|-|-|-|-|-|-}
                          & \cellcolor{\grayCol} Trouser pocket &\cellcolor{\grayCol} 42 &\cellcolor{\grayCol} N/A & \cellcolor{\grayCol} 0.2 & \cellcolor{\grayCol} 1.4 & \cellcolor{\grayCol} 5.9 & \cellcolor{\grayCol} 0.3\\ \hhline{|-|-|-|-|-|-|-|-}
 \multirow{3}{*}{Walking} & \cellcolor{\whiteCol} Shirt pocket & \cellcolor{\whiteCol} 42 & \cellcolor{\whiteCol}
                            \cellcolor{\whiteCol} 1.9& \cellcolor{\whiteCol} 128.6 & \cellcolor{\whiteCol} 155.2
                            & \cellcolor{\whiteCol} 186.0 & \cellcolor{\whiteCol} 31.0\\ \hhline{~|-|-|-|-|-|-|-}
                          & \cellcolor{\grayCol} Waist belt & \cellcolor{\grayCol} 42 & \cellcolor{\grayCol} 2.0& \cellcolor{\grayCol} 151.8 & \cellcolor{\grayCol} 180.3 & \cellcolor{\grayCol} 200.3  & \cellcolor{\grayCol} 36.0\\ \hhline{~|-|-|-|-|-|-|-}
                          & \cellcolor{\whiteCol} Trouser pocket & \cellcolor{\whiteCol} 42 & \cellcolor{\whiteCol} 2.0& \cellcolor{\whiteCol} 163.4 & \cellcolor{\whiteCol} 202.4 & \cellcolor{\whiteCol} 274.5 & \cellcolor{\whiteCol} 40.4\\ \hhline{|-|-|-|-|-|-|-|-}
 \multirow{3}{*}{Running}& \cellcolor{\grayCol} Shirt pocket & \cellcolor{\grayCol} 42 & \cellcolor{\grayCol} 2.8& \cellcolor{\grayCol} 724.2 & \cellcolor{\grayCol} 813.3 & \cellcolor{\grayCol} 910.0 & \cellcolor{\grayCol} 162.6\\ \hhline{~|-|-|-|-|-|-|-}
                          & \cellcolor{\whiteCol} Waist belt  & \cellcolor{\whiteCol} 41 & \cellcolor{\whiteCol} 2.8 & \cellcolor{\whiteCol} 623.5 & \cellcolor{\whiteCol} 678.3 & \cellcolor{\whiteCol} 752.8 & \cellcolor{\whiteCol} 135.6\\ \hhline{~|-|-|-|-|-|-|-}
                          & \cellcolor{\grayCol} Trouser pocket & \cellcolor{\grayCol} 42 & \cellcolor{\grayCol}2.8 & \cellcolor{\grayCol} 542.3 & \cellcolor{\grayCol} 612.7 & \cellcolor{\grayCol} 727.4 & \cellcolor{\grayCol} 122.5\\ \hline
 \multirow{3}{*}{Cycling} & \cellcolor{\whiteCol} Shirt pocket & \cellcolor{\whiteCol} 30 & \cellcolor{\whiteCol}3.5
                           & \cellcolor{\whiteCol} 37.4 & \cellcolor{\whiteCol} 52.0 & \cellcolor{\whiteCol} 72.3 & \cellcolor{\whiteCol} 10.4\\ \hhline{~|-|-|-|-|-|-|-}
                          & \cellcolor{\grayCol} Waist belt &  \cellcolor{\grayCol} 29 & \cellcolor{\grayCol} 3.8 & \cellcolor{\grayCol} 36.3 & \cellcolor{\grayCol} 45.4 & \cellcolor{\grayCol} 59.2 & \cellcolor{\grayCol} 9.1\\ \hhline{~|-|-|-|-|-|-|-}
                          & \cellcolor{\whiteCol} Trouser pocket & \cellcolor{\whiteCol} 30 & \cellcolor{\whiteCol} 1.1& \cellcolor{\whiteCol} 35.6 & \cellcolor{\whiteCol} 41.3 & \cellcolor{\whiteCol} 59.5 & \cellcolor{\whiteCol} 8.3\\ \hhline{--------}
\end{tabular}
\ifJSACsubmission
\vspace*{-0.7 cm}
\else
\vspace*{-0.5 cm}
\fi
\end{table*}

\subsection{Study Summary} \label{sect:PerActivityStudyDetails}
The dataset we examine~\cite{xue2010naturalistic} contains 
motion samples for 7 common human activities -- relaxing, walking, fast walking, running, cycling, going upstairs, and going downstairs, -- performed by \emph{over 40 participants} and recorded from the 3 sensing unit placements, shown
 in Fig.~\ref{fig:BoardPlacementsTag}. For each 20-second motion sample, we
 use the acceleration, $a(t)$, trace to calculate $D$, $f_m$, $\overline{P}$,
 and $r$. To obtain $\overline{P}$, we use the exhaustive search harvester
 optimization algorithm 
\ifJSACsubmission
described in 
\else
from
\fi
Section~\ref{sect:ObtainingOptimalHarvester}. By determining the best harvester for each motion, we can offer important insights into the harvester design.


\ifJSACsingle
\noindent
\begin{minipage}[c]{\textwidth}

\begin{minipage}[l]{0.47\textwidth}
\begin{figure}[H]
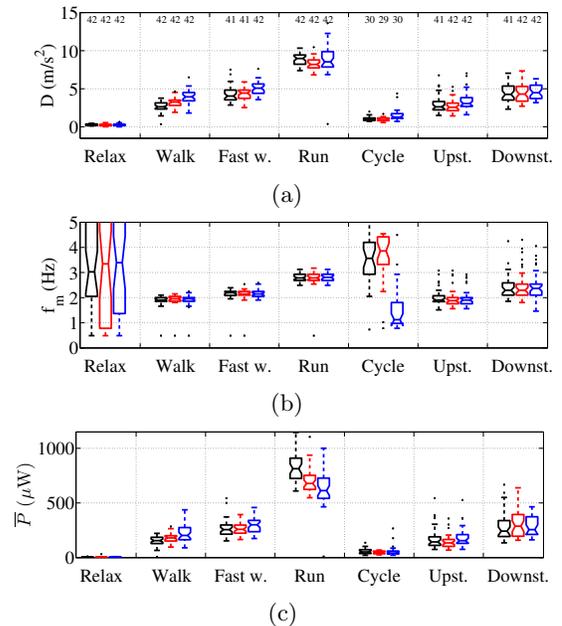
 
\centering
\subfigure[\label{fig:DisplacementChineseData}]
{\includegraphics[width=1.0\columnwidth]{fig/acceleration/ChinesePerActivityAbsoluteDeviation.eps}}
\subfigure[\label{fig:FreqChineseData}]
{\includegraphics[width=1.0\columnwidth]{fig/acceleration/ChinesePerActivityFrequency.eps}}
\subfigure[\label{fig:PowerChineseData}]
{\includegraphics[width=1.0\columnwidth]{fig/acceleration/ChinesePerActivityPower.eps}}
\caption{\label{fig:ChineseDataResults} Characterization of kinetic energy for common human activities, based on a 40-participant study: (a) average absolute deviation of acceleration, $D$, (b) dominant motion frequency, $f_m$, and (c) power harvested by an optimized inertial harvester, $\overline{P}$.
}
\end{figure}
\end{minipage}
\hfill
\begin{minipage}[r]{0.47\textwidth}
\begin{figure}[H] 
\centering
\subfigure[]{ 
\includegraphics[width=0.27\textwidth]{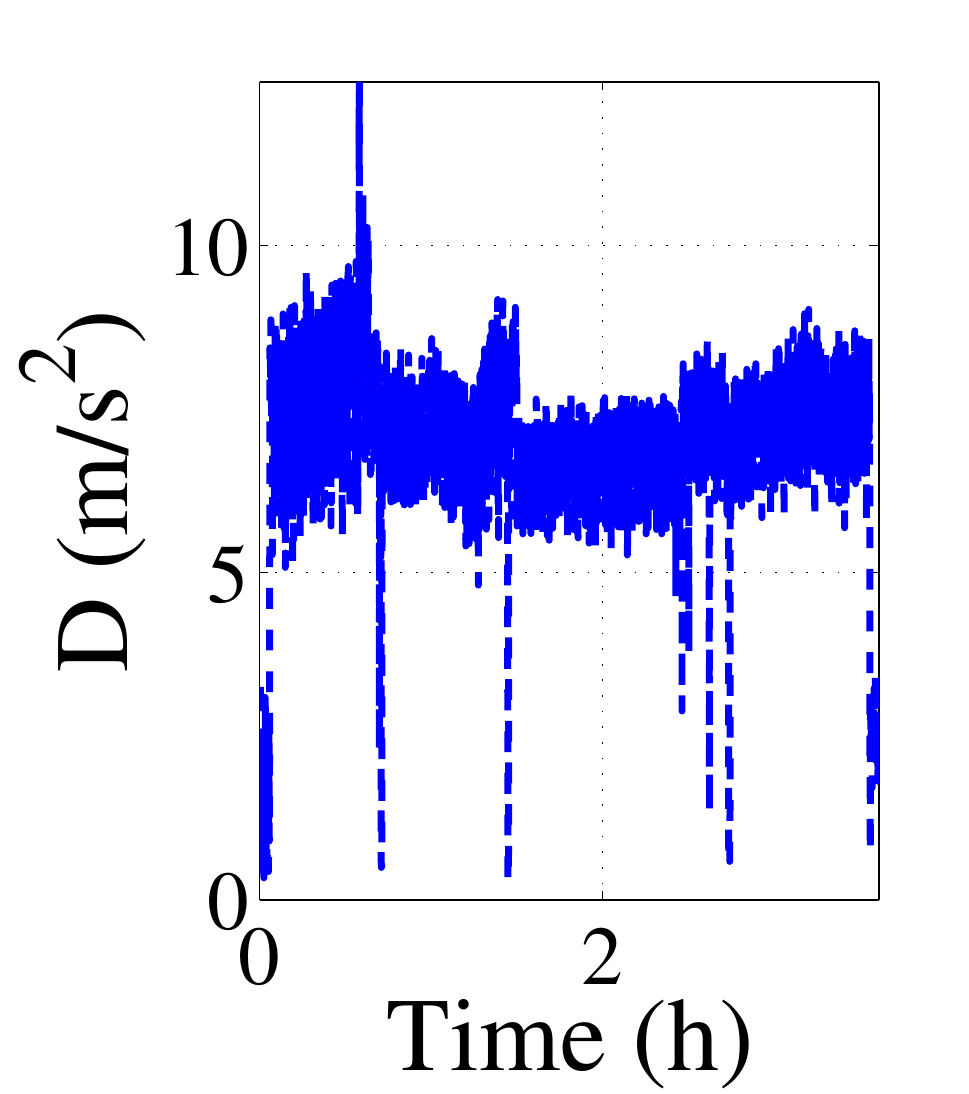}}
\subfigure[\label{fig:LongRunPeakFrequency}]{
\includegraphics[width=0.27\textwidth]{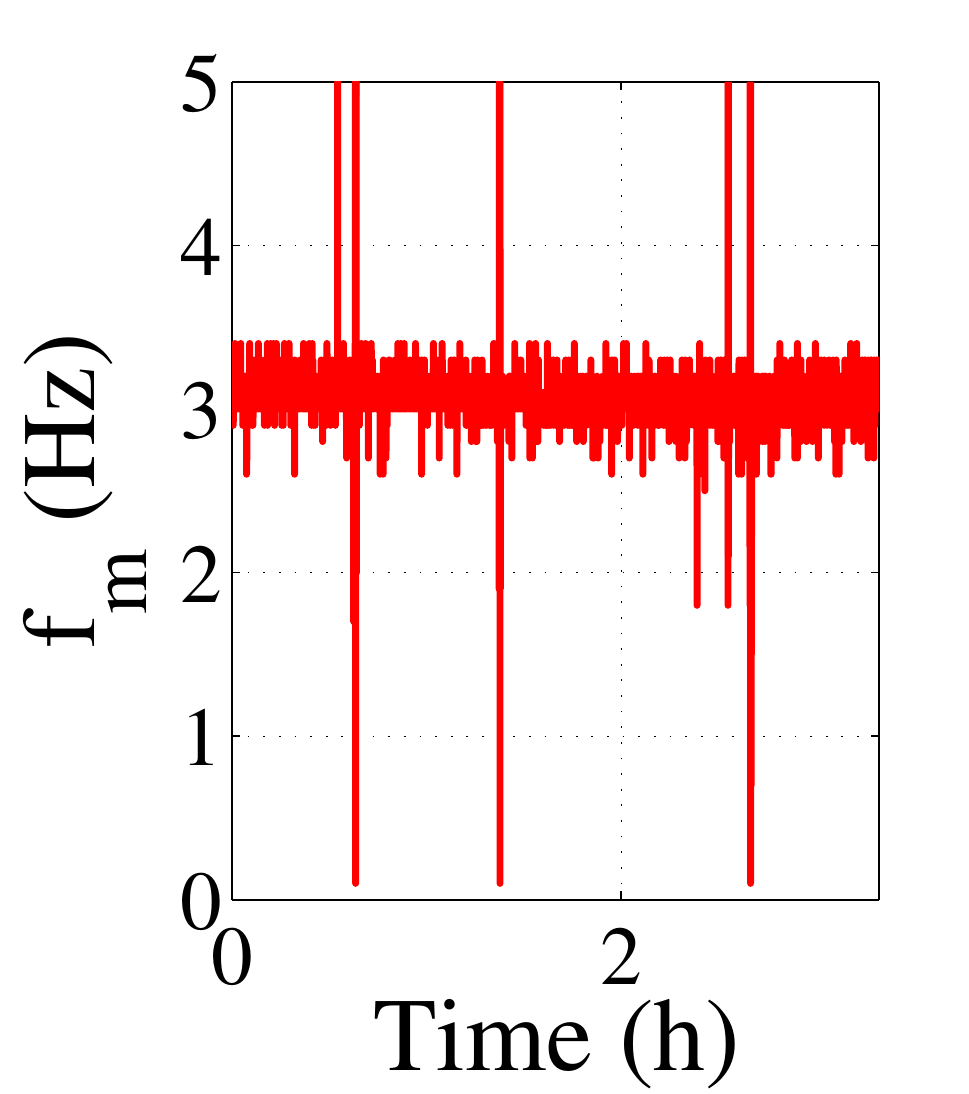}}
\subfigure[\label{fig:MotionAsAfunctionOfTimeHist}]{
\includegraphics[width=0.27\textwidth]{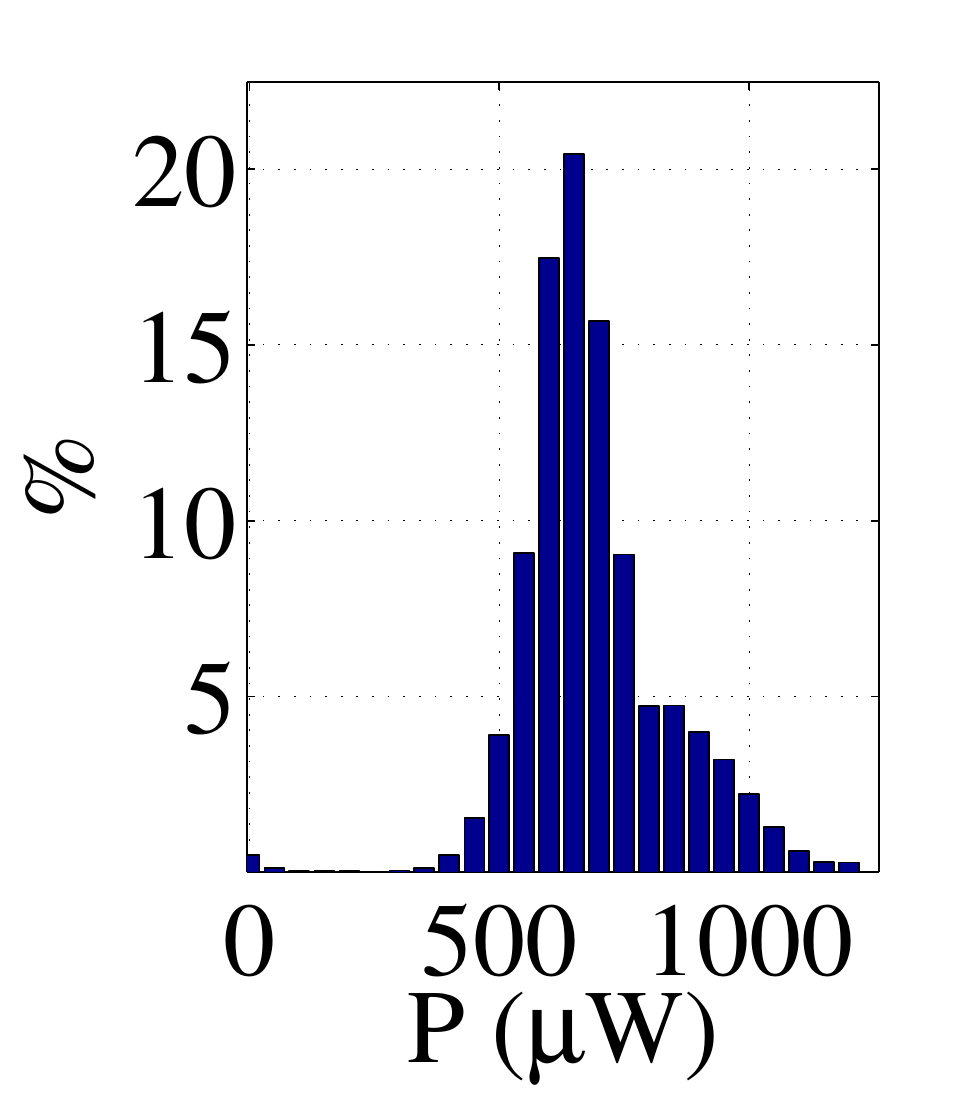}}
\caption{\label{fig:MotionAsAfunctionOfTime} Motion energy characterization for a 3~hour run: (a) the absolute deviation of acceleration, $D$, and (b) dominant motion frequency, $f_m$, as functions of time, and (c) the distribution of the corresponding power harvested, $P(t)$. 
}
\end{figure}
\vspace{-5pt}
\begin{figure}[H]
\centering
\subfigure[]{
\includegraphics[width=0.78\textwidth]{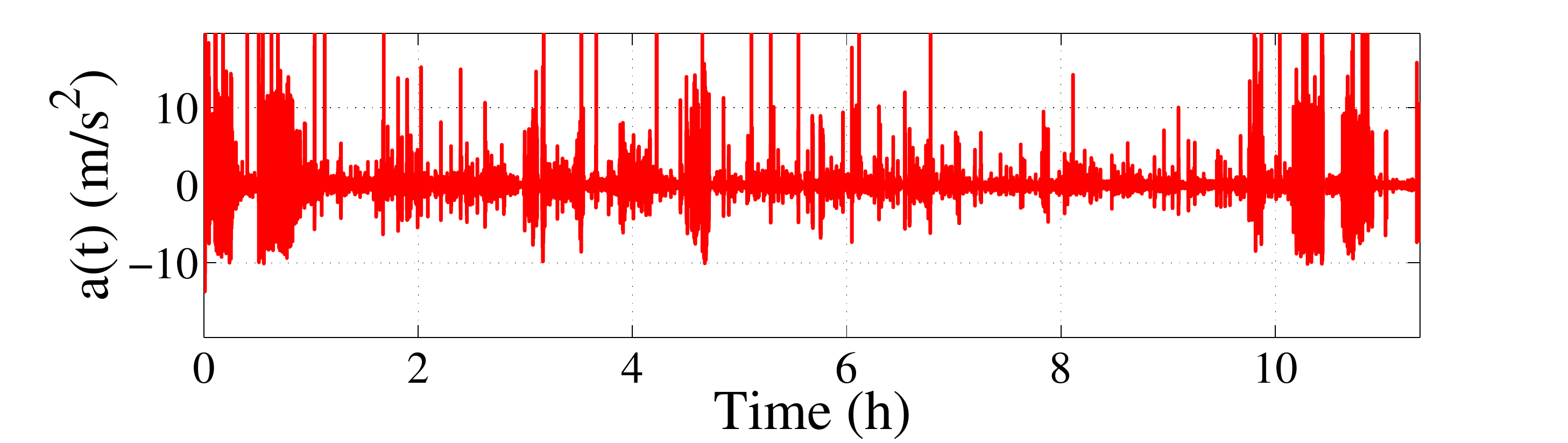}}
\subfigure[]{
\includegraphics[width=0.78\textwidth]{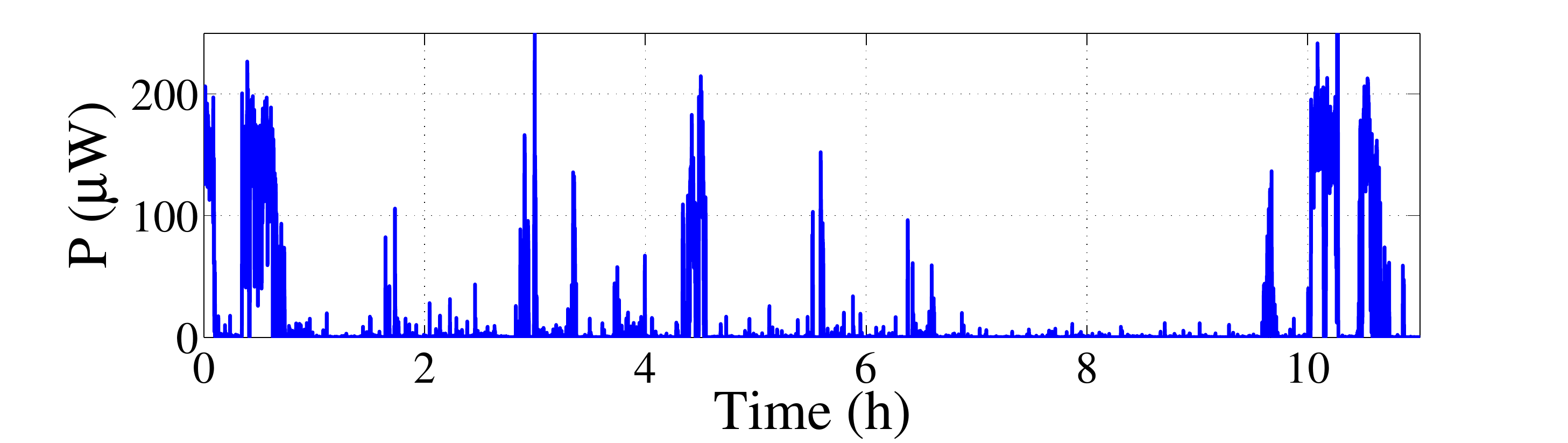}}
\caption{Kinetic energy for normal daily human routine:
(a) acceleration, $a(t)$, recorded over 11~hours for participant $M5$,
and (b) the power harvested, $P(t)$.\label{fig:EnergyGenerationProcessKinetic}}
\end{figure} 

\end{minipage}
\end{minipage}
\vspace*{3.5pt}
\fi

To validate the data from~\cite{xue2010naturalistic}, we replicated the measurements using our sensing units. 
The results of our measurements were consistent with the provided data.
\ifJSACsubmission
\else
 Namely, both the $D$ and $f_m$ values in our experiments were well within range of the data in~\cite{xue2010naturalistic}.
\fi
We note that the $f_m$ values calculated for the different motions in the dataset
are consistent with the physiology of human motion. For example, the range of the calculated $f_m$ values for running motion samples in the dataset corresponds to the typical foot strike cadence for running 
(180 foot strikes per minute, i.e., $f_m = 3$Hz, is considered an
optimal running cadence \cite{galloway1998pick}). 

The statistics of the calculated $D$, $f_m$, and $\overline{P}$ are summarized in the boxplots in Fig.~\ref{fig:ChineseDataResults}. For each of the 7 motions the leftmost (black), middle (red), and rightmost (blue) boxes correspond to the 
\emph{shirt pocket}, \emph{waist belt}, and \emph{trouser pocket} sensing unit placements, respectively.
For each motion and sensing unit placement, the number of participants that had $a(t)$ samples appears on the top of Fig.~\ref{fig:DisplacementChineseData}. At each box, the central mark is the median, the edges are the 25$^{\textrm{th}}$ and 75$^{\textrm{th}}$ percentiles, the ``whiskers'' cover $2.7\sigma$ of the data, and the outliers are plotted individually. 
In Table~\ref{table:InstantaneosDataRates} we separately summarize the results 
for 4 important motions. 


\subsection{Energy for Different Activities} \label{sect:StudySummary}





\par \noindent \textbf{Relaxing}: As expected, almost no energy can be harvested
\ifJSACsingle
for a stationary person
\else
when a person is not moving 
\fi
($\overline{P}<5$~$\mu$W). 

\par \noindent \textbf{Walking and fast walking}: Walking is the predominant periodic motion in normal human lives and thus particularly important for motion energy harvesting.
For walking, the median $\overline{P}$ is 155~$\mu$W for shirt pocket sensing unit placement, 180~$\mu$W for waist belt placement, and 202~$\mu$W for trouser pocket placement. These $\overline{P}$ values
are in agreement with previous 
studies of 
energy harvesting for human walking ~\cite{huang2011human,von2006optimization}.
In comparison, \emph{indoor light energy} availability is on the order of 50--100~$\mu$W/cm$^2$. Considering harvester energy conversion efficiency estimates~\cite{yun2011design,Gorlatova_TMC2013}, \emph{a similarly sized harvester would harvest more energy from walking than from indoor light}.
Fast walking (identified as ``fast'' by the participants themselves) has higher $D$ and $f_m$ than walking at a normal pace (Fig.~\ref{fig:ChineseDataResults}) and generates up to twice as much $\overline{P}$. 

\par \noindent \textbf{Running}: Running, an intense repetitive activity, is associated with high $D$ and $f_m$ (Fig.~\ref{fig:ChineseDataResults}(a,b)), and hence results in 612 $\le$ $\overline{P}$ $\le$ 813~$\mu$W. 

\par \noindent \textbf{Cycling:} For the examined unit placements,
cycling generates relatively little energy -- the median $\overline{P}$ values are 41--52~$\mu$W, 3.7--3.9~times less than the $\overline{P}$ for walking. 
While the high cadence of cycling motion results in relatively high $f_m$ (Fig.~\ref{fig:ChineseDataResults}(b)), a harvester not on the legs will be subject to only small displacements, resulting in small values of $D$ (Fig.~\ref{fig:ChineseDataResults}(a)) and $\overline{P}$ (Fig.~\ref{fig:ChineseDataResults}(c)).
For cycling-specific IoT applications, \emph{harvester placements on the lower legs should be considered}. 

\par \noindent \textbf{Walking upstairs and downstairs:} 
Comparing the $\overline{P}$ values for 
relaxing, walking, and running, one may conclude that higher exertion (perceived effort and energy expenditure) corresponds to higher energy harvesting rates. Our examination of walking upstairs and downstairs 
demonstrates that this is not the case. 
While people exert themselves more going upstairs, the $\overline{P}$  for going downstairs is \emph{substantially higher than for going upstairs}, 
with the median $\overline{P}$ values differing by 1.65--2.1 times depending on the sensing unit placement. 
Although counterintuitive, going downstairs is associated with higher magnitudes of motion and higher motion frequencies ({Fig.~\ref{fig:ChineseDataResults}(a,b)}),
which leads to the higher $\overline{P}$. We observed the disconnect between perceived effort and energy harvesting rates in other measurements as well. For example, in our measurements 
\ifJSACsingle
\else
highly 
\fi
strenuous 
\mbox{push-ups} and \mbox{sit-ups} resulted in lower $\overline{P}$ than non-strenuous walking at a normal pace.

\subsection{Consistency of Dominant Motion Frequency} \label{sect:HumanMotion}


To maximize power output, the resonant frequency of a harvester, $f_r$, should ``match'' the dominant frequency of motion, $f_m$. In this section, we comment on the variability in $f_m$ and provide important observations for harvester design. Due to space constraints, we leave the study of harvester sensitivity to different design parameters to future work. 

\par \noindent \textbf{Consistency among sensing unit placements}: 
The same motion will result in a different $f_m$ depending on the sensing unit's placement on the human body~\cite{huang2011human,yun2011design}. We observed this in measurements that we conducted, especially for sensing units attached to the lower legs and lower arms. However, for the sensing unit placements examined in this section (shirt, waist, and trousers), the same motion resulted in similar $f_m$ values, as can be seen in Fig.~\ref{fig:ChineseDataResults}(b). 
These placements are on or near the torso, and are subjected to similar stresses. Cycling is an exception; the $f_m$ for the trouser placement is different from the other placements. 
Because the body is in a sitting position, the stresses experienced by the legs and the torso are different and $f_m$ differs for the different placements.

The uniformity of $f_m$ offers valuable hints for energy harvesting node designers. People are likely to keep many objects that will become IoT nodes (keys, wallets, and cell phones) in pockets located in places that correspond to the placements we examine. This suggests that \emph{a harvester tuned to a particular $f_m$ will perform well regardless of where a person chooses to carry such an object}. 


\par \noindent \textbf{Inter-participant consistency}: For common periodic motions, such as walking and running, 
the $f_m$ values are relatively consistent among the different participants. The 25$^{\textrm{th}}$ and 75$^{\textrm{th}}$ percentiles of the participants' $f_m$ values are separated by only 0.15~Hz for walking and by only 0.3~Hz for running. For less commonly practiced motions (cycling, going upstairs, going downstairs), the values of $f_m$ are less consistent, but are still somewhat similar.
This consistency indicates that \emph{an all-purpose harvester designed for human walking or running will work reasonably well for a large number of different people}. The next section examines whether harvesters can be tuned to particular human parameters.

\subsection{Dependency on Human Height and Weight} \label{sect:HeightAndWeight}

We examine the dependency of energy availability on human physiological parameters. 
We correlate $D$, $f_m$, and $\overline{P}$ obtained
for different motions and different participants with their height and weight data from~\cite{xue2010naturalistic}.\footnote{The dataset~\cite{xue2010naturalistic} is also annotated with participants' age and gender.
However, the age range (20 to 23 years) and the number of females (10 participants) are insufficient for obtaining statistically significant correlations.} The participants' heights range was 155--182~cm, and their weights range was 44--65~kg. We verified that, in agreement with general human physiology studies, the participants' height and weight are strongly positively correlated ($\rho=0.7$, $p<0.001$).

As indicated in the previous subsection, for many activities $f_m$ is consistent among different participants. Yet, we additionally observed $f_m$ dependencies on human physiology. For many of the activities we examined, we determined \emph{negative correlations} \emph{of $f_m$} \emph{with the participants'} \emph{height} \emph{and weight}. When walking, running, and going upstairs and downstairs, heavier and taller people took fewer steps per time interval than lighter and shorter people.

For example, for going upstairs with waist unit placement, $f_m$ and the participant's height are correlated as $\rho = -0.34$ ($p = 0.03$, $n = 39$). When going upstairs, the taller half of the participants made, on average, 9 fewer steps per minute ($0.15$~Hz) than the shorter half ($f_m =1.85$ and $2.05~$Hz, correspondingly).
For running, with trouser placement, $f_m$ and the participant's weight are correlated as $\rho= -0.46$ ($p < 0.01$, $n = 39$). When running, the heavier half of the participants made, on average, 18 fewer steps per minute ($0.3$~Hz) than the lighter half. This suggests that future harvester designs may benefit from \emph{targeting harvesters} \emph{with different $f_r$ values} \emph{for human groups} \emph{with different physiological parameters}. For example, \emph{different harvesters} \emph{may be integrated in} \emph{clothing of different sizes}. 



Generally, motion energy availability increases as $f_m$ increases~\cite{Mitcheson2008}. However, in human motion, other dependencies may additionally come into play. In our study, for running with trouser unit placement, we determined a positive correlation between $D$ and participants' height ($\rho = 0.35$, $p = 0.03$, $n = 38$) 
and a positive correlation between $\overline{P}$ and participants' height ($\rho = 0.38$, $p = 0.01, n = 38$). 
\emph{For the taller half of the participants}, \emph{the average $\overline{P}$ is} \emph{20\% higher} \emph{than for the shorter half} (704 and 582~$\mu$W, respectively). Studies with larger number of participants, wider participant demographics, and wider range of participant parameters will most likely identify many additional dependencies. This
will allow harvester designers to \emph{develop harvesters for different demographics},
as well as to provide guarantees on the performance of different harvesters based on different human parameters.

\section{Long-term Human Mobility}
\label{sect:DailyEnergy}

\ifJSACsingle
\else

\begin{table*}[t]
\centering
\scriptsize
\caption{Energy budgets, variability, and data rates based on collected traces for daily human routines.\label{table:longTermResults}}
\begin{tabular}{|p{0.6cm}|p{3.5cm}|p{0.4cm}|p{0.5cm}|p{2.1cm}|p{2.1cm}|p{0.65cm}|p{2.1cm}|p{2.1cm}|}
  \hline
   Par-& Occupation and commute& \#   & Total  & \multicolumn{2}{c|}{Optimized harvester} & $r_d$,  & $\overline{P}^{H4}$ ($\mu$W), & \% ON, \\ \cline{5-6}
   tici-pant   &  & days & dur. (h) & $\overline{P}$ ($\mu$W), min/avg/max  & $\overline{P_d}$ ($\mu$W), min/avg/max  & avg (Kb/s) & min/avg/max   & min/avg/max  \\\hline

 \textbf{$M1$}& Undergraduate student, male, living on campus,  always goes to the lab  & 5 & 60.4 & 6.9 / 13.8 / 17.3 & 4.8 / 6.5 / 8.1 & 1.3 & 5.0 / 8.5 / 10.9 & 5.4 / 9.9 / 12.2 \\ \hline
 \textbf{$M2$}& Undergraduate student, male,  commuting to campus,  always goes to the lab  & 
 3 & 27.7 & 23.3 / 29.0 / 38.2 & 8.4 / 11.5 / 17.7 & 2.3 & 17.1 / 19.6 / 24.5 & 13.6 / 16.1 / 18.4 \\ \hline 
 \textbf{$M3$}& Undergraduate student, female, living on campus, sometimes works from home & 
 9 & 62.0 & 2.4 / 7.16 / 13.4 & 0.6 / 2.02 / 3.6 & 0.4 & 2.0 /  5.8 / 12.2 & 3.6 / 6.0 /9.95\\ \hline 
 \textbf{$M4$}& Graduate student, female, commuting to campus, sometimes works from home& 
 7 & 80.1 & 1.4 / 11.98 / 25.3 & 0.6 / 5.6 / 10.7 &  1.1 & 1.4 / 11.98 / 25.3 & 2.8 / 12.7 / 18.1 \\ \hline
  \textbf{$M5$}& Software developer, male, commuting to office, always goes to the office & 
 1 & 11.0 & 16.3 & 7.5 & 1.5 & 15.9 & 11.5 \\ \hline
\end{tabular}
\vspace*{-0.45 cm}
\end{table*}
\fi

The results presented in the previous section are based on short motion samples from an activity recognition dataset. In this section, we present results of our own, \emph{longer-term, motion measurements}. 
We describe our set of \mbox{day-long} human mobility measurements and discuss energy budgets and generation process properties.

\subsection{Prolonged Activities} \label{sect:ActivityDuration}

\ifJSACsingle
\else

\begin{figure}[t]
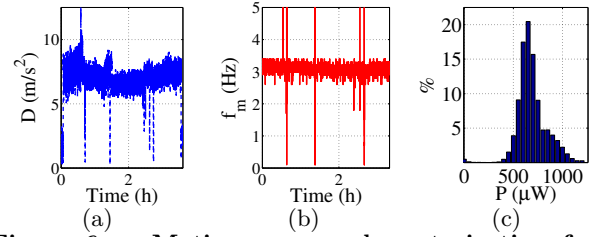
 
\centering
\subfigure[]{ 
\includegraphics[width=0.3\columnwidth]{fig/acceleration/LongRunAbsDeviationsShort.eps}}
\subfigure[\label{fig:LongRunPeakFrequency}]{
\includegraphics[width=0.3\columnwidth]{fig/acceleration/LongRunPeakFrequenciesShort.eps}}
\subfigure[\label{fig:MotionAsAfunctionOfTimeHist}]{
\includegraphics[width=0.3\columnwidth]{fig/acceleration/LongRunPowerHistShort.eps}}
\ifJSACdouble
\else
\vspace*{-0.35cm} 
\fi
\caption{\label{fig:MotionAsAfunctionOfTime} Motion energy characterization for a 3~hour run: (a) the absolute deviation of acceleration, $D$, and (b) dominant motion frequency, $f_m$, as functions of time, and (c) the distribution of the corresponding power harvested, $P(t)$. 
}
\end{figure}
\fi

To study motion energy properties over time, we collected a set of measurements of longer activity durations (over 20~minutes). We considered long
walks, bike rides, runs, and other activities, performed in normal
environments (i.e., not on a  treadmill or a stationary bike). To the best of
our knowledge, the properties of 
\ifJSACsingle
longer motion samples were not
\else
motion of longer samples have not been 
\fi
analyzed before. 

The measurements demonstrate that for prolonged activities, $D$, $f_m$, and $P(t)$ vary substantially over time.
This variability is related to physiological parameters, such as changes in cadence or posture over time due to fatigue, and changes in the surrounding environment, such as traffic lights, terrain changes, or pedestrian traffic. For example, Fig.~\ref{fig:MotionAsAfunctionOfTime} shows $D$, $f_m$, and $P$ corresponding to a 3~hour run, calculated for 1-second $a(t)$ intervals. 
In this trace, the average $D$ changes subtly over time (Fig.~\ref{fig:MotionAsAfunctionOfTime}(a)),
and $f_m$ varies continuously in the 2.6--3.4~Hz range (Fig.~\ref{fig:MotionAsAfunctionOfTime}(b)). Correspondingly, while the mean $P(t)$ is 550~$\mu$W, the 10$^{\textrm{th}}$--90$^{\textrm{th}}$ percentiles of the $P(t)$ span the range of 459--710~$\mu$W (Fig.~\ref{fig:MotionAsAfunctionOfTime}(c)).

The variability of $P(t)$ 
throughout an activity suggests that node energy management policies are essential even for specifically targeted IoT applications, such as nodes for fitness runners or cyclists.
In the following section we demonstrate even more variability in $P(t)$ 
for the 
regular everyday human mobility patterns. 



\ifJSACsingle
\begin{table*}[t]
\centering
\scriptsize
\caption{Energy budgets, variability, and data rates based on collected traces for daily human routines.\label{table:longTermResults}}
\ifJSACsingle
\begin{tabular}{|p{0.6cm}|p{3.3cm}|p{0.35cm}|p{0.6cm}|p{2.1cm}|p{2.1cm}|p{0.8cm}|p{1.4cm}|p{1.4cm}|}
  \hline
\else
\begin{tabular}{|p{0.6cm}|p{2.9cm}|p{0.5cm}|p{1.0cm}|p{2.1cm}|p{2.1cm}|p{0.8cm}|p{2.1cm}|p{2.1cm}|}
  \hline
\fi
   Par-& Occupation and com-& \#   & Total  & \multicolumn{2}{c|}{Optimized harvester} & $r_d$,  & $\overline{P}^{H4}$ ($\mu$W), & \% ON, \\ \cline{5-6}
   tici-pant   & mute & days & dur.(h) & $\overline{P}$ ($\mu$W), min/avg/max  & $\overline{P_d}$ ($\mu$W), min/avg/max  & avg (Kb/s) & min/avg/max   & min/avg/max  \\\hline

 \textbf{$M1$}& Undergraduate student, male, living on campus, always goes to the lab & 5 & 60.4 & 6.9 / 13.8 / 17.3 & 4.8 / 6.5 / 8.1 & 1.3 & 5.0 / 8.5 / 10.9 & 5.4 / 9.9 / 12.2 \\ \hline
 \textbf{$M2$}& Undergraduate student, male,  commuting to campus,  always goes to the lab & 
 3 & 27.7 & 23.3 / 29.0 / 38.2 & 8.4 / 11.5 / 17.7 & 2.3 & 17.1 / 19.6 / 24.5 & 13.6 / 16.1 / 18.4 \\ \hline 
 \textbf{$M3$}& Undergraduate student, female, living on campus, sometimes works from home & 
 9 & 62.0 & 2.4 / 7.16 / 13.4 & 0.6 / 2.02 / 3.6 & 0.4 & 2.0 /  5.8 / 12.2 & 3.6 / 6.0 /9.95\\ \hline 
 \textbf{$M4$}& Graduate student, female, commuting to campus, sometimes works from home & 
 7 & 80.1 & 1.4 / 11.98 / 25.3 & 0.6 / 5.6 / 10.7 &  1.1 & 1.4 / 11.98 / 25.3 & 2.8 / 12.7 / 18.1 \\ \hline
  \textbf{$M5$}& Software developer, male, commuting to office, always goes to the office & 
 1 & 11.0 & 16.3 & 7.5 & 1.5 & 15.9 & 11.5 \\ \hline
\end{tabular}
\vspace*{-0.25 cm}
\end{table*}
\fi

\subsection{Day-Long Human Mobility} \label{sect:EnergyBudgets}
\ifJSACsingle
\else

\begin{figure}[t]
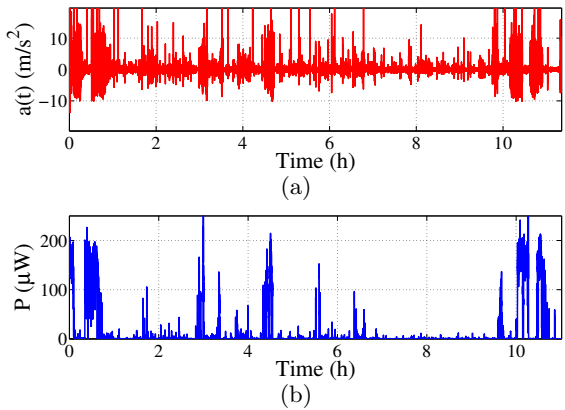

\centering
\subfigure[]{
\includegraphics[width=\columnwidth]{fig/acceleration/ExampleEnergyGenerationProcess_Acceleration2.eps}}
\subfigure[]{
\includegraphics[width=\columnwidth]{fig/acceleration/ExampleEnergyGenerationProcess.eps}}
\vspace*{-0.35cm}
\caption{Kinetic energy for normal daily human routine:
(a) acceleration, $a(t)$, recorded over 11~hours for participant $M5$,
and (b) the power harvested, $P(t)$.\label{fig:EnergyGenerationProcessKinetic}}
\end{figure} 
\fi


To determine the daily energy available to an IoT node with an inertial harvester, we collected acceleration traces from different participants during their normal daily routines. We obtained over 200~hours of acceleration information for 5~participants for a total of 25~days. 
The participants (see Table~\ref{table:longTermResults}) were instructed to
carry a sensing unit in any convenient way. Thus, the measurements correspond
to the motion that a participant's keys, mobile phone, or wallet would
experience.
\ifJSACsubmission
\else
Typical daily activities of the participants included commuting to office or lab (by train, bus, or subway), working on the computer or on an electronics bench, attending meetings, and walking outside for lunch and coffee breaks.
\fi

Fig.~\ref{fig:EnergyGenerationProcessKinetic} shows the $a(t)$ for a day-long trace of participant $M5$, and the corresponding $P(t)$. 
For all the collected traces,
the dominant motion frequency, $f_m$, 
range is 1.92--2.8~Hz, corresponding to human walking.

The calculated energy budgets 
are summarized in Table~\ref{table:longTermResults}. We calculated $\overline{P}$, the average power a harvester would generate over the length of the trace, as well as $\overline{P_d}$, the average power a harvester would generate over a 24-hour interval. To calculate $\overline{P_d}$ we assumed that when the sensing unit did not record data (e.g., before the participants got dressed for school or work), it was stationary and that a harvester would not generate energy during these intervals.  Specifically, for a $T$~hour-long measurement trace, $\overline{P_d} = \overline{P}\cdot T /24$. For each of the participants,  Table~\ref{table:longTermResults} summarizes the minimum, average, and maximum $\overline{P}$ and $\overline{P_d}$ over the different measurement days, and the data rate $r_d$ that a node would be able to maintain 
continuously throughout a day when powered by the harvested $\overline{P}_d$. 
For completeness, for all participants we additionally calculate 
$\overline{P}^{H4}$, the average power a particular harvester, same for all participants (in this case, the harvester calculated based on the traces for participant $M4$), would harvest. 
 An extensive examination of the sensitivity of power harvested 
 to different harvester design parameters is subject of ongoing work.

\subsubsection{Power Budgets}
For most participants, an inertial harvester can provide sufficient power to continuously maintain a data rate of at least 1~Kb/s (i.e., $\overline{P_d}>5~\mu$W). This is comparable with the data rates estimated in~\cite{Gorlatova_TMC2013} for nodes with a similarly sized \emph{light} harvester in indoor environments (not exposed to outdoor light).

The majority of inter-participant and inter-day differences seem to relate to
the \emph{participants' amount of walking}. For example, participant $M2$,
whose $\overline{P}$ and $\overline{P_d}$ values are higher than the others,
has a relatively long walk to the office, and walks frequently between two
different offices in the same building. Other factors (unit placement, amount
of daily activity as perceived by the participants) appear to be only of
secondary importance. We note that the majority of traces that correspond to
$\overline{P_d}<5$~$\mu$W (and thus $r_d<$1~Kb/s) correspond to participants
\emph{working from home}\footnote{As
      indicated by the minimum value of $\overline{P_d}$ in
      Table \ref{table:longTermResults}, several individual traces with
      $\overline{P_d}<5$~$\mu$W were considered for participants M1, M3, and
      M4.}. Overall, daily routines that involve a lot of walking correspond
to relatively high levels of energy
availability. 




\ifJSACsingle

\begin{figure}[t] 
\centering        
\subfigure[] {\includegraphics[width=0.45\textwidth]{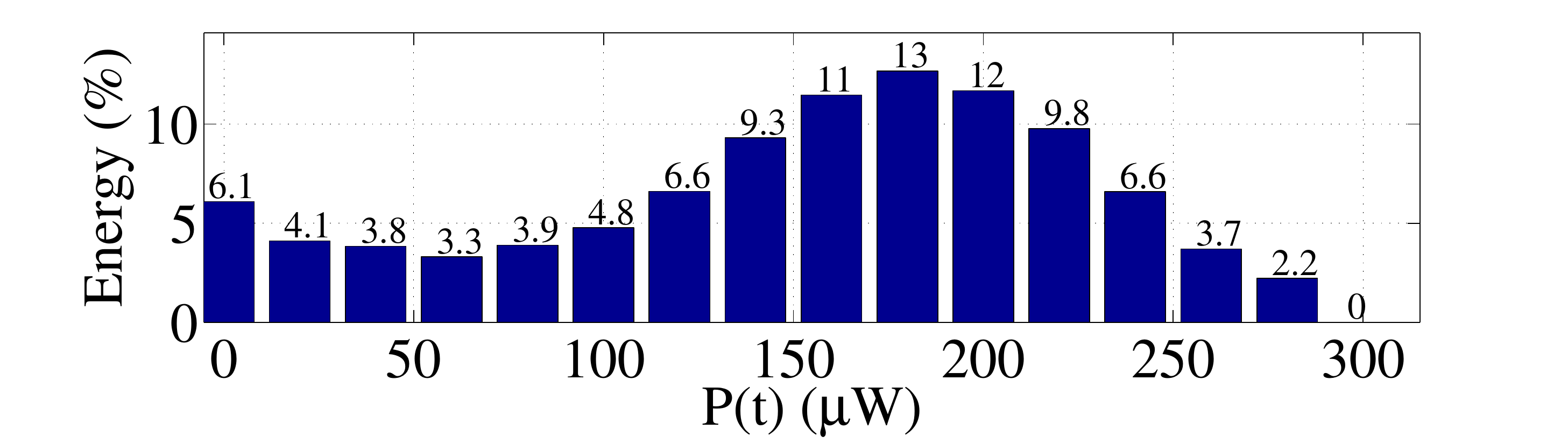}
\label{fig:PercPowerAtPowerLevels}}
\subfigure[]
{\includegraphics[width=0.45\textwidth]{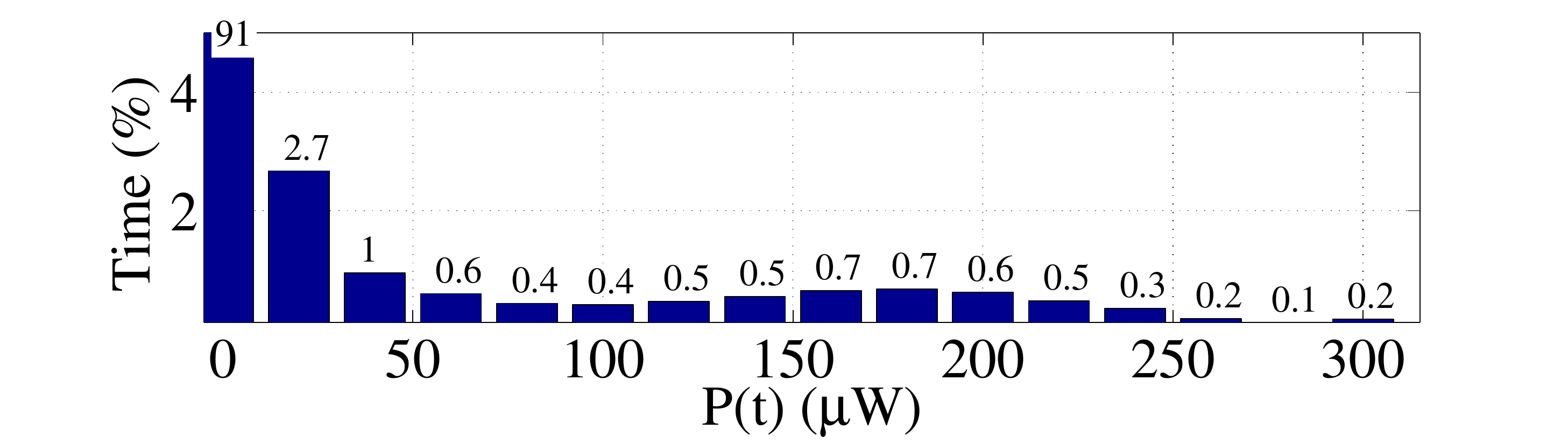}
\label{fig:PercTimeAtPowerLevels}}
\vspace*{-0.35cm}
\caption{Motion energy harvesting process variability for participant $M1$: (a) the percentage of the total energy harvested at different power levels $P(t)$, and (b)
the \emph{percentage of time} the power is harvested at the different $P(t)$ (notice that for $0\leq P(t) \leq 15$, the value is 91\%).\label{fig:PercAtLevels}}
\end{figure}

\else
\begin{figure}[t] 
\centering        
\subfigure[] {\includegraphics[width=0.95\columnwidth]{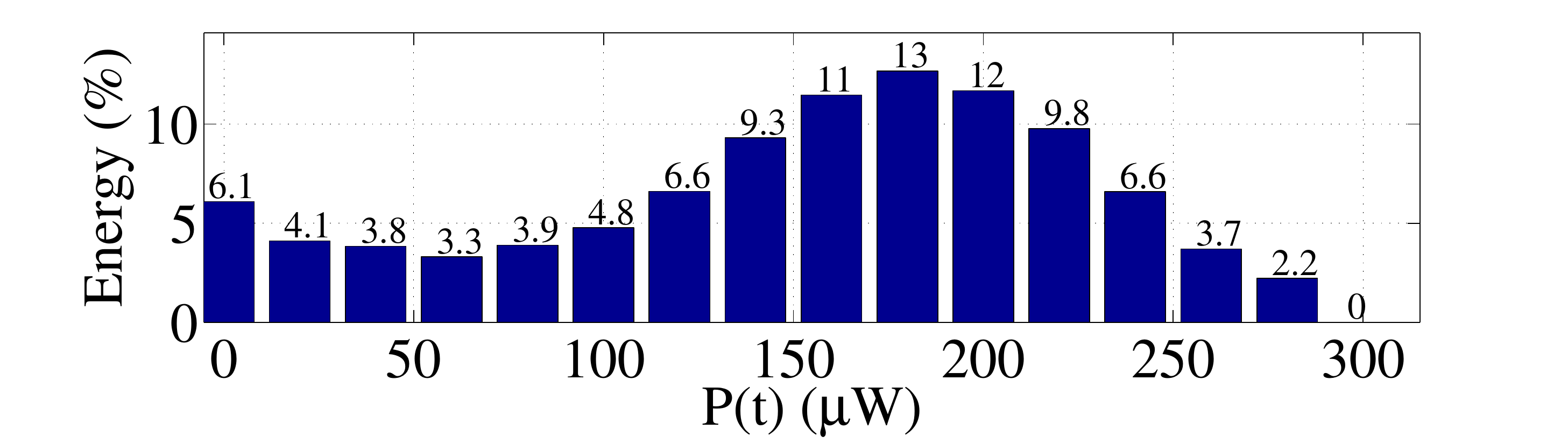}
\label{fig:PercPowerAtPowerLevels}}
\subfigure[]
{\includegraphics[width=0.95\columnwidth]{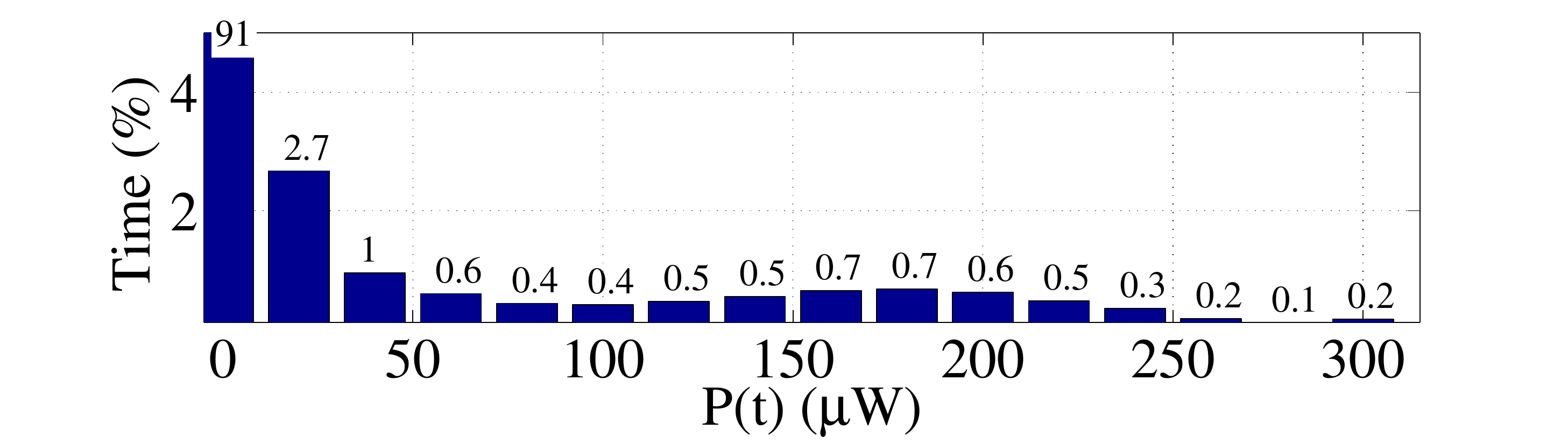}
\label{fig:PercTimeAtPowerLevels}}
\vspace*{-0.35cm}
\caption{Motion energy harvesting process variability for participant $M1$: (a) the percentage of the total energy harvested at different power levels $P(t)$, and (b)
the \emph{percentage of time} the power is harvested at the different $P(t)$ (notice that for $0\leq P(t) \leq 15$, the value is 91\%).\label{fig:PercAtLevels}}
\end{figure}
\fi

\subsubsection{Harvesting Process Variability \& Properties} \label{sect:EHprocessVariability}

The amount of energy that can be harvested varies widely throughout the day.
As shown in Section~\ref{sect:perActivityEnergy}, walking generates substantial amounts of energy, while being stationary generates little.
Physiological studies (e.g.,~\cite{orendurff2008humans})
have shown that people are at rest the majority of the time.
Correspondingly, in our measurements, $P(t)$ is low for most of the day and over 95\% of the total energy is collected during only 4--7\% of a day. For example, Fig.~\ref{fig:PercAtLevels} shows, 
for participant $M1$,
the percentage of the total energy that would be harvested over different ranges of $P(t)$
and 
the \emph{percentage of the time} that the harvester would generate these $P(t)$ values. For this participant, the harvester would generate $P(t)<15$~$\mu$W 91\% of the time, and only 6.1\% of the total energy would be harvested during this time.

Consider an \emph{ON/OFF representation} of the energy harvesting process, $P_{\textrm{onoff}}(t)$, where $P_{\textrm{onoff}}(t) \gets ON$ 
if $P(t) >\gamma$, and $P_{\textrm{onoff}}(t) \gets OFF$ 
otherwise.
For the analysis below, we empirically set $\gamma =$ 10~$\mu$W; the results are similar for 10 $\le$ $\gamma$ $\le$ 40~$\mu$W. For all participants, $P_{\textrm{onoff}}$ is ON for less than 
20\% of the time (Table~\ref{table:longTermResults}). 
The participants do not lead sedentary lifestyles; their activity patterns are in line with general health guidelines. However, the generally recommended 30~minutes of physical activity per day correspond to only 9\% of an \mbox{11-hour} trace.
Additionally, the typical \emph{duration} of ON intervals is short -- \emph{on the order of seconds}. While some of the ON intervals are long (over 200~seconds), the vast majority of the ON intervals (78.5--89.0\%) are shorter than 30~seconds; the \emph{median ON intervals are 5--9.5~seconds}. The longer ON intervals correspond to commuting (e.g., walking from a public transit station to a campus building), and represent only
1--3\% of the ON intervals.
These results are consistent with the overall results for
  \emph{walking intervals} examined in a physiological study of human mobility
  \cite{orendurff2008humans}.

In summary, $P(t)$ is low for the majority of the time, and when it does become high, it stays high for only a brief period of time. This 
emphasizes the need for energy harvesting-adaptive algorithms. 

\subsubsection{Harvesting Process vs.~I.i.d.~and Markov Processes} \label{sect:PolicyPerformanceSimplifiedRepr}

\ifJSACsingle

\else

\begin{figure}[t]
\centering 
\subfigure[]{\includegraphics[width=0.485\columnwidth]{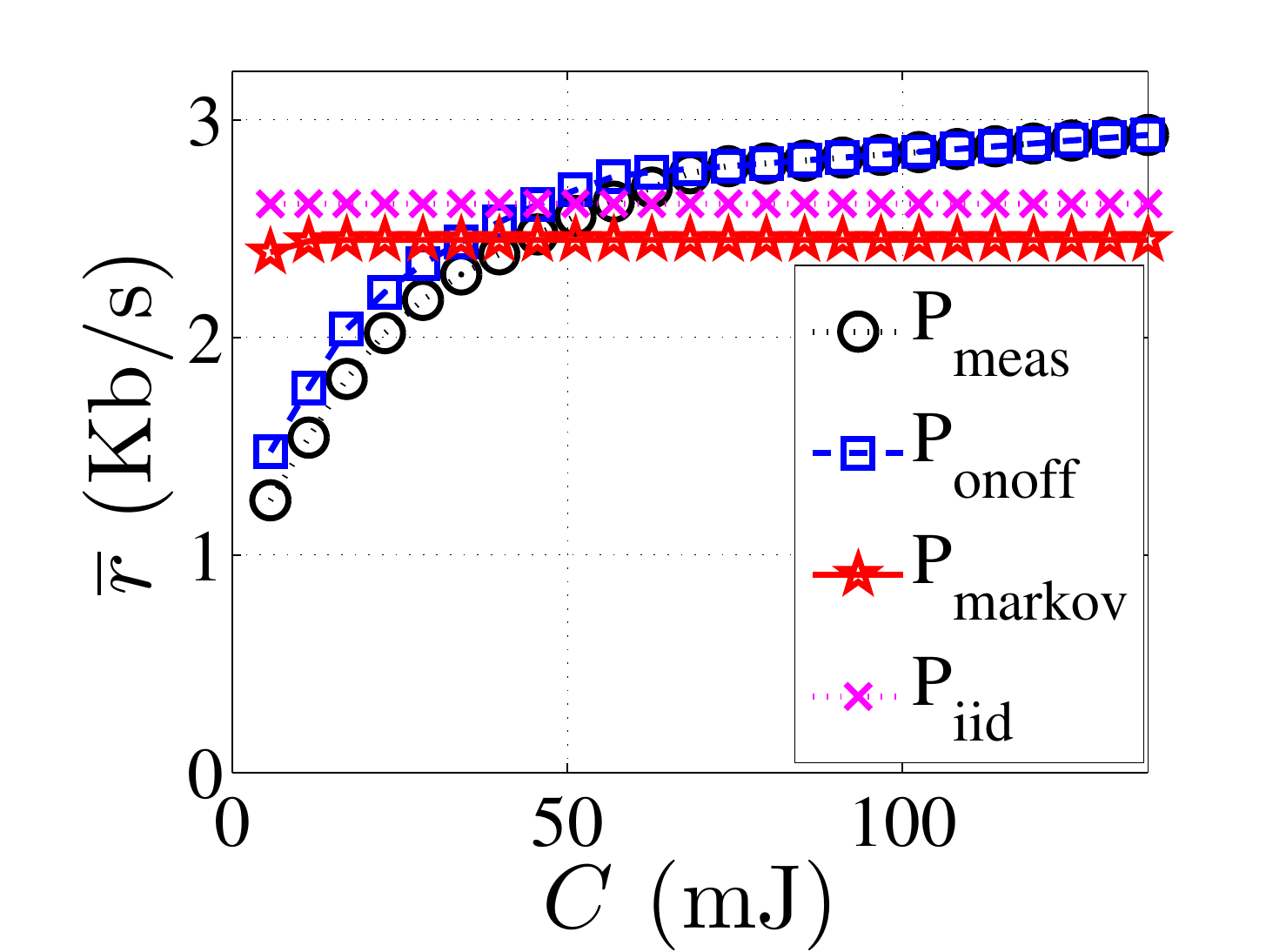}}
\subfigure[\label{fig:MarkovIidReal}]
{\includegraphics[width=0.485\columnwidth]{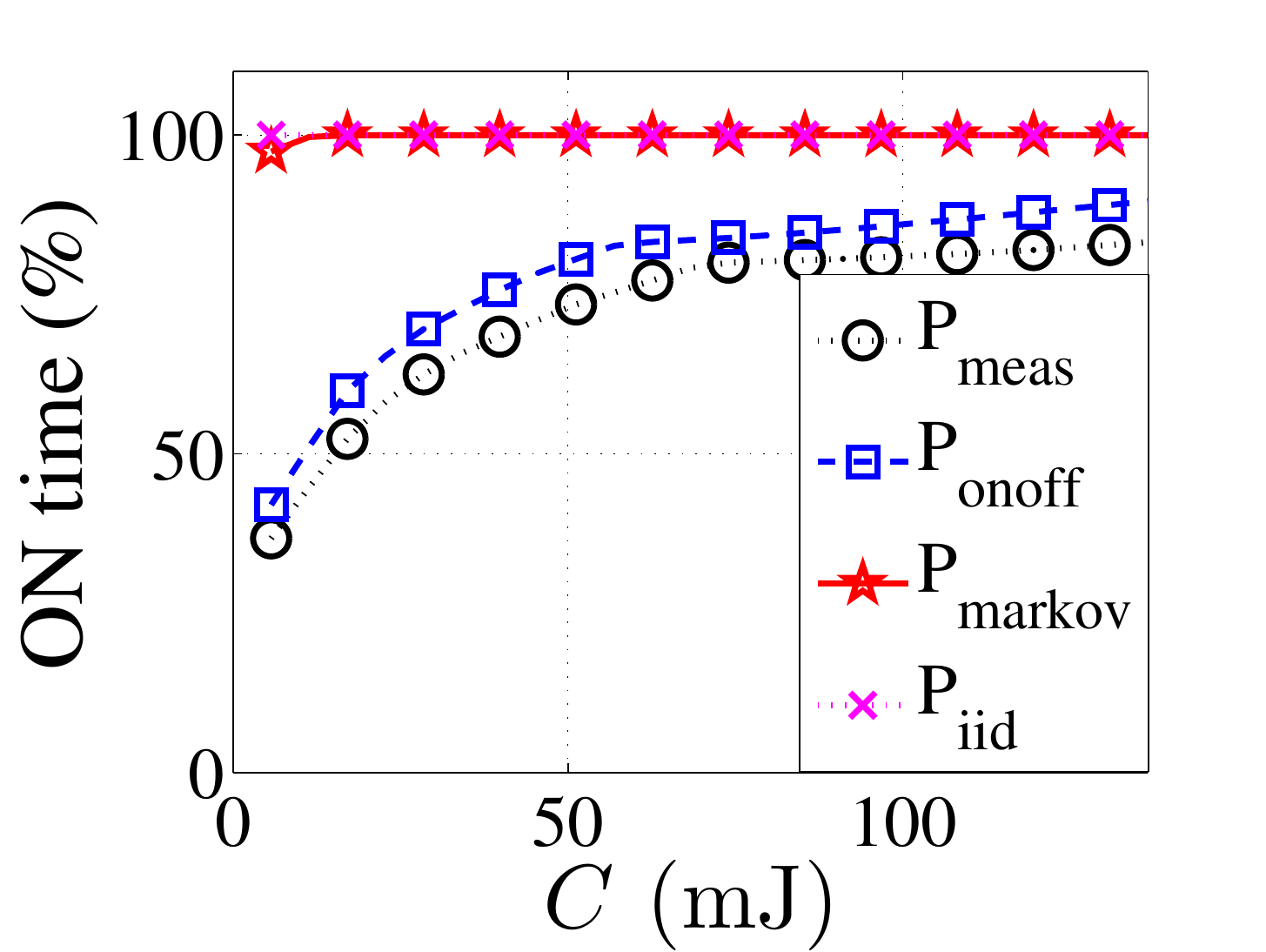}}
\ifJSACdouble
\else
\vspace{-0.35cm}
\fi
\caption{\label{fig:policyPerformanceSimplRepres} \mbox{Scheme-LB} policy performance using energy traces ($P_{\textrm{meas}}$) for participant $M1$ and using the corresponding ON/OFF ($P_{\textrm{onoff}}$), Markov ($P_{\textrm{markov}}$), and \mbox{i.i.d.}\ ($P_{\textrm{iid}}$) processes:
(a) average data rates, $\overline{r}$, and (b) node ON times.}
\end{figure}
\fi


Several energy harvesting adaptive algorithms were developed under the assumption
that the energy harvesting process is Markov, or has independent identically distribu\-ted (i.i.d.) per-slot energy inputs~\cite{Gorlatova_TMC2013,huang2010utility,Wang2013WhenSimplicity}.
However, such assumptions, 
realistic in certain scenarios~\cite{Gorlatova_TMC2013}, \emph{do not hold for our motion energy traces}.
We use a slotted representation of the energy harvesting processes, $P_{\textrm{meas}}$,
setting the time slot length $T_{\textrm{int}}=1$~second, and determining the $P_{\textrm{meas}}(i)$ by computing the average value of the $P(t)$ for each $T_{\textrm{int}}$. For all day-long traces, $P_{\textrm{meas}}$ is \emph{clearly not i.i.d.\ or Markovian}. For example, for the $P_{\textrm{meas}}$ for participant $M1$ for $\gamma=20$~$\mu$W,
$p(P_{\textrm{meas}}(i)>\gamma| P_{\textrm{meas}}(i-1)>\gamma)= 0.84$, while
$p(P_{\textrm{meas}}(i)>\gamma| P_{\textrm{meas}}(i-1)>\gamma,$ $P_{\textrm{meas}}(i-2)<\gamma)= 0.45$.

To demonstrate the differences between the traces and i.i.d.\ and Markov processes,
we examine the performance of the \mbox{Scheme-LB} policies~\cite{Chen2012Asymptotically} with the different processes. In the Scheme-LB policies~\cite{Chen2012Asymptotically}, 
$s(i) \gets (1-\epsilon) \hat{Q}(i)$ if $B(i)+Q(i)\geq (1-\epsilon) \hat{Q}(i)$, and $s(i) \gets B(i)+Q(i)$ otherwise, where $\hat{Q}(i)$ is the running average of $Q(i)$ ($\hat{Q}(i) \gets$ $\sum_{j = 0}^{i-1} Q(j)/i$), and $\epsilon$ is a small constant (we use $\epsilon=0.01$). For a process $P_{\textrm{meas}}$, we generate a corresponding i.i.d.\ process, $P_{\textrm{iid}}$, by randomly permuting the values of $P_{\textrm{meas}}$
(we use the Wald-Wolfowitz runs test to verify the independence of the $P_{\textrm{iid}}$ values). To generate a Markov process, $P_{\textrm{markov}}$, we calculate the empirical state transition probabilities of the $P_{\textrm{onoff}}$ process (defined in Section~\ref{sect:EHprocessVariability}) and 
generate a Markov process with states $\{$ON,OFF$\}$ and the calculated transition probabilities. 
We set the $P_{\textrm{markov}}$ values for ON and OFF states to the average values of  $P_{\textrm{meas}}(i)$ for which $P_{\textrm{onoff}}(i)= ON$, and for which $P_{\textrm{onoff}}(i)=OFF$, respectively.
This ensures that \emph{the processes have the same first-order statistics}.\footnote{
For each of the processes, we calculate $Q(i)$ as $Q(i) \gets \eta_h \cdot T_{\textrm{int}} \cdot P(i)$, where $\eta_h=20$\%~\cite{yun2011design}.
We rely on a \emph{battery} node model and set 
$B_0 = 0.5C$. We calculate the data rate as $r(i) \gets s(i)/c_{\textrm{tx}}$.}

The policy performance observed using i.i.d.\ and Mar\-kov processes differs dramatically from the policy performance observed using the traces. For example, Fig.~\ref{fig:policyPerformanceSimplRepres} shows the $\overline{r}$ and the ON times obtained under the Sche\-\mbox{me-LB}~po\-licy for the different processes based on a trace of participant $M1$. Using the process $P_{\textrm{onoff}}$, the performance is similar to the performance obtained using $P_{\textrm{meas}}$ -- the $\overline{r}$ values differ by at most 17\% (0.23~Kb/s),
and the ON times differ by at most 7\%. However, the performance observed using $P_{\textrm{iid}}$ and $P_{\textrm{markov}}$ differs greatly from the performance observed using $P_{\textrm{meas}}$. The differences in $\overline{r}$ values reach over~105\% (1.35~Kb/s), and the differences in ON times reach~63\%.

Moreover, using i.i.d.\ and Markov processes results in \emph{different performance trends}.
Using $P_{\textrm{meas}}$, the performance strongly depends on $C$, with $\overline{r}$ for the different values of $C$ differing by over 2.3~times, and with the ON percentages differing by over 45\%. However, using $P_{\textrm{iid}}$ and $P_{\textrm{markov}}$, both $\overline{r}$ and ON times are nearly independent of $C$.
Additionally, evaluating policy performance using $P_{\textrm{meas}}$ shows that the ON times are an important metric because they can be low for small values of $C$ (Fig.~\ref{fig:policyPerformanceSimplRepres}(b)). However,
when evaluating using $P_{\textrm{iid}}$ and $P_{\textrm{markov}}$, the ON times are nearly 100\% for all values of $C$, including values as low as 15~mJ (i.e., less than 15\% of the average energy harvested per day).
This \emph{emphasizes the need to evaluate energy harvesting-adaptive policies for wireless nodes equipped with an inertial harvester using real traces}.

\ifJSACsingle
\noindent
\begin{minipage}[c]{\textwidth}
\begin{minipage}[l]{0.6\textwidth}
\begin{figure}[H]
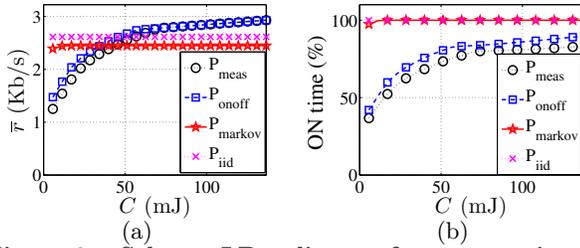

\centering 
\subfigure[]{\includegraphics[width=0.4\textwidth]{fig/simulations/RMarkovIid_V2.eps}}
\subfigure[\label{fig:MarkovIidReal}]
{\includegraphics[width=0.4\textwidth]{fig/simulations/OutageMarkovIid_V2.eps}}
\vspace{-0.35cm}
\caption{\label{fig:policyPerformanceSimplRepres} \mbox{Scheme-LB} policy performance using energy traces ($P_{\textrm{meas}}$) for participant $M1$ and using the corresponding ON/OFF ($P_{\textrm{onoff}}$), Markov ($P_{\textrm{markov}}$), and \mbox{i.i.d.}\ ($P_{\textrm{iid}}$) processes:
(a) average data rates, $\overline{r}$, and (b) node ON times.}
\end{figure}
\end{minipage}
\hfill
\begin{minipage}[r]{0.4\textwidth}
\begin{table}[H]
\centering
\scriptsize
\caption{Object motion measurements. \label{table:objectMotionExperiments}}
\begin{tabular}{|p{3.0cm}|p{1.25cm}|} \hline
Scenario & $\overline{P}$ \\ \hline  
Taking a book off a shelf & $<$10~$\mu$W \\
Putting on reading glasses & $<$10~$\mu$W \\
Reading a book & $<$10~$\mu$W \\
Writing with a pencil & 10--15~$\mu$W \\
Opening a drawer & 10--30~$\mu$W \\
Spinning in a swivel chair & $<$10~$\mu$W \\
Opening a building door & $<$1~$\mu$W \\
Shaking an object & $>$3,000~$\mu$W \\ \hline
\end{tabular}
\end{table}
\end{minipage}
\end{minipage}
\fi

\section{Object Motion Energy}
\label{sect:Intuition}

While Sections~\ref{sect:perActivityEnergy} and \ref{sect:DailyEnergy} focus on \emph{human motion}, in this section we also provide some brief observations regarding the energy availability associated with the \emph{motion of objects}. We conducted extensive experiments, recording $a(t)$ and calculating $\overline{P}$ for a wide range of motions. Our experiments included performing everyday activities with a variety of everyday objects (see Table \ref{table:objectMotionExperiments}), shipping a FedEx box with a sensing unit in it from Houston, TX to New York, NY, transporting sensing units in \mbox{carry-on} and checked airport luggage, and taking sensing units on cars, subways, and trains.
Below, we present observations based on our measurements. To put the $\overline{P}$ values in perspective,
we note that, as we demonstrated in Section~\ref{sect:perActivityEnergy}, \emph{human walking} typically corresponds to 120 $\le$ $\overline{P}$ $\le$ 280~$\mu$W.

\ifJSACsingle
\else

\begin{table}[t]
\centering
\small
\caption{Object motion measurements. \label{table:objectMotionExperiments}}
\begin{tabular}{|p{5.6cm}|p{1.8cm}|} \hline
Scenario & $\overline{P}$ \\ \hline  
Taking a book off a shelf & $<$10~$\mu$W \\
Putting on reading glasses & $<$10~$\mu$W \\
Reading a book & $<$10~$\mu$W \\
Writing with a pencil & 10--15~$\mu$W \\
Opening a drawer & 10--30~$\mu$W \\
Spinning in a swivel chair & $<$10~$\mu$W \\
Opening a building door & $<$1~$\mu$W \\
Shaking an object & $>$3,000~$\mu$W \\ \hline
\end{tabular}
\end{table}

\fi


Expectedly, for the vast majority of common object motion the energy availability is low. Due to the filter properties of inertial harvesters (see Section~\ref{sect:InertialHarvModel}), a motion needs to be \emph{periodic} to be ``harvestable''. The vast majority of common object motion is not periodic, and hence the corresponding energy availability is low. For example, we attached a sensing unit to a book and observed that when the book is being taken off the shelf, read, or put back on the shelf, $\overline{P} < 10$~$\mu$W. For a sensing unit attached to a pencil used by a student to write homework, 10 $\le$ $\overline{P}$ $\le$ 15~$\mu$W. 
Even high-acceleration non-periodic motions, such as a plane landing and taking off, and an accelerating or decelerating car, correspond to only limited energy availability ($\overline{P}$ $<$ 5~$\mu$W). For example, when a unit was placed in a bag  checked in on a 3:13 hour flight the recorded $a(t)$ showed that the luggage was subjected to varying high-acceleration motions, but the $\overline{P}$ did not exceed 5~$\mu$W even during the most turbulent intervals of the flight. Furthermore, substantially more energy could be harvested from a human walking around the airport with the luggage (i.e., periodic motion of a human walk) than from the motion associated with the entire flight. 


Our study additionally demonstrated low levels of energy availability for many \emph{high-amplitude and high-periodicity} motions. The motion of many objects in our environment is \emph{damped} for human comfort (e.g., by door dampers, cabinet drawer dampers, and springs in swirling chairs). In such cases, most of the motion energy is absorbed in the dampers and only small amounts can be harvested (e.g., by sticker form factor harvesters~\cite{Gorlatova_Enhants_wircom}). 
Opening and closing a drawer, spinning a swivel chair, and opening and closing a building door corresponded to 10 $\le$ $\overline{P}$ $\le$ 30~$\mu$W, 1 $\le$ $\overline{P}$ $\le$ 6.5~$\mu$W, and $\overline{P} < 1$~$\mu$W, respectively.
This suggest that IoT nodes embedded in objects such as doors and drawers should \emph{integrate motion energy harvesters with the mechanical dampers}.

Finally, our study confirmed that purposeful object motion can be extremely energy rich. Periodic shaking of objects can generate a relatively large amount of energy in a short time (as demonstrated by ``shake" flashlights). In our experiments, purposeful shaking corresponded to $\overline{P}$ of up to 3,500~$\mu$W, that is, \emph{12--29~times more than the power for walking}. In IoT applications with mobile nodes, this can be useful for quickly recharging battery-depleted nodes.





\section{Energy-aware Algorithms}
\label{sect:EnergyAlgorithms}

We now formulate an optimization problem of energy allocation for ultra-low-power
energy harvesting IoT node and prove it to be NP-hard.
As mentioned in Section~\ref{sect:ModelDesignConsiderations}, the formulation captures realistic
constraints that 
have not been jointly
considered before: (i) \emph{discrete}, rather than continuous, energy
spending rates; (ii) \emph{general}, rather than concave or linear, utility
functions; and (iii) use of a \emph{capacitor}, rather than a battery, as an energy storage component.

In Section~\ref{sect:PolicyPerformanceSimplifiedRepr} we demonstrated that the environmental energy available
to the node in each slot $i$, $e(i)$, cannot be represented by a Markov
or an i.i.d.\ process. Therefore, there is need to develop algorithms 
that do not make an assumption on the distribution of $e(i)$.  Since the energy allocation problem
is NP-hard, solving it is difficult even if $e(i)\ \forall i$ is known in
advance.  We distinguish between two types of energy allocation
algorithms:
(i) \emph{offline}, where $e(i)\, \forall i$ is part of the input;
an offline algorithm can be used as a benchmark since it provides 
an upper bound on the utility a node can achieve in practice, and (ii) \emph{online}, where a decision
in slot $i$ is made based only on $e(i')\ \forall i'<i$; an online algorithm
can be used by a real node to determine spending rate $s(i)$ in each slot.
We develop optimal and approximate offline algorithms. We then develop an
online algorithm and prove it to be optimal for some cases.
We also evaluate the performance of the algorithms with
the collected motion energy traces.
\ifIncludeAppendix
\textcolor{blue}{The proofs for this section appear in Appendix~I}. 
\else
Due to space constraints, the proofs are omitted and appear in \cite{MoversShakersReport}.
\fi


\subsection{Energy Allocation Problem}
\label{sect:ResourceAllocationProblem}


We start by formulating the energy allocation problem for a wireless IoT node:


\par \noindent {\bf Energy Allocation (EA) Problem:}
\ifJSACsingle
\begin{align}
  \underset{s(i)}\max \left\{ \sum_{i=0}^{K-1}U(s(i))\right\}\  \textrm{s.t.:}\  \ \ &\  \ \ \ \ \ \ \notag \\
\frac{s(i)}{\eta(i,B(i))}\leq B(i),\ s(i) \in \mathcal{S}\cup\{0\} \  & \forall\ i \ \ \ \ \ \label{eq:EA-1}\\
B(i)\leq B(i\text{--}1) \text{+} Q(e(i-1), B(i-1)) - 
L(i-1,B(i- 1))-\frac{s(i-1)}{\eta(i-1, B(i-1))}\ & \forall\ i\geq1 \label{eq:EA-2} \\
0 \leq B(i) \leq C\ \forall\ i; \ B(0) = B_0; B(K) & \geq B_K   \label{eq:EA-3}
\end{align}
\else
\begin{align}
  \underset{s(i)}\max \left\{ \sum_{i=0}^{K-1}U(s(i))\right\}\  \textrm{s.t.:}\  \ \ &\  \ \ \ \ \ \ \notag \\
\frac{s(i)}{\eta(i,B(i))}\leq B(i),\ s(i) \in \mathcal{S}\cup\{0\} \  & \forall\ i \ \ \ \ \ \label{eq:EA-1}\\
B(i)\leq B(i-1) + Q(e(i-1), B(i-1)) - \notag \\
L(i-1,B(i- 1))-\frac{s(i-1)}{\eta(i-1, B(i-1))}\ & \forall\ i\geq1 \label{eq:EA-2} \\
0 \leq B(i) \leq C\ \forall\ i; \ B(0) = B_0; B(K) & \geq B_K   \label{eq:EA-3}
\end{align}
\fi

This is an integer optimization problem, namely, all the coefficients
and function values are integers. Constraint~\eqref{eq:EA-1} ensures that a node
does not spend more energy than it has stored and that the spending rate, $s(i)$, is 
from a fixed set,~\eqref{eq:EA-2}
represents the energy storage evolution dynamics, and~\eqref{eq:EA-3} imposes
the storage component capacity constraints and sets the
initial and final energy levels to $B_0$ and $B_K$.
To simplify the notation, we omit the dependency of $\eta(i,B(i))$, $Q(e(i), B(i))$, and $L(i,B(i))$, on $B(i)$ in the rest of the section. However, unless mentioned otherwise, \emph{the proofs and the algorithms are also valid when the dependency on $B(i)$ is considered}.

The proof of the following theorem 
demonstrates the NP-hardness of the EA Problem even for ``simple'' cases (e.g., $B_0=B_K=0$ and linear $U(s(i))$).
\begin{theorem} \label{th:NP-hard}
The EA Problem is NP-hard.
\end{theorem}

\subsection{Energy Allocation Algorithms}
\label{sect:ResourceAllocationAlgs}

For solving the EA Problem, we 
present 
a dynamic pro\-gram\-ming-ba\-sed
pseudopolynomial algorithm\footnote{A \emph{pseudopolynomial algorithm} is an algorithm whose running time is polynomial if the input is encoded in unary
  format. 
  }, a Fully P\-ol\-ynomial Time Approximation Scheme
(FPTAS)\footnote{An \emph{FPTAS} is an algorithm which takes an instance of an
  optimization problem and a parameter $\epsilon > 0$ and, in polynomial time
  in both the problem size and $1 / \epsilon$, produces a solution that is
  within a $1 - \epsilon$ factor of the optimal solution.}, and 
a greedy online algorithm which is optimal in particular scenarios. 


We first present an optimal offline dynamic programming algorithm for
  solving the EA Problem. Thus, the algorithm jointly considers realistic constraints
  that have not been jointly considered before and uses similar
  ideas to the dynamic programming algorithm
  from \cite{Gorlatova_TMC2013}. However, compared
  to \cite{Gorlatova_TMC2013}, the dynamic programming procedure's parameters
  and return value switch places. This difference is used to develop the FPTAS
  we present later in this section.


\par \noindent \textbf{Dynamic programming algorithm:}
We determine $M(i,$ $U)$ which is the maximum battery level
when obtaining utility $U$ in the beginning of slot $i$. We set $M(0,0)=B_0$
and $M(0,U)=-\infty\ \forall\ U>0$. For $i>0$, $M(i,U)$ is
calculated as 
$ M(i,U)=\max_{s(i-1)\in \mathcal{S}\cup \{0\}} \{M(i-1, U-
 U(s(i-1))) +Q(i-1) - s(i-1) / \eta(i-1) -
 L(i-1) \} $.  Let the optimal solution utility be $U^*$, and let $U^H\geq U^*$ be an upper
bound. We calculate $M(i,U)$ for $1\leq i\leq K$ and $0\leq U\leq U^H$. Then,
$U^* = \operatorname{arg\,max} \{M(K, U)$ s.t. $M(K,U) \geq
B_K\}$. The optimal energy spending values $s^*(i)$ are found by maintaining
an array $A(i,U)$ that stores the $s(i-1)$ values chosen when calculating
$M(i, U)$. Then, $s^*(K-1) = A(K,U^*)$. We can obtain $s(K-2)$ using $A(K-1,
U^*-U(s^*(K-1)))$. This process is repeated to find $s^*(i)$ for $0\leq i\leq K-1$.

The space complexity of the algorithm is $O(K\cdot U^H)$ for storing $A(i,U)$. Since in every calculation of
$M(i,U)$ we go over $\mathcal{S}$, the time complexity is $O(K\cdot |\mathcal{S}|\cdot U^H)$. Let $s_{\max}$ be the maximum item in $\mathcal{S}$, clearly $U^H=K\cdot U(s_{\max})$ is an upper bound, for which we obtain space and time complexities of $O(K^2\cdot U(s_{\max}))$ and $O(K^2\cdot U(s_{\max})\cdot |\mathcal{S}|)$, respectively.


\par \noindent \textbf{FPTAS:} For large values of $U(s_{\max})$, the time and
space complexities render the dynamic programming algorithm
impractical. The\-re\-fore, we develop an approximation scheme. It relies
on a lower bound $U^L=U(s_{\max})$, which is a lower bound, since
if spending only $s_{\max}$ energy at some slot is
always infeasible, $s_{\max}$ can be removed from $\mathcal{S}$. We define a
scaling factor $\mu = \epsilon \cdot U(s_{\max})/K$ and a new
utility function $\tilde{U}(s)= \left\lfloor U(s)/\mu
\right\rfloor$. Next, we invoke the dynamic programming algorithm for
$\tilde{U}()$ to compute $M(i,\tilde{U})$ for $0\leq i\leq K$ and
$0\leq \tilde{U}\leq U^H/\mu$. The algorithm returns the energy
spending rates $\tilde{s}(i)$ found by the dynamic programming algorithm.
Below we show that the algorithm is an FPTAS.
\begin{theorem} \label{th:FPTAS}
  The above algorithm runs in times poly($1/\epsilon$, $K$), and
  the solution $\tilde{s}(i)$ is a $(1-\epsilon)$-app\-ro\-xi\-ma\-tion.
\end{theorem}

\label{sect:OptimalFixedPacket}


\par \noindent \textbf{Greedy online algorithm}: In every time slot, the algorithm tries to maximize the  utility while not letting the energy storage level go below
$B_K$. Namely, in each slot $i$ the algorithm spends
$s(i)=\max\{U(s)\ |\ s\in\mathcal{S}\cup\{0\}\wedge (B(i)-s/\eta) \geq B_K\}$.


We first focus on the \emph{battery} node model and on a scenario where (i) for $x,y$, $U(x+y)=U(x)+U(y)$, and (ii) the set $\mathcal{S}$ is $\{ j\cdot s\text{ , } j=1,\ldots,|\mathcal{S}|\}$ and $s>0$. 
Such conditions hold, for example, when a node uses a fixed power level and changes its transmission rate by transmitting a different number of equal-sized packets.

\begin{theorem} \label{th:GreedyBattery}
  For battery energy storage model, for $B_K=0$, if conditions (i) and (ii) hold, the greedy algorithm is optimal.
\end{theorem}


In Section~\ref{sect:AlgsPerformance} we evaluate the performance of the
greedy online algorithm under the capacitor model and for cases where $B_K>0$.

We additionally demonstrate a set of cases where no performance is guaranteed
for any online algorithm. Since we showed in Section~\ref{sect:PolicyPerformanceSimplifiedRepr} that
$e(i)$ cannot be represented by a Markov or an i.i.d process, for these cases
any online algorithm may perform arbitrary worse and it should be
evaluated with collected traces in order to assess its performance.

\begin{theorem} \label{th:NegativeOnline}
  For $B_K>0$, the performance of any online algorithm that guarantees a feasible solution can be arbitrarily worse for
  $K\geq 2$. 
\end{theorem}


\subsection{Trace-based Performance Evaluation}
\label{sect:AlgsPerformance}

In this section, we evaluate the algorithms using the motion energy
traces we collected, 
for both \emph{battery} and \emph{capacitor} node models defined in Section~\ref{sect:ModelDesignConsiderations}. We refer to the algorithm and model combinations as follows:

\ifJSACsingle
\vspace{10pt}
\noindent
\begin{minipage}[c]{\textwidth}
\begin{spacing}{1}

\begin{minipage}[l]{0.48\textwidth}
\par \noindent \textbf{Algorithms invoked for the battery model}
\begin{itemize}
\item {\bf ALG-OB:} The optimal dynamic programming algorithm.
\item \textcolor{blue}{{\bf ALG-FB:} The FPTAS.}
\item {\bf ALG-GB:} The greedy online algorithm.
\end{itemize}
\end{minipage}
\hfill
\begin{minipage}[r]{0.48\textwidth}
\par \noindent \textbf{Algorithms invoked for the capacitor model}
\begin{itemize}

\item {\bf ALG-OC:} The optimal dynamic programming algorithm.


\item {\bf ALG-FC:} The FPTAS.

\item {\bf ALG-GC:} The greedy online algorithm.
\end{itemize}
\end{minipage}
\end{spacing}
\end{minipage}

\else

\par \noindent \textbf{Algorithms invoked for the battery model:}
\ifJSACdouble
\begin{itemize}
\item {\bf ALG-OB:} The optimal dynamic programming algorithm.
\item \textcolor{blue}{{\bf ALG-FB:} The FPTAS.}
\item {\bf ALG-GB:} The greedy online algorithm.
\end{itemize}
\else
\begin{myitemize}
\item {\bf ALG-OB:} The optimal dynamic programming algorithm.
\item {\bf ALG-FB:} The FPTAS.
\item {\bf ALG-GB:} The greedy online algorithm.
\end{myitemize}
\fi
\par \noindent \textbf{Algorithms invoked for the capacitor model:}
\ifJSACdouble
\begin{itemize}

\item {\bf ALG-OC:} The optimal dynamic programming algorithm.


\item {\bf ALG-FC:} The FPTAS.

\item {\bf ALG-GC:} The greedy online algorithm.
\end{itemize}
\else
\begin{myitemize}

\item {\bf ALG-OC:} The optimal dynamic programming algorithm.


\item {\bf ALG-FC:} The FPTAS.

\item {\bf ALG-GC:} The greedy online algorithm.
\end{myitemize}
\fi

\fi

We consider an IoT node that changes its data rate $r(i)$ by changing the
number of packets it sends in a time slot (where the length of a time slot is
$T_{\textrm{int}}=1$~second). The maximal $r(i)$ is 250~Kb/s, the packet size
is 127 bytes\footnote{These parameters correspond to IEEE 802.15.4/Zigbee nodes~\cite{CC2420}.}, and $c_{\text{tx}}=1$~nJ/bit
(i.e., it takes 1,016~nJ to transmit 1 packet). 
Thus, $\mathcal{S} = \{1016\cdot j\text{, } j=1,\ldots , 246
\}$, and $s_{\min}=\min\{s\in \mathcal{S}\} = 1016$. We set
$L(i,B(i))=0$. We use the day-long motion energy traces (see Table~\ref{table:longTermResults})\footnote{
    From the traces, we calculate $Q(i)$ as
  $Q(i) \gets \eta_h \cdot T_{\textrm{int}} \cdot P_{\textrm{meas}}(i)$, where
  $\eta_h=20\%$. To evaluate ALG-GB, ALG-FC, and ALG-GC, we compare their
  performance with the optimal algorithms ALG-OC and ALG-OB.
}. We evaluated the algorithms for traces of different users and
  for different days. We observed that the performance trends of the algorithms are very
  similar for all the considered day-long traces. Therefore, only the graphs
  corresponding to a day-long trace of participant $M1$ are shown.  Since for the
day-long traces $K$ is very large, to draw a single point in the graphs we run
the algorithms over 66 consecutive 10-minute intervals of $Q(i)$ and average
the results. Unless specified otherwise, the evaluation results are shown for
$B_0=B_K=0$ and for $10\cdot s_{\min} \leq C \leq 100\cdot
s_{\min}$. 

We first explain in detail how to compute the conversion efficiency $\eta()$.
Recall that 
$\eta()$ depends on the node's fixed operating voltage $V_{\textrm{op}}$ 
and energy storage voltage $V_{\text{out}}(i)$ (see Section~\ref{sect:ModelDesignConsiderations}). 
For the \emph{battery} model, we assume $V_{\text{out}}(i)=V_{\textrm{op}}$ and set $\eta=1$. 
For the \emph{capacitor} model, approximating voltage converter properties~\cite{LTM4607}, we compute:
\ifJSACsingle
\\$
\else
\begin{equation*}
\fi
\eta(i,B(i)) = \left \{
\begin{array}{ll}
\frac {V_{\text{out}}(i)}{V_{\text{op}}} \text{, } & V_{\min}\leq V_{\text{out}}(i)
\leq V_{\text{op}} \\
1 - \frac {V_{\text{out}}(i)-V_{\text{op}}} {2\cdot (V_{\max}-V_{\text{op}})}\text{, } &
V_{\text{op}} < V_{\text{out}}(i) \leq V_{\max}\\
0 & \text{otherwise},
\end{array}
\ifJSACsingle
\right\}
\else
\right.
\fi
\ifJSACsingle
$,
\else
\end{equation*}
\fi where $V_{\max}=2.8$~V is the maximum voltage of the capacitor,
$V_{\text{out}}(i)$ is node's voltage in a time slot $i$ ($V_{\text{out}}(i)= \sqrt{B(i)/C}\cdot V_{\max}$),
$V_{\textrm{op}}=2.5$~V, and $V_{\min}=0.7$~V. 


\ifJSACsubmission
\else
For the evaluations, we implemented an improved iterative version of the dynamic programming algorithm 
using
the following definition:
\begin{definition}
  An entry $M(i,U)$ dominates an entry $M(i,$ $U')$ if $U\geq U'$ and
  $M(i,U) > M(i,U')$.
\end{definition}
A dominated entry will never be considered for the optimal solution and can be
removed during the calculation~\cite{Book:Kellerer04}. 
\fi

\ifJSACsubmission
We first examine the performance of ALG-FC
as a function of its approximation ratio, $1-\epsilon$ 
(see Theorem~\ref{th:FPTAS}). Fig.~\ref{fig:capacitorAlgsResults}(a) shows the
ratio of the ALG-FC 
performance to the optimal (ALG-OC) for $C=100\cdot s_{\min}$. Even for small $1-\epsilon$, 
the ALG-FC performance is close to the optimal (much closer than the theoretical bound).
Similar results were obtained for ALG-FB.
\else
We first examine the performance of ALG-FC and ALG-FB
as a function of their approximation ratio, $1-\epsilon$ 
(see Theorem~\ref{th:FPTAS}). Fig.~\ref{fig:capacitorAlgsResults}(a) shows the
ratio of the ALG-FC and ALG-FB 
performance to the optimal (ALG-OC and ALG-OB) for $C=100\cdot s_{\min}$. Even for small $1-\epsilon$, 
the ALG-FC and ALG-FB performance is close to the optimal (much closer than the theoretical bound).
\fi

\ifJSACsubmission

\ifJSACsingle
\noindent
\begin{minipage}[c]{\textwidth}
\begin{minipage}[l]{0.47\textwidth}
\begin{figure}[H]
\centering
\subfigure[]{\includegraphics[width=0.485\columnwidth]{fig/algs_simulation/capacitor_fptas_vs_epsilon}}
\subfigure[]{\includegraphics[width=0.485\columnwidth]{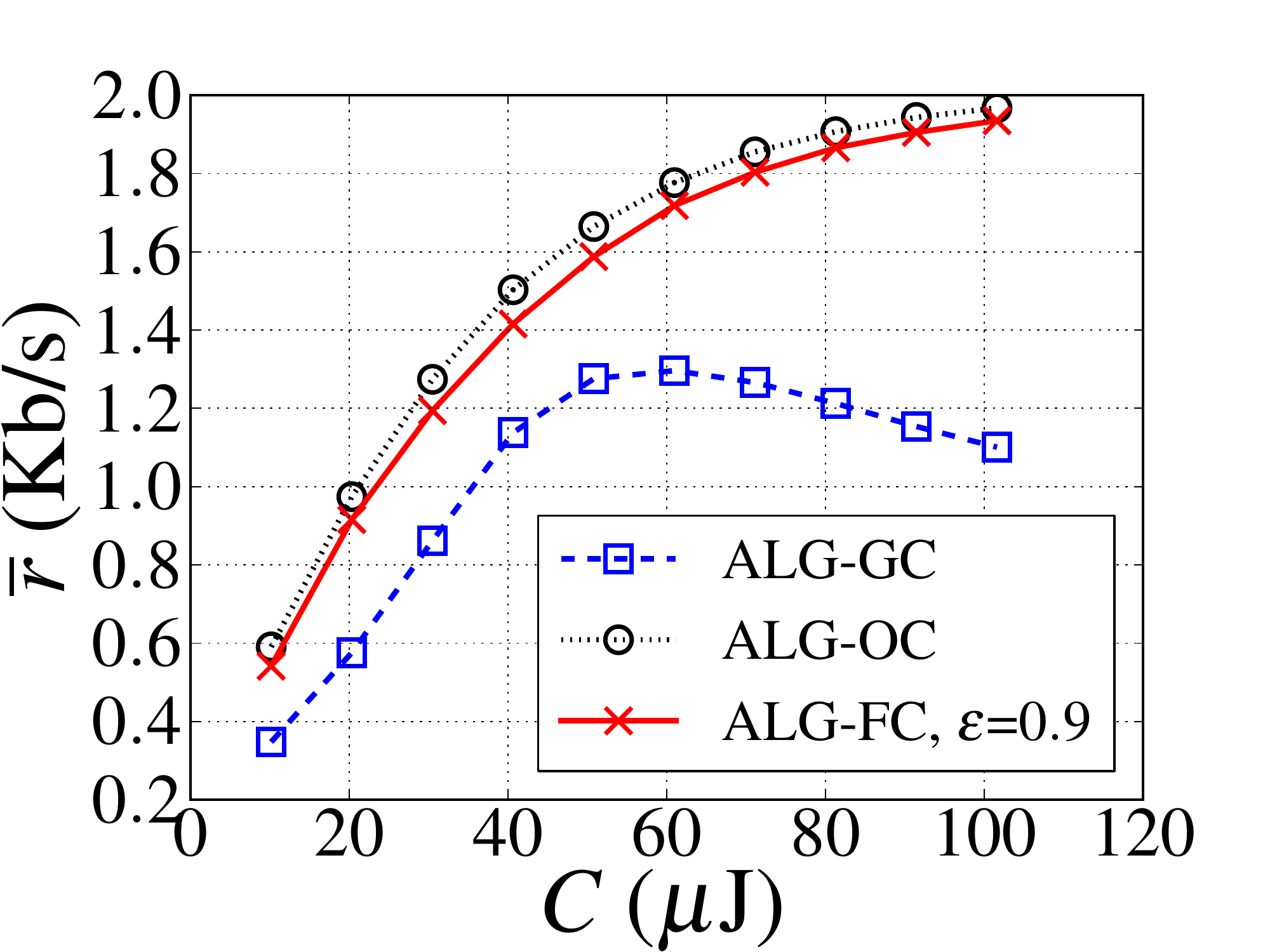}}
\vspace{-0.55cm}
\caption{Algorithm performance using energy traces for participant $M1$, for the \emph{capacitor} model: (a)
  performance ratio between ALG-FC and ALG-OC, and (b) average data rate, $\overline{r}$, achieved by different algorithms. \label{fig:capacitorAlgsResults}}
\end{figure}
\end{minipage}
\hfill
\begin{minipage}[r]{0.47\textwidth}
\begin{figure}[H]
\centering
\subfigure[]{\includegraphics[width=0.485\columnwidth]{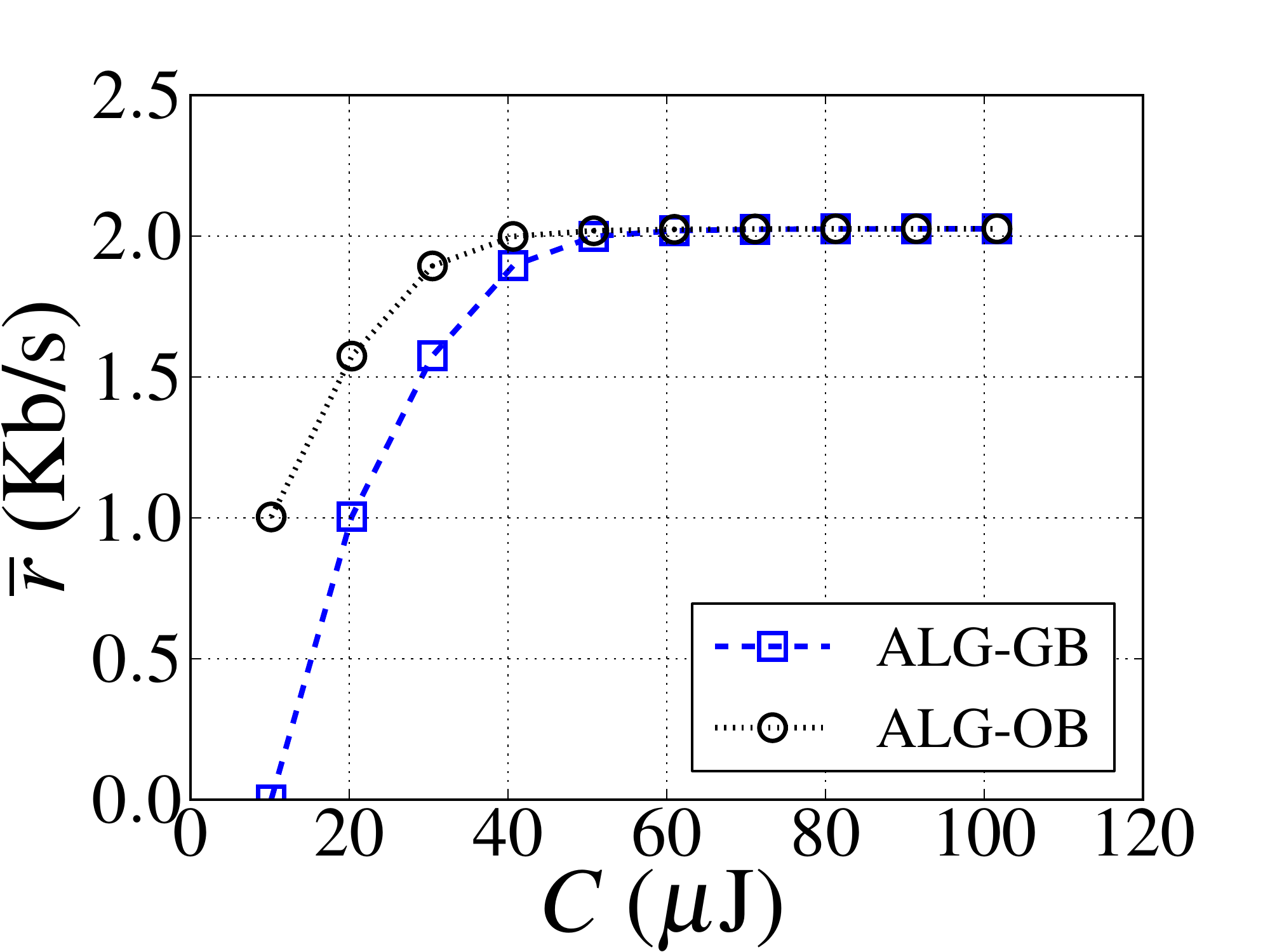}}
\subfigure[]{\includegraphics[width=0.485\columnwidth]{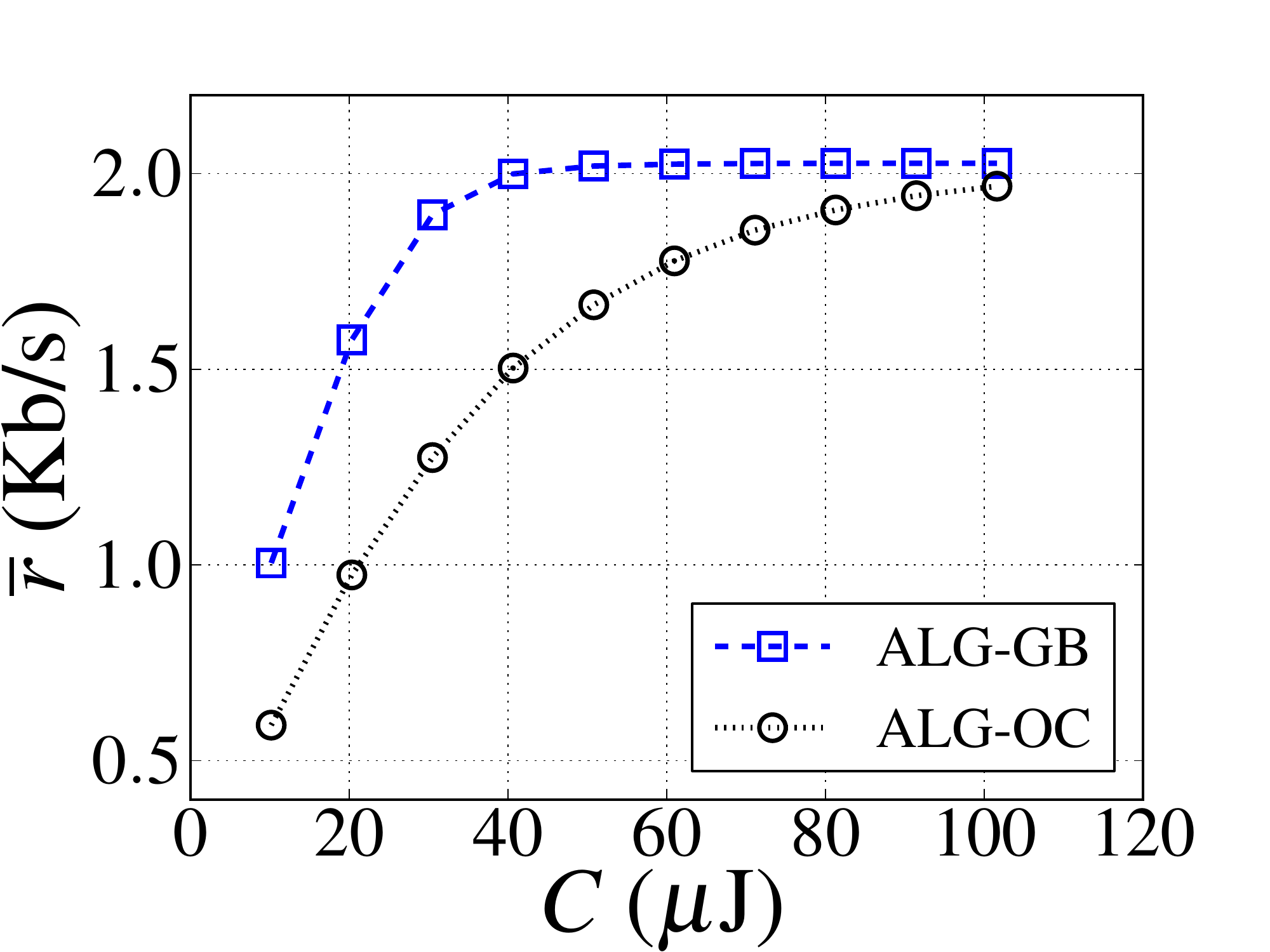}}
\vspace{-0.55cm}
\caption{The average data rate, $\overline{r}$, achieved by the
  algorithms using energy traces for participant $M1$, for (a) the battery
  model, for $B_K=B_0=10\cdot s_{\text{min}}$, and (b) the battery and capacitor models. \label{fig:battryAlgsResults}}
\end{figure}
\end{minipage}

\end{minipage}
\vspace*{3.5pt}
\else

\begin{figure}[t]
\centering
\subfigure[]{\includegraphics[width=0.485\columnwidth]{fig/algs_simulation/capacitor_fptas_vs_epsilon}}
\subfigure[]{\includegraphics[width=0.485\columnwidth]{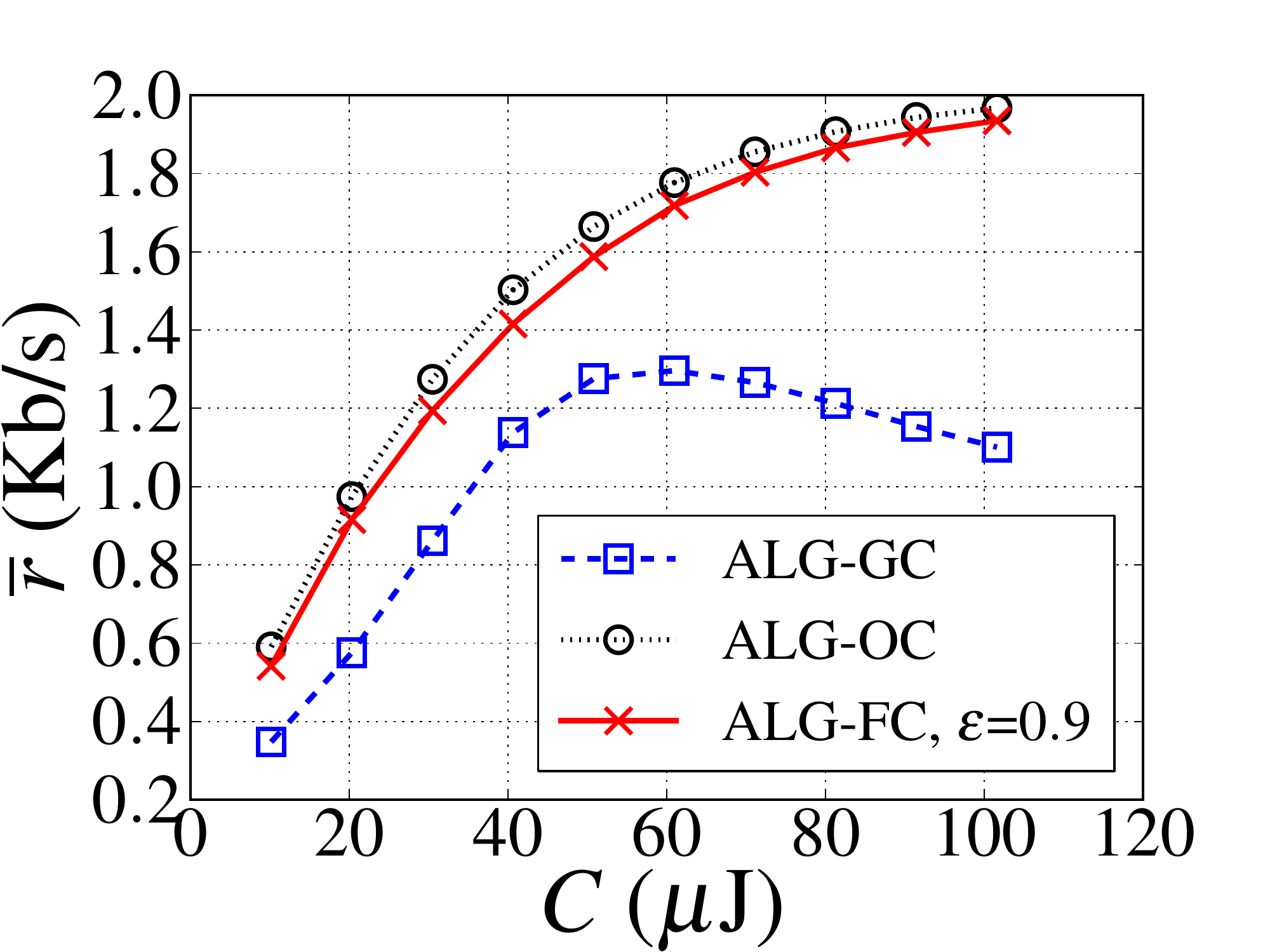}}
\ifJSACdouble
\else
\vspace{-0.35cm}
\fi
\caption{Algorithm performance using energy traces for participant $M1$, for the \emph{capacitor} model: (a)
  performance ratio between ALG-FC and ALG-OC, and (b) average data rate, $\overline{r}$, achieved by different algorithms. \label{fig:capacitorAlgsResults}}
\end{figure}
\fi

\else
\begin{figure}[t]
\centering
\subfigure[]{\includegraphics[width=0.485\columnwidth]{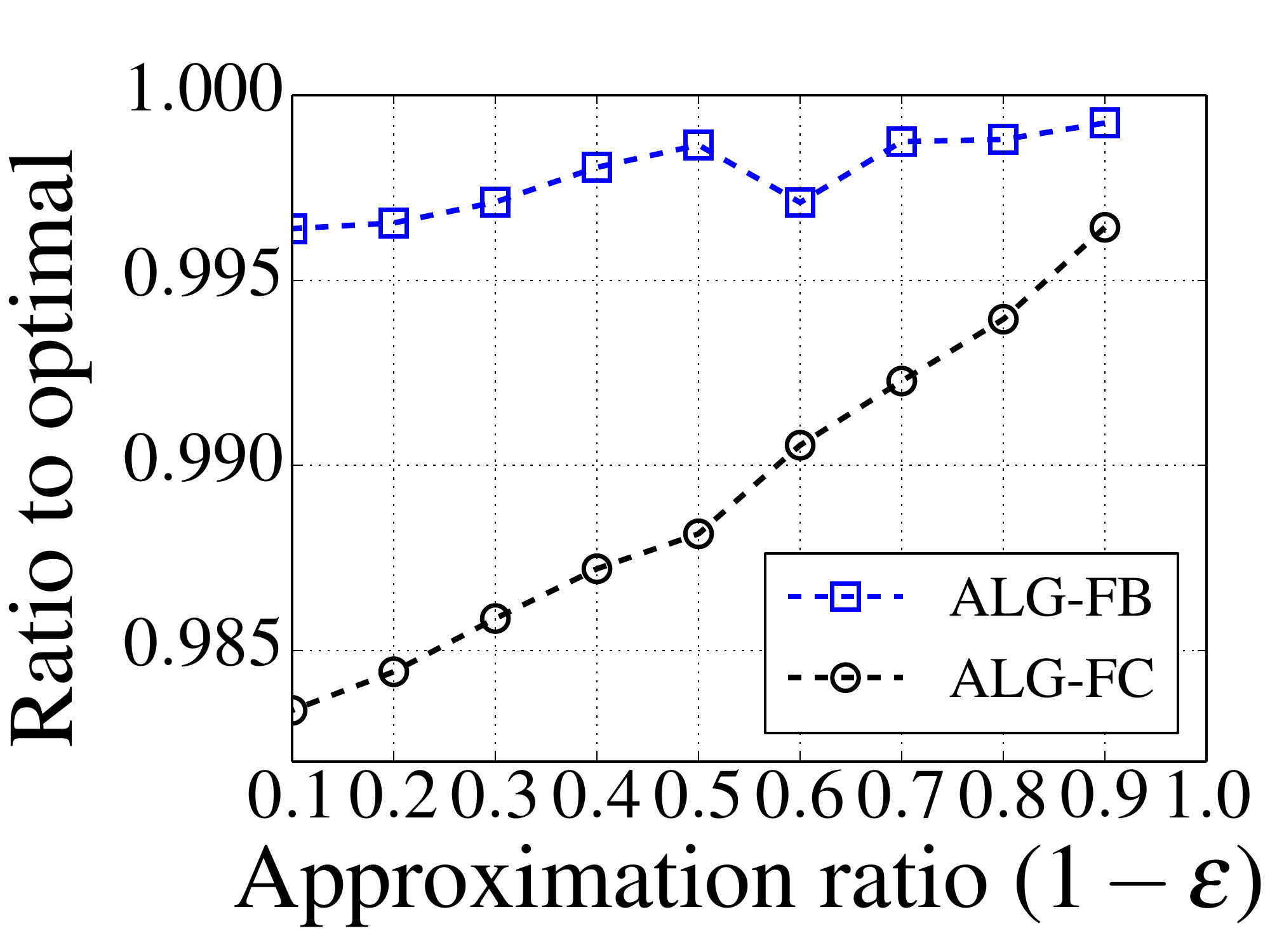}}
\subfigure[]{\includegraphics[width=0.485\columnwidth]{fig/algs_simulation/capacitor_utility_vs_capacity}}
\vspace{-0.35cm}
\caption{Algorithm performance using energy traces for participant $M1$, for:
  (a) battery and capacitor models, performance ratio between
  ALG-FC (ALG-FB) and ALG-OC (ALG-OB), and (b) the capacitor model, average data rate, $\overline{r}$, achieved by different algorithms. \label{fig:capacitorAlgsResults}}
\end{figure}
\fi

Next, we examine the performance of the ALG-GC, ALG-OC, and ALG-FC 
for the capacitor mode. Fig.~\ref{fig:capacitorAlgsResults}(b) shows the average data rates
$\overline{r}$ obtained by the algorithms. 
The performance of ALG-FC is close to that of ALG-OC. 
The performance of ALG-GC gets worse compared to ALG-OC for larger
$C$ because it obtains lower $V_{\text{out}}(i)$ (recall that $V_{\text{out}}(i)= \sqrt{B(i)/C}\cdot V_{\max}$),
resulting in lower $\eta()$. Furthermore, for $C>60\ \mu$J, its obtained $\overline{r}$
\emph{decreases} as $C$ increases.


We also examine the performance of the ALG-GB and ALG-OB algorithms for the
battery model. Since for $B_K=0$ ALG-GB is optimal (see
Theorem~\ref{th:GreedyBattery}), 
we consider 
$B_K=B_0=10\cdot
s_{\min}$. Fig.~\ref{fig:battryAlgsResults}(a) shows the $\overline{r}$
values obtained by ALG-GB and ALG-OB.
Since ALG-GB cannot take advantage of the initial energy 
(because $B_0=B_K$), for a particular $C$ value the capacity
available to ALG-GB is $C-B_0$. Correspondingly, since consecutive plotted
points differ by $B_0$ in their $C$ value, the plotted points $(C,\
\overline{r})$ for ALG-OB and $(C$$+$$B_0,\ \overline{r})$ for ALG-GB appear in
the figure.

To compare the performance for the battery and the capacitor models,
Fig.~\ref{fig:battryAlgsResults}(b) shows the data rates obtained by ALG-GB and ALG-OC. 
For ALG-OC, for larger $C$ 
there is a
wider range of charge level for which $\eta()$ is close to 1. Correspondingly,
ALG-OC can keep $\eta()$ close to 1,
thus its performance approaches that of ALG-GB. 

In summary, the evaluations demonstrate that the algorithms perform well and showcase that for the \emph{capacitor} node model, having a larger energy storage may worsen the overall performance.



\ifJSACsingle
\else

\begin{figure}[t]
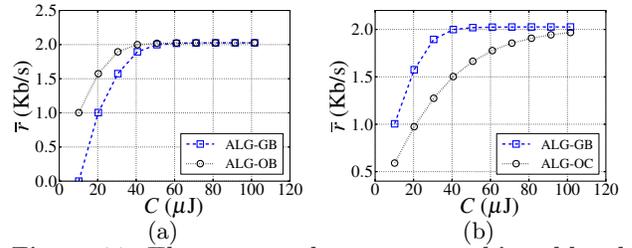

\centering
\subfigure[]{\includegraphics[width=0.485\columnwidth]{fig/algs_simulation/battery_utility_vs_capacity_final_energy_10160000}}
\subfigure[]{\includegraphics[width=0.485\columnwidth]{fig/algs_simulation/battery_vs_capacitor_vs_capacity}}
\ifJSACdouble
\else
\vspace{-0.35cm}
\fi
\caption{The average data rate, $\overline{r}$, achieved by the
  algorithms using energy traces for participant $M1$, for (a) the battery
  model, for $B_K=B_0=10\cdot s_{\text{min}}$, and (b) the battery and capacitor models. \label{fig:battryAlgsResults}}
\end{figure}

\fi

\section{Conclusions}
\label{sect:Conclusions}

This paper considers motion (kinetic) energy availability for Internet of Things (IoT) applications.
We thoroughly study \emph{human motion} and provide observations regarding \emph{object motion}. For human motion, we use the results of our measurement campaign that include 200~hours of acceleration traces from day-long human activities. Moreover, we use a dataset of 7~common human motions performed by over~40 participants \cite{xue2010naturalistic}. We consider a wireless energy harvesting node model that captures several practical IoT node design considerations.
We design optimal, approximation, and online \emph{energy allocation algorithms} and evaluate their performance using the 
collected motion energy traces. 

In future work we will expand our measurement study to include additional motions and additional human participants. 
We will jointly measure light and motion energy (available to the same device)
to obtain insight into the use of multipurpose harvesters. 
\ifJSACsubmission
\else
We
will evaluate the performance and energy consumption of  the energy allocation algorithms in real devices under various utility functions.  Finally, we will develop algorithms for more
complex networking scenarios (e.g., multihop networks of IoT nodes) and
evaluate their performance using the traces.
\fi

\ifJSACsubmission
\else
\section*{Acknowledgments}
This work was supported in part by Vodafone Americas Foundation Wireless
Innovation Project and NSF grants CCF-09-64497 and CNS-10-54856. We thank
Sonal Shetkar for her contributions to the development of the measurement
setup and study methodology, and Craig Gutterman for his contributions to
preliminary data analysis. We additionally thank Chang Sun and Kanghwan Kim
for their contributions. We also thank our shepherd Ranveer Chandra for
valuable feedback.
\balance 
\fi

\ifJSACsingle
\begin{spacing}{1}
\fi
{
\scriptsize
\bibliographystyle{abbrv}
\ifJSACsingle
\vspace{-3pt}
\else
\vspace{-5pt}
\fi
\bibliography{EnergyMeasurements2012,EnergyAlgs}

\begin{thebibliography}{10}

\bibitem{nPowerPEG}
{nPower Personal Energy Generator}.
\newblock \url{http://www.npowerpeg.com}.

\bibitem{LTM4607}
Linear technology. {LTM4607} 36{$V_{\text{IN}}$}, 24{$V_{\text{OUT}}$} high
  efficiency buck-boost {DC/DC} $\mu$module regulator.
\newblock 2008.
\newblock \url{http://cds.linear.com/docs/en/datasheet/4607fb.pdf}.

\bibitem{LTC3588}
Linear technology. {LTC3588} piezoelectric energy harvesting power supply
  datasheet.
\newblock 2010.
\newblock \url{http://cds.linear.com/docs/en/datasheet/35881fa.pdf}.

\bibitem{Piezo}
Piezo systems. piezo energy harvester datasheet.
\newblock 2011.
\newblock \url{http://www.piezo.com/catalog8.pdf%20files/Cat8.20&21.pdf}.

\bibitem{CC2420}
Chipcon. {CC2420} 2.4g{GH}z {IEEE}802.15.4/{ZigBee}-ready {RF} transceiver
  datasheet.
\newblock Nov. 2013.
\newblock \url{http://www.ti.com/lit/ds/symlink/cc2420.pdf}.

\bibitem{Mide}
Mide. volture piezoelectric energy harvesters datasheet.
\newblock Jan. 2013.
\newblock \url{http://www.mide.com/pdfs/Volture_Datasheet_001.pdf}.

\bibitem{SmartMaterial}
Smart material. macro fiber composite - mfc datasheet.
\newblock 2013.
\newblock
  \url{http://www.smart-material.com/media/Datasheet/MFC-V2.1-2013-web.pdf}.

\bibitem{KineticMeasurements}
{EnHANTs} kinetic energy measurements.
\newblock Mar. 2014.
\newblock \url{http://enhants.ee.columbia.edu/kinetic-datasets}.

\bibitem{beeby2006energy}
S.~Beeby, M.~Tudor, and N.~White.
\newblock Energy harvesting vibration sources for microsystems applications.
\newblock {\em Measurement science and technology}, 17(12):R175, 2006.

\bibitem{buren2003kinetic}
T.~Buren, P.~Lukowicz, and G.~Troster.
\newblock Kinetic energy powered computing: experimental feasibility study.
\newblock In {\em Proc. IEEE ISWC'03}, Feb. 2003.

\bibitem{Chen2012Asymptotically}
S.~Chen, P.~Sinha, N.~Shroff, and C.~Joo.
\newblock A simple asymptotically optimal energy allocation and routing scheme
  in rechargeable sensor networks.
\newblock In {\em Proc. IEEE INFOCOM'12}, Mar. 2012.

\bibitem{CrawdadMoversShakers}
M.~Cong, K.~Kim, M.~Gorlatova, J.~Sarik, I.~Kymissis, and G.~Zussman.
\newblock {CRAWDAD} data set columbia/kinetic.
\newblock http://crawdad.cs.dartmouth.edu/columbia/kinetic, Apr. 2014.

\bibitem{Gunduz2012}
B.~Devillers and D.~Gunduz.
\newblock A general framework for the optimization of energy harvesting
  communication systems with battery imperfections.
\newblock {\em IEEE J. Commun. Netw.}, 14(2):130--139, 2012.

\bibitem{Domingo2011}
M.~C. Domingo.
\newblock Packet size optimization for improving the energy efficiency in body
  sensor networks.
\newblock {\em ETRI Journal}, 33(3):299--309, June 2011.

\bibitem{galloway1998pick}
J.~Galloway.
\newblock Pick up the beat.
\newblock In {\em Runners World}, July 2007.
\newblock {\url{http://www.runnersworld.com/running-tips/pick-beat}}.

\bibitem{Gorlatova_Enhants_wircom}
M.~Gorlatova, P.~Kinget, I.~Kymissis, D.~Rubenstein, X.~Wang, and G.~Zussman.
\newblock Energy-harvesting active networked tags ({EnHANTs}) for ubiquitous
  object networking.
\newblock {\em IEEE Wireless Commun.}, 17(6), Dec. 2010.

\bibitem{Gorlatova2013Prototyping}
M.~Gorlatova, R.~Margolies, J.~Sarik, G.~Stanje, J.~Zhu, B.~Vigraham,
  M.~Szczodrak, L.~Carloni, P.~Kinget, I.~Kymissis, and G.~Zussman.
\newblock Prototyping energy harvesting active networked tags ({EnHANTs}).
\newblock In {\em Proc. IEEE INFOCOM'13 mini-conference}, Apr. 2013.

\bibitem{Gorlatova_TMC2013}
M.~Gorlatova, A.~Wallwater, and G.~Zussman.
\newblock Networking low-power energy harvesting devices: Measurements and
  algorithms.
\newblock {\em IEEE Trans. Mobile Comput.}, 13(9):1853--1865, Sept. 2013.

\bibitem{Gummeson10}
J.~Gummeson, S.~S. Clark, K.~Fu, and D.~Ganesan.
\newblock On the limits of effective micro-energy harvesting on mobile {CRFID}
  sensors.
\newblock In {\em Proc. ACM MobiSys'10}, June 2010.

\bibitem{huang2011human}
H.~Huang, G.~Merrett, and N.~White.
\newblock Human-powered inertial energy harvesters: the effect of orientation,
  location and activity on obtainable power.
\newblock In {\em Proc. Eurosensors'11}, Sept. 2011.

\bibitem{huang2010utility}
L.~Huang and M.~Neely.
\newblock Utility optimal scheduling in energy harvesting networks.
\newblock {\em IEEE/ACM Trans. Netw.}, 21(4):1117--1130, Aug. 2013.

\bibitem{Kansal06Journ}
A.~Kansal, J.~Hsu, S.~Zahedi, and M.~B. Srivastava.
\newblock Power management in energy harvesting sensor networks.
\newblock {\em ACM Trans. Embedded Comput. Syst.}, 6(4):32:1--32:38, 2007.

\bibitem{Book:Kellerer04}
H.~Kellerer, U.~Pferschy, and D.~Pisinger.
\newblock {\em Knapsack Problems}.
\newblock Springer, 2004.

\bibitem{Krupenkin2011reverse}
T.~Krupenkin and J.~A. Taylor.
\newblock Reverse electrowetting as a new approach to high-power energy
  harvesting.
\newblock {\em Nature Communications}, 2:448, 2011.

\bibitem{kymissis98}
J.~Kymissis, C.~Kendall, J.~Paradiso, and N.~Gershenfeld.
\newblock Parasitic power harvesting in shoes.
\newblock In {\em Proc. IEEE ISWC'98}, Oct. 1998.

\bibitem{ApplePatentShake}
G.~Lin, P.~Rahul, M.~Rosenblatt, T.~Nakajima, B.~Germansderfer, and
  S.~Dasgupta.
\newblock Harnessing power through electromagnetic induction utilizing printed
  coils.
\newblock United States Patent Application \#20120235510, Sept. 2012.

\bibitem{liu_infocom2010}
R.-S. Liu, P.~Sinha, and C.~E. Koksal.
\newblock Joint energy management and resource allocation in rechargeable
  sensor networks.
\newblock In {\em Proc. IEEE INFOCOM'10}, Mar. 2010.

\bibitem{merrett2008empirical}
G.~V. Merrett, A.~S. Weddell, A.~P. Lewis, N.~R. Harris, B.~M. Al-Hashimi, and
  N.~M. White.
\newblock An empirical energy model for supercapacitor powered wireless sensor
  nodes.
\newblock In {\em Proc. IEEE ICCCN'08}, Aug. 2008.

\bibitem{NKM13}
N.~Michelusi, K.~Stamatiou, and M.~Zorzi.
\newblock Transmission policies for energy harvesting sensors with
  time-correlated energy supply.
\newblock {\em IEEE Trans.\ Commun.}, 61(7):2988--3001, July 2013.

\bibitem{Mitcheson2008}
P.~Mitcheson, E.~Yeatman, G.~Rao, A.~Holmes, and T.~Green.
\newblock Energy harvesting from human and machine motion for wireless
  electronic devices.
\newblock {\em Proc. of the IEEE}, 96(9):1457 --1486, Sept. 2008.

\bibitem{noh-efficient}
D.~Noh and T.~Abdelzaher.
\newblock {Efficient flow-control algorithm cooperating with energy allocation
  scheme for solar-powered WSNs}.
\newblock {\em Wireless Comm. and Mobile Comput.}, 2010.

\bibitem{orendurff2008humans}
M.~Orendurff, J.~Schoen, G.~Bernatz, A.~Segal, and G.~Klute.
\newblock How humans walk: bout duration, steps per bout, and rest duration.
\newblock {\em J. Rehabil. Res. Dev.}, 45(7):1077--89, 2008.

\bibitem{ODE13}
O.~Orhan, D.~Gunduz, and E.~Erkip.
\newblock Optimal packet scheduling for an energy harvesting transmitter with
  processing cost.
\newblock In {\em Proc. IEEE ICC'13}, June 2013.

\bibitem{OKYSA11}
O.~Ozel, K.~Tutuncuoglu, J.~Yang, S.~Ulukus, and A.~Yener.
\newblock Transmission with energy harvesting nodes in fading wireless
  channels: Optimal policies.
\newblock {\em IEEE J.\ Sel.\ Areas Commun.}, 29(8):1732--1743, Sept. 2011.

\bibitem{paradiso2005esm}
J.~Paradiso and T.~Starner.
\newblock {Energy scavenging for mobile and wireless electronics}.
\newblock {\em IEEE Perv. Comput.}, 4(1):18--27, 2005.

\bibitem{starner1996human}
T.~Starner.
\newblock Human-powered wearable computing.
\newblock {\em IBM systems J.}, 35(3.4):618--629, 1996.

\bibitem{von2006optimization}
T.~Von~Buren, P.~Mitcheson, T.~Green, E.~Yeatman, A.~Holmes, and G.~Troster.
\newblock Optimization of inertial micropower generators for human walking
  motion.
\newblock {\em IEEE Sensors J.}, 6(1):28--38, 2006.

\bibitem{Vullers2009684}
R.~Vullers, R.~van Schaijk, I.~Doms, C.~Van~Hoof, and R.~Mertens.
\newblock Micropower energy harvesting.
\newblock {\em Solid-State Electron.}, 53(7):684 -- 693, Apr. 2009.

\bibitem{Wang2013WhenSimplicity}
Q.~Wang and M.~Liu.
\newblock When simplicity meets optimality: Efficient transmission power
  control with stochastic energy harvesting.
\newblock In {\em Proc. IEEE INFOCOM'13}, Apr. 2013.

\bibitem{xue2010naturalistic}
Y.~Xue and L.~Jin.
\newblock A naturalistic {3D} acceleration-based activity dataset and benchmark
  evaluations.
\newblock In {\em Proc. IEEE SMC'10}, Oct. 2010.

\bibitem{YS12}
J.~Yang and S.~Ulukus.
\newblock Optimal packet scheduling in an energy harvesting communication
  system.
\newblock {\em IEEE Trans. Commun.}, 60(1):220--230, Jan. 2012.

\bibitem{Yang2012power}
Z.~Yang, E.~Halvorsen, and T.~Dong.
\newblock Power generation from conductive droplet sliding on electret film.
\newblock {\em Appl. Phys. Lett.}, 100(21):213905, 2012.

\bibitem{yerva2012grafting}
L.~Yerva, B.~Campbell, A.~Bansal, T.~Schmid, and P.~Dutta.
\newblock Grafting energy-harvesting leaves onto the sensornet tree.
\newblock In {\em Proc. IEEE IPSN'12}, Apr. 2012.

\bibitem{yun2011design}
J.~Yun, S.~Patel, M.~Reynolds, and G.~Abowd.
\newblock Design and performance of an optimal inertial power harvester for
  human-powered devices.
\newblock {\em IEEE Trans. Mobile Comput.}, 10(5):669 --683, May 2011.

\bibitem{zafer2009calculus}
M.~Zafer and E.~Modiano.
\newblock A calculus approach to energy-efficient data transmission with
  quality-of-service constraints.
\newblock {\em IEEE/ACM Trans. Netw.}, 17(3):898--911, 2009.

\bibitem{ting-mobisys09}
T.~Zhu, Z.~Zhong, Y.~Gu, T.~He, and Z.-L. Zhang.
\newblock Leakage-aware energy synchronization for wireless sensor networks.
\newblock In {\em Proc. ACM Mobi{S}ys'09}, June 2009.

\end{thebibliography}
}

\ifJSACsingle
\end{spacing}
\fi

\ifJSACsingle
\else
\newpage
\newpage
\fi
\setcounter{section}{0}
\setcounter{theorem}{0}
\setcounter{table}{0}
\normalsize

\renewcommand\thetable{\Alph{table}}

\ifJSACsingle
\vspace{-2.5pt}
\fi

\ifIncludeAppendix
\section*{APPENDIX I}

\ifJSACsingle
\noindent
{\bf Proof of Theorem~\ref{th:NP-hard}: }
\else
\subsection*{Proof of Theorem~\ref{th:NP-hard}}
\fi
%
We prove that the EA Problem is NP-hard using a reduction from a well-known NP-hard 
problem~\cite{Book:Kellerer04}. The reduction performs several transformations, all of which are polynomial in time and space. We start with two definitions:

\begin{definition}
 An instance of the EA Problem is defined using the integers $K \geq 0$, $C \geq 0$, $B_0\geq 0$, and $B_K\geq 0$, the set $\mathcal{S}$, the functions $Q()$, $\eta()$, $U()$, and $L()$, and the value of $e(i)$ for every
  slot $i=0,\ldots, K-1$.
\end{definition}

\begin{definition}
  Given an instance of the EA Problem, a vector $s(i)$, $i=0,\ldots, K-1$ is 
  feasible if constraints~\eqref{eq:EA-1}-\eqref{eq:EA-3} hold with respect to
  it.
\end{definition}

 The {\em decision version} of the EA Problem (EA-D) is defined using the
  same values as those defining the EA Problem, as well as an additional
  integer $U\geq 0$. A solution to the EA-D Problem is a ``yes'' or ``no''
  answer, where ``yes'' is returned if and only if there is a feasible vector
  $s(i)$ with $\sum_{i=0}^{K-1} U(s(i)) \geq U$. It is easy to see that given
  a polynomial-time solver to EA-D, one can solve the EA Problem using
  binary search on the values of $U$. Therefore, in order to prove that the EA
  Problem is NP-hard, 
\ifJSACsingle
we show
\else
it is sufficient to show 
\fi
that the EA-D Problem is
  NP-hard.

  We show a polynomial time reduction from the decision form of subset sum
  Problem (SSP-D), which is known to be an NP-hard
  Problem~\cite{Book:Kellerer04}. The SSP-D Problem is defined as
  follows: 
\ifJSACsingle
  $\text{SSP-D}(w,c) = \left\{
    \exists x \text{ such that:}
    \sum_{j=1}^n w_jx_j=c;\ 
    x_j\in \{0,1\}\ \forall j, \right.$
\else
\begin{equation} \notag
  \text{SSP-D}(w,c) = \left\{
  \begin{array}{l}
    \exists x \text{ such that:}\\
    \sum_{j=1}^n w_jx_j=c;\ 
    x_j\in \{0,1\}\ \forall j,
  \end{array}
\right.
\end{equation}
\fi
where $w=(w_1,\ldots,w_n)$ is a vector of size $n$. We assume that $c$ and all
coefficients $w_j$ are integers.

It is clear that in any solution to SSP-D, the inequality $\sum_{j=1}^nx_j
\leq n$ holds. Therefore, we can add this as an additional constraint to
SSP-D. We also introduce slack variables $y_j$ and obtain the following
formulation equivalent to SSP-D, denoted SSP-D$_1$:
\ifJSACsingle
$  \text{SSP-D}_1(w,c) = \left\{
    \exists x \text{ such that:}\\
    \sum_{j=1}^n w_jx_j=c,\ 
    \sum_{j=1}^nx_j+y_j\leq n\\
    x_j+y_j=1,\ 
    x_j,y_j\in \mathbb{N}_0 \  \forall j.
\right.$
\else
\begin{equation} \notag
  \text{SSP-D}_1(w,c) = \left\{
\begin{array}{l}
    \exists x \text{ such that:}\\
    \sum_{j=1}^n w_jx_j=c,\ 
    \sum_{j=1}^nx_j+y_j\leq n\\
    x_j+y_j=1,\ 
    x_j,y_j\in \mathbb{N}_0 \  \forall j.
  \end{array}
\right.
\end{equation}
\fi

We now follow the same technique as used in~\cite{Book:Kellerer04} to
merge the equation $x_1+y_1=1$ with the equation  $\sum_{j=1}^n w_jx_j=c$,
obtaining the new equation $x_1+y_1+2\sum_{j=1}^n w_jx_j=2c+1$. As shown
in~\cite{Book:Kellerer04}, this does not change the set of feasible
solutions. Repeating the process of merging with $x_j+y_j=1$ for
$j=2,\ldots,n$, we get the following formulation, denoted SSP-D$_2$:
\ifJSACsingle
$ \text{SSP-D}_2(w,c) = \left\{
  \exists x,y \text{ s.t.: }
      2^n \sum_{j=1}^n 2^{-j}(y_j + x_j) +    w_jx_j=2^nc+2^n-1
    \sum_{j=1}^nx_j+y_j\leq n;\ 
    x_j,y_j\in \mathbb{N}_0\ \forall j.
\right.$
\else
\begin{equation} \notag
  \text{SSP-D}_2(w,c) = \left\{
\begin{array}{l}
    \exists x,y \text{ such that:}\\
    2^n \sum_{j=1}^n 2^{-j}(y_j + x_j) + \\
     \ \ \ \ \ \ w_jx_j=2^nc+2^n-1\\
    \sum_{j=1}^nx_j+y_j\leq n;\ 
    x_j,y_j\in \mathbb{N}_0\ \forall j.
  \end{array}
\right.
\end{equation}
\fi

Setting $\tilde{w}_j=2^{n-j}+2^nw_j$, $\overline{w}_j=2^{n-j}$, and
$\overline{c}=2^nc+2^n-1$, we reach the equivalent formulation, denoted
SSP-D$_3$:
\ifJSACsingle
$ \text{SSP-D}_3(\tilde{w},\overline{w},c) = \left\{
    \exists x,y \text{ s.t.: }
    \sum_{j=1}^n \tilde{w}_jx_j + \overline{w}_jy_j=\overline{c},
    \sum_{j=1}^nx_j+y_j\leq n\ ;\right.$ 
    $x_j,y_j\in \mathbb{N}_0 \ \forall j.$
\else
\begin{equation} \notag
  \text{SSP-D}_3(\tilde{w},\overline{w},c) = \left\{
\begin{array}{l}
    \exists x,y \text{ such that:}\\
    \sum_{j=1}^n \tilde{w}_jx_j + \overline{w}_jy_j=\overline{c},\\
    \sum_{j=1}^nx_j+y_j\leq n;\ 
    x_j,y_j\in \mathbb{N}_0 \ \forall j.
  \end{array}
\right.
\end{equation}
\fi
Let $n_b(w,c)$ be the number of bits required to represent $(w,c)$.  It is shown
in~\cite{Book:Kellerer04} that the new coefficients $\tilde{w}_j$,
$\overline{w}_j$, and $\overline{c}$, are polynomial in $n_b(w,c)$. Therefore,
the transformation can be performed in polynomial time.

We now show how to reduce SSP-D$_3$ into an instance of EA-D, which will
complete the proof. As input for EA-D we set $B_0=B_K=0$, $K=n+1$,
$C=U=e(0)=\overline{c}$, $\mathcal{S}= \{\tilde{w}_j\} \cup
\{\overline{w}_j\}$; $L(i)=0$, $\eta(i)=1$ $\forall i$; and $e(i)=0 \forall
i\geq 1$. We set $U()$ and $Q()$ as the identity function: $U(x)=x$ and $Q(x)=x$.  Clearly, generating
this input can be performed in polynomial time.

We now show that the reduction holds, namely, that the generated EA-D is a
``yes'' instance if and only if SSP-D$_3$ is a ``yes'' instance. Note that
since $B_0=0$, we get $s(0)=0$. In addition,  $\forall i\geq 1\ Q(i)=0$, $B(1)=e(0)=\overline{c}$, and $B_K=0$. Therefore, the considered EA-D
instance is a ``yes'' instance if and only if 
\ifJSACsingle
$\exists s(i)$
\else
there exist $s(i)$
\fi 
such that
$\sum_{i=1}^ns(i)\leq \overline{c}$ and $\sum_{i=1}^n U(s(i))=\sum_{i=1}^n
s(i)=\overline{c}$.

If the SSP-D$_3$ is a ``yes'' instance, there exist $x_j, y_j$ such that
$\sum_{j=1}^nx_j+y_j\leq n$. A feasible vector $s(i)$
for EA-D can be obtained as follows: for $j=1,\ldots,n$, use $x_j$ slots
by spending $\tilde{w}_j$ amount of energy in each such slot and use $y_j$
slots by spending $\overline{w}_j$ amount of energy in each such
slot. Clearly, such energy spending is feasible and obtains the total utility of
$U=\overline{c}$. Therefore, the EA-D is a ``yes'' instance. The other
direction, namely, that if the EA-D instance is a ``yes'' instance, the
SSP-D$_3$ instance is a ``yes'' instance, can be proved in a similar way. $\Box$

\ifJSACsingle
\noindent
{\bf Proof of Theorem~\ref{th:FPTAS}: }
\else
\subsection*{Proof of Theorem~\ref{th:FPTAS}}
\fi
The total profit of the solution returned by the algorithm is
$\sum_{\tilde{s}(i)} U(\tilde{s}(i))$, and, due to the definition of
$\tilde{U}()$:
\ifJSACsingle
$  \sum_{i=0}^{K-1} U(\tilde{s}(i)) \geq \sum_{i=0}^{K-1} \mu \cdot \tilde{U}(\tilde{s}(i)).$
\else
\begin{equation} \notag
  \sum_{i=0}^{K-1} U(\tilde{s}(i)) \geq \sum_{i=0}^{K-1} \mu \cdot \tilde{U}(\tilde{s}(i)).
\end{equation}
\fi
Since the dynamic programming returns the optimal solution with respect to
$\tilde{U}()$,
\ifJSACsingle
$\mu \sum_{i=0}^{K-1} \tilde{U}(\tilde{s}(i))  \geq \mu \sum_{i=0}^{K-1} \tilde{U}(s^*(i))  \geq \sum_{i=0}^{K-1}
  \mu  \left (\frac {U(s^*(i))} {\mu} - 1 \right )  \geq U^*-K\cdot \mu$.
\else
\begin{align}
\mu \sum_{i=0}^{K-1} \tilde{U}(\tilde{s}(i)) & \geq \mu \sum_{i=0}^{K-1} \tilde{U}(s^*(i))  \geq \sum_{i=0}^{K-1}
  \mu  \left (\frac {U(s^*(i))} {\mu} - 1 \right ) \notag \\
  \sum_{i=0}^{K-1}
  \mu \cdot & \left (\frac {U(s^*(i))} {\mu} - 1 \right ) \geq U^*-K\cdot \mu \notag
\end{align}
\fi
Since $\mu=\frac {\epsilon \cdot U(s_{\max})} {K}$ and $U^L=U(s_{\max})$,
using the above equations, we get $ \notag
 \sum_{i=0}^{K-1} U(\tilde{s}(i)) \geq (1-\epsilon) U^*,
$
which proves the approximation ratio.

Due to the invocations the dynamic programming with utility function $\tilde{U}()$,
the space and time complexities are $O(K^2\cdot \tilde{U}(s_{\max}))$ and
$O(|\mathcal{S}|\cdot K^2\cdot \tilde{U}(s_{\max}))$, respectively.
Replacing $\tilde{U}(s_{\max})$ with $\frac{U(s_{\max})} {\mu}$, we obtain the
space and time complexities 
\ifJSACsingle
\else
of 
\fi
$O(\frac {K^3} {\epsilon})$ and
$O(|\mathcal{S}| \cdot \frac {K^3} {\epsilon})$, respectively. $\Box$



\ifJSACsingle
\noindent
{\bf Proof of Theorem~\ref{th:GreedyBattery}: }
\else
\subsection*{Proof of Theorem~\ref{th:GreedyBattery}}
\fi
We first make the following observation~\cite{merrett2008empirical,ting-mobisys09}:
\begin{observation} \label{ob:leakage}
Let $i_1$ and $i_2$ be two slots. If $B(i_1)\geq B(i_2)$, then $L(i_1)\geq L(i_2)$.
\end{observation} 

Since condition (i) holds, $\sum_{i=0}^{K-1} U(s(i)) =
U(\sum_{i=0}^{K-1} s(i))$ and the total energy spent is $\eta_h
\sum_{i=0}^{K-1} s(i)$. Therefore, maximizing the utility is equivalent to
maximizing the total
\ifJSACsingle
energy.
\else
energy spent over the $K$ slots.
\fi

To complete the proof we now show a transformation from an optimal solution
$s^*(i)$ to the greedy algorithm's solution $s^g(i)$, which does not decrease
the total amount of energy spent over the $K$ slots.
\ifJSACsingle
\else

\fi
Let $i'$ be the earliest slot for which $s^*(i') \neq s^g(i')$, clearly
$s^*(i')<s^g(i')$. Since $s^*(i)$ obtains maximal energy spending, there
must be a set $S'$ of slots after slot $i'$ in which the total energy spent
is at least $s^g(i')-s^*(i')$. Also note that, due to condition
(ii), for some $j>0$, $s^g(i')-s^*(i')=j\cdot s$. In each of the slots in $S'$
the energy spent is a multiple of $s$. Therefore, we can reduce the amount of
energy spent in $S'$ by $s^g(i')-s^*(i')$ and set $s^*(i')=s^g(i')$. Due to
Observation~\ref{ob:leakage} we get
a feasible energy spending. Furthermore, at least the $i'+1$ first slots are identical
the greedy algorithm's solution. We repeat the process until we obtain the
energy spending $s^g(i)$ for $i=0,\ldots, K-1$. $\Box$

\ifJSACsingle
\noindent
{\bf Proof of Theorem~\ref{th:NegativeOnline}: }
\else
\subsection*{Proof of Theorem~\ref{th:NegativeOnline}}
\fi
We set $U(s(i))=s(i)$, $\eta=1$, $L(i)=0$, $C=s_{\min}$ where
$s_{\min}=\min\{s\in \mathcal{S}\}$, and $B_0=B_K=C$.  It is sufficient to
consider instances in which $e(i)>0$ only for $i \geq K-2$. Therefore, without
loss of generality we assume $K=2$.

Assume that $e(0)=0$. The online algorithm can: (i)
set $s(0)=0$, or (ii) set $s(0)=s_{\min}$. If the first option is used and
$e(1)<s_{\min}$, the solution is infeasible. Thus, to ensure feasibility the
online algorithm will set $s(0)=0$ and similarly $s(1)=0$, obtaining
no utility. Therefore, the performance gap $s_{\min}$ can be arbitrarily
large. $\Box$





\fi

\end{document}